\documentclass[binding=0.6cm,PhD]{sapthesis}

\usepackage[final]{pdfpages}
\usepackage{microtype}
\usepackage{float}
\floatstyle{ruled}
\newfloat{algorithm}{tbp}{loa}
\floatname{algorithm}{Routine}
\usepackage{hyperref}

\hypersetup{colorlinks=true,
	linkcolor=blue,
	anchorcolor=blue,
	citecolor=red,
	urlcolor=brown,
	pdftitle={The Renormalization Group for Disordered Systems},
	pdfauthor={Michele Castellana}}

\usepackage{epstopdf}
\usepackage{dcolumn}   
\usepackage{bm}        
\usepackage{amssymb}   
\usepackage{feynmp}
\usepackage{feyn}
\usepackage{yfonts}
\usepackage{verbatim}
\usepackage{mathrsfs}
\usepackage{latexsym} 
\usepackage{bbm}
\usepackage{amsmath}

\usepackage{amssymb}
\usepackage{eufrak}
\usepackage{pstricks}
\usepackage{color}

\providecommand{\tabularnewline}{\\}
\providecommand{\be}{\begin{equation}}
\providecommand{\ee}{\end{equation}}
\providecommand{\bea}{\begin{eqnarray}}
\providecommand{\eea}{\end{eqnarray}}
\providecommand{\beas}{\begin{eqnarray*}}
\providecommand{\eeas}{\end{eqnarray*}}

\providecommand{\beni}{\begin{equation*}}
\providecommand{\eeni}{\end{equation*}}

\providecommand{ }{\begin{widetext}}
\providecommand{ }{\end{widetext}}
 \providecommand{\no}{\nonumber}

\title{The Renormalization Group \\for Disordered Systems}
\author{Michele Castellana}

\IDnumber{694948}
\course[Physics]{Fisica} 
\courseorganizer{
Thesis codirected between Dipartimento di Fisica,  Universit\`a  La Sapienza, Rome, Italy and Laboratoire de Physique Th\'eorique et Mod\`eles Statistiques, Universit\'e Paris Sud, Orsay, France. \vspace{1cm}\\
Scuola Dottorale in Scienze Astronomiche, Chimiche, Fisiche, Matematiche e della Terra ``Vito Volterra''.\\
\'Ecole Doctorale ED517 ``Particules, Noyaux et Cosmos''.}
\cycle{XXIII }
\submitdate{October 2010}
\copyyear{2011}
\advisor{Professor Giorgio Parisi}
\advisor{Professor Marc M\'ezard}
\authoremail{michele.castellana@gmail.com}

\examdate{January 31, 2012}
\examiner{Professor Federico Ricci-Tersenghi}
\examiner{Professor Alain Billoire}
\examiner{Professor Marc M\'ezard}
\examiner{Professor Giorgio Parisi}
\examiner{Professor Emmanuel Trizac}
\examiner{Professor Francesco Zamponi}

\versiondate{\today}

\begin{document}

\frontmatter
\maketitle

\dedication{a Laureen} 

\begin{abstract} 
In this thesis we investigate  the Renormalization Group (RG) approach in finite-dimensional glassy systems, whose critical features are still not well-established, or simply unknown.  We focus on spin and structural-glass models built on hierarchical lattices, which are the simplest non-mean-field systems where the RG framework emerges in a natural way. The resulting critical properties shed light on the critical behavior of spin and structural glasses beyond mean field, and suggest future directions for understanding the criticality of  more realistic glassy systems. 
\end{abstract}

\begin{acknowledgments}[Acknowledgments]

Thanks to all those who supported me and  stirred up my enthusiasm and my willingness to work. \\

I am glad to thank my italian thesis advisor Giorgio Parisi, for sharing with me his experience and scientific knowledge, and especially for  teaching  me a scientific frame of mind which is characteristic of a well-rounded scientist. I am glad to thank my french thesis advisor Marc M\'ezard, for showing  me how important  scientific open-mindedness is, and in particular for teaching  me the extremely precious skill of tackling scientific problems by considering only their fundamental features first, and of separating them from details and technicalities which would avoid an overall view.    \\

Thanks to my family for supporting me during this thesis and for pushing me to pursue my own aspirations, even though this implied that I would be far from home. \\

My heartfelt thanks to Laureen for having been constantly by my side with mildness and wisdom, and for having constantly pushed me to pursue my ideas, and  to keep the optimism about the future and the good things that it might bring. \\

Finally, I would like to thank two people who closely followed me in my personal and scientific development over these three years. Thanks to my friend Petr \v{S}ulc for the good time we spent together, and in particular for our sharing of that dreamy way of doing Physics that is the only one that can keep passion and imagination alive. Thanks to Elia Zarinelli, collaborator and friend, for sharing the experience of going abroad, leaving the past behind, and looking at it with new eyes. Thanks to him also  for the completely free and pleasing environment where our scientific collaboration took place, and which should lie  at the bottom of any scientific research.

\end{acknowledgments}

\tableofcontents

\mainmatter

\part{Introduction}\label{introduction}

\chapter{Historical outline}\label{historical_outline}

Paraphrasing P. W. Anderson \cite{anderson1995through}, ``\textit{the deepest and most interesting unsolved problem in solid state theory is probably the 
nature of glass and the glass transition}''. Indeed, the complex and rich behavior of simplified models for real, physical glassy systems has interested theoreticians for its challenging complexity and difficulty, and opened new avenues in a large number of other problems such as computational optimization and neural networks.\\

When speaking of glassy systems, one can distinguish between two physically different classes of systems: spin glasses and structural glasses. \\

Spin glasses have been originally \cite{edwards1975theory} introduced  as  models to study disordered uniaxial magnetic materials, like a dilute solution of, say, $\text{Mn}$ in $\text{ Cu}$,  modeled by an array of spins on the $\text{Mn}$ arranged at random in the matrix of $\text{Cu}$, interacting with a potential which oscillates as a function of the separation of the spins. Typical examples of spin-glass systems are $\textrm{FeMnTiO}_3$ \cite{ito1986time,gunnarson1991static,ItoArugakatori94,bert2004spin}, $(\text{H}_3\text{O})\text{Fe}_3(\text{SO}_4)_2(\text{OH})_6$ \cite{dupuis2002aging}, $\text{CdCr}_{1.7}\text{In}_{0.3}\text{S}_4$ \cite{jonason1998memory,vincent2000aging}, $\text{Eu}_{0.5}\text{Ba}_{0.5}\text{MnO}_{3}$ \cite{nair2007critical} and several others. \\

\begin{centering}
  \begin{figure}[htb] 
\centering
\includegraphics[width=10cm]{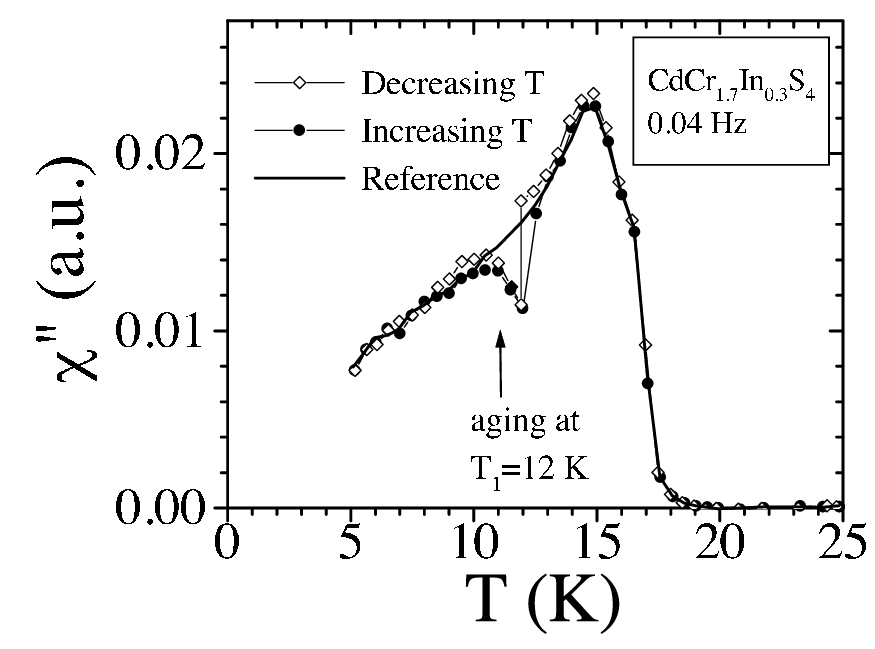}
\caption{Out out phase susceptibility $\chi''$ of $\text{CdCr}_{1.7}\text{In}_{0.3}\text{S}_4$ as a function of the temperature. The solid curve is the reference curve. The open diamonds-curve is obtained by cooling the system and stopping the cooling process at $T_1 = 7K$ for seven hours. The solid circles-curve is obtained upon re-heating the system after the above cooling process. Data is taken from \cite{jonason1998memory}. }
\label{fig18}
\end{figure}
  \end{centering}

Spin glasses exhibit a very rich phenomenology. Firstly, the very first magnetization measurements  of $\textrm{FeMnTiO}_3$ in a magnetic field showed \cite{ito1986time} the existence of  a  cusp in the susceptibility as a function of the temperature. Occurring at a finite temperature $T_{sg}$, this experimental observation is customary interpreted as the existence of a phase transition.\\
  Later on, further experimental works confirmed this picture \cite{gunnarson1991static}, and revealed some very rich and interesting features of the low-temperature phase: the chaos and memory effect. 
Consider a sample of $\text{CdCr}_{1.7}\text{In}_{0.3}\text{S}_4$ in a low-frequency magnetic field \cite{jonason1998memory}. The system is cooled from above $T_{sg} = 16.7 K$ down to $5 K$, and is then heated back with slow temperature variations. The curve for the out-of-phase susceptibility $\chi''$ as a function of the temperature obtained upon reheating will be called the reference curve, and is depicted in Fig. \ref{fig18}.

One repeats the cooling experiment but stops it at $T_1 = 12 K$. Keeping the system at $T_1$, one waits $7$ hours. In this lapse of time $\chi''$ relaxes downwards, i. e. the system undergoes an aging process. When the cooling process is restarted, $\chi''$ merges back with the reference curve just after a few Kelvins. This immediate merging back is the chaos phenomenon: aging at $T_1$ does not affect the dynamics of the system at lower temperatures. From a microscopic viewpoint, the aging process  brings the system at an equilibrium configuration at $T_1$. When cooling is restarted, such an equilibrium configuration behaves as a completely random configuration at lower temperatures, because the susceptibility curve immediately merges the reference curve. The effective randomness of the final aging configuration reveals a chaotic nature of  the free-energy landscape. \\

The memory effect is even more striking. When the system is reheated at a constant rate, the susceptibility curve retraces the curve of the previous stop at $T_1$. This is quite puzzling, because even if the configuration after aging at $T_1$ behaves as a random configuration at lower-temperatures, the memory of the aging at $T_1$ is not erased. \\

Such a  rich phenomenology challenged the theoreticians for decades. The theoretical description of such models, even in the mean-field approximation, revealed a complex structure of the low-temperature phase that could be responsible for such a rich phenomenology. Still, such a complex structure has been shown to be correct only in the mean-field approximation, and the physical features of the low-temperature phase beyond mean field are still far from being understood. \vspace{1cm}\\

\begin{centering}
\begin{figure}[htb] 
\centering
\includegraphics[width=10cm]{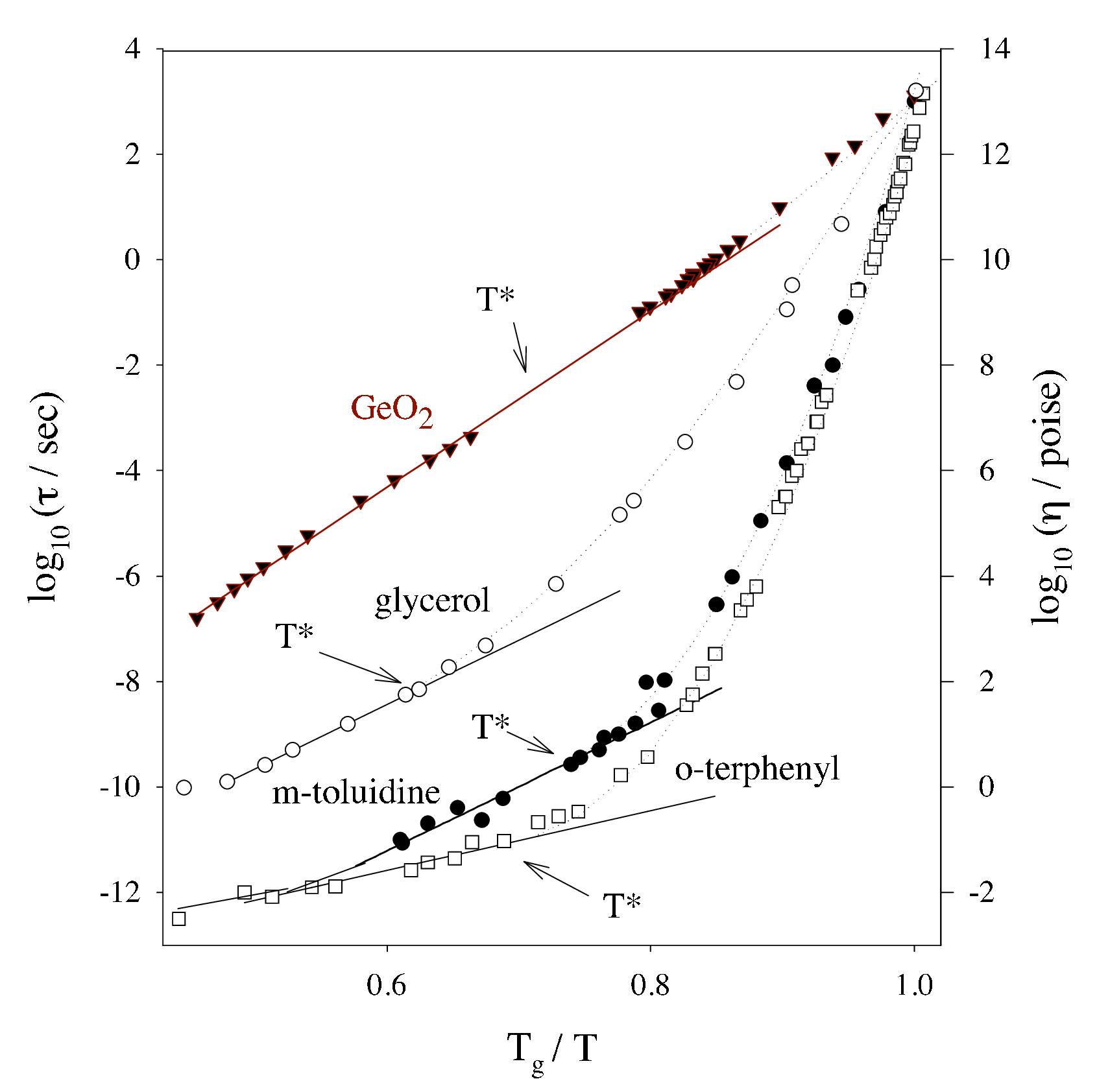}
\caption{Base-$10$ logarithm of the relaxation time $\tau$ and of the viscosity $\eta$ as a function of the logarithm of the inverse reduced temperature $T_g/T$, where $T_g$ is the glass-transition temperature, for several glass formers. Data is taken from \cite{tarjus2001viscous}.  }
\label{fig19}
\end{figure}
\end{centering}

Structural glasses, also known as glass-forming liquids or glass-formers, are liquids that have been cooled fast enough to avoid crystallization \cite{biroli2009random,tarjus2001viscous}. 
When cooling  a sample of $\textrm{o}-\textrm{Terphenyl}$  \cite{laughlin1972viscous}, or $\textrm{Glycerol}$  \cite{menon1992wide}, the viscosity $\eta$ or the relaxation time $\tau$ can change of fifteen order of magnitude when decreasing the temperature of a factor two, as shown in Fig. \ref{fig19}. 

This striking increase of the viscosity can be interpreted in terms of a particle jamming process, and suggests that a  phase transition occurs at a finite temperature $T_g$. 
The physical features of this transition are strikingly more complex than the ordinary first order transitions yielding a crystal as the low-temperature state. 
 
Indeed, crystals break the translational symmetry at low temperatures, the particles being arranged on a periodic structure. Ergodicity is broken as well, because the only accessible microscopic configuration of the particles is the crystal. The sharp increase in the viscosity of a glass below the glass-transition point yields also evidence of ergodicity breaking: elementary particle moves become extremely slow, and energetically expensive, in such a way that the system is stuck in a mechanically-stable state. Differently from the crystal, this state has the same symmetry properties as the liquid: no evident symmetry breaking occurs, and there is no static order parameter to signal the transition. 
Moreover, at $T_g$ the excess entropy $S_e$ of the glass over the crystal is remarkably high, suggesting that there is  a big degeneracy of mechanically-stable states a glass can get stuck in below the transition point. 

Once the system is frozen in one of these exponentially many configurations, there is no way to keep it equilibrated below $T_g$. Accordingly, the equilibrium properties in the whole temperature phase cannot be investigated in experiments.  Still, interesting properties of the low-temperature phase result from the pioneering works of Kauzmann \cite{kauzmann1948nature}, 
who first realized that if the excess entropy of a glass former \cite{richert1998dynamics} is extrapolated from above $T_g$ down in the low-temperature phase, there is a finite temperature $T_K$, the Kauzmann temperature, where this vanishes. This is rather startling because, if the  geometry of the crystal is not too different from that of the liquid, one expects the entropy of the liquid to be always larger than that of the crystal. There have been countless speculations on the solution of this paradox \cite{kauzmann1948nature,biroli2009random,tarjus2001viscous}, and the existence of a Kauzmann temperature  in a real glass-former is nowadays a still hotly-debated and untamed problem  from both an experimental and theoretical viewpoint.\\

Despite the triking difference between these two kinds of systems
spin glasses and structural glasses have some deep common features. Indeed, according to a wide part of the community, spin-glass models with quenched disorder are good candidates to mimic the dynamically-induced disorder of  glass-forming liquids \cite{biroli2009random}, even if some people are still critical about this issue \cite{langer2007mysterious}. There are several points supporting the latter statement. For instance, it has been shown that hard particle lattice models \cite{biroli2001lattice} describing the phenomenology of structural glasses, display the phenomenology of spin systems with quenched disorder like spin glasses. Accordingly, there seems to be an underlying universality between the dynamically-induced disorder of glass-formers and the quenched disorder of spin glasses, in such a way that the theoretical description of spin glasses and that of structural glasses shared an important interplay in the last decades. More precisely, in the early $80$'s the solution of mean-field versions of spin \cite{parisi1980order} and structural \cite{derrida1980random} glasses were developed, and new interesting features of the low temperature phase were discovered. Since then, a huge amount of efforts has been done to develop a theoretical description of real, non-mean-field spin and structural glasses. A contribution in this direction through the implementation of the Renormalization Group (RG) method would hopefully shed light on the critical behavior of such systems. \\

Before discussing how the RG framework could shed light on  the physics of finite-dimensional spin and structural glasses, we give a short outline of the mean-field theory of spin and structural glasses, and on the efforts that have been done to clarify their non-mean-field regime.  

\subsubsection*{The Sherrington-Kirkpatrick model}

The very first spin-glass model, the Edwards-Anderson (EA) model, was introduced in the middle $70$'s \cite{edwards1975theory} as a model describing disordered uniaxial magnetic materials. Later on, Sherrington and Kirkpatrick (SK) \cite{SherringtonKirkpatrick75} introduced a mean-field version of the EA model, which is defined as a system of $N$ spins $S_i = \pm 1$ with Hamiltonian
\be \label{h_sk}
H[\vec S] = - \sum_{i>j=1}^N J_{ij} S_i S_j,
\ee
with $J_{ij}$ independent random variables distributed according to a Gaussian distribution with zero mean and variance $1/N$. 

 The model can be solved  with the replica method \cite{MPV}: given $n$ replicas $\vec S_1, \vec S_2, \ldots, \vec S_n$ of the system's spins, the order parameter is the $n\times n$ matrix $Q_{ab} \equiv 1/N \sum_{i=1}^NS_{a,i} S_{b,i} $ representing the overlap between replica $a $ and replica $b$. The free energy is computed as an integral over the order parameter, and thermodynamic quantities are calculated with the saddle-point approximation, which is exact in the thermodynamic limit. 
SK first proposed a solution for the saddle point $Q^\ast_{ab}$, which was later found to be inconsistent, since it yields a negative entropy at low temperatures. This solution is called the replica-symmetric (RS) solution, because the matrix $Q^\ast_{ab}$ has a uniform structure, and there is no way to discern between two distinct replicas. 
Some mathematically non-rigorous aspects of the replica approach had been blamed \cite{vanhemmen1979replica} to explain the negative value of the entropy at low-temperatures. Amongst these issues, there is the continuation of the replica index $n$ from integer to non-integer values, and the exchange of the $n \rightarrow 0 $-limit with the thermodynamic limit $N\rightarrow \infty$. Still, no alternative approach was found to avoid these issues. \\

In the late $70$'s Parisi started investigating more complicated saddle points. In the very first work \cite{parisi1979toward}, an approximate saddle point was found, yielding a still negative but small value of the entropy at low temperatures. The solution was called replica-symmetry-broken solution, because $Q^\ast_{ab}$ was no more uniform, but presented a block structure. Notwithstanding the negative values of the entropy at low temperatures, the solution was encouraging, since it showed a good agreement with Monte Carlo (MC) simulations \cite{kirkpatrick1978infinite}, whereas the replica-symmetric solution showed a clear disagreement with MC data. Later on, better approximation schemes for the saddle point were considered \cite{parisi1980sequence}, where the matrix $Q^\ast_{ab}$ was given by a hierarchical structure of blocks, blocks into blocks, and so on. The step of this hierarchy is called the replica-symmetry-breaking (RSB) step $K$. The final result of such works was presented in the papers of 1979 and 1980 \cite{parisi1979infinite,parisi1980order}, where the full-RSB ($K = \infty$) solution was presented. According to this solution, the saddle point $Q^\ast_{ab}$ is uniquely determined in terms of a function $q(x)$ in the interval $0 \leq x \leq  1$, being the order parameter of the system.
Parisi's solution resulted from a highly nontrivial ansatz for the saddle point  $Q^\ast_{ab}$, and there was no proof of its exactness.  Still, the entropy of the system resulting from Parisi's solution is always non-negative, and vanishes only at zero temperature, and the quantitative results for thermodynamic quantities such as the internal energy showed a good agreement \cite{parisi1979infinite} with the Thouless-Almeida-Palmer (TAP) solution \cite{thouless1977solution} at low temperatures. These facts were rather encouraging, and gave a strong indication that Parisi's approach gave a significant improvement over the original solution by SK.\\

Still, the physical interpretation of the order parameter stayed unclear until 1983 \cite{parisi1983order}, when it was shown that the function $q(x)$ resulting from the baffling mathematics of Parisi's solution is related to the probability distribution $P(q)$  of the overlap $q$ between two real, physical copies of the system, through the relation $x(q) = \int_{-\infty}^q dq' P(q')$. Accordingly, in the high temperature phase  the order parameter $q(x)$ has a trivial form, resulting in a $P(q) = \delta(q)$, while in the low-temperature phase the  nontrivial form of $q(x)$ predicted by Parisi's solution implies a nontrivial structure of the function $P(q)$. In particular, the smooth form of $P(q)$ implies the existence of many pure states. \\

Further investigations in 1984 \cite{mezard1984replica} and 1985 \cite{mezard1985microstructure} gave a clear insight into the way these pure states are organized: below the critical temperature the phase space is fragmented into several ergodic components, and each component is also fragmented into sub-components, and so on. The free-energy landscape could be qualitatively represented as an ensemble of valleys, valleys inside the valleys, and so on. 
Spin configurations can be imagined as the leaves of a hierarchical tree \cite{MPV}, and the  distance between  two of them is measured in terms of number of levels $k$ one has to go up in the tree to find a common root to the two leaves. To each hierarchical level $k$ of the tree one associates a value of the overlap $q_k$, where the  set of possible values $q_k$ of the overlap is encoded into the function $q(x)$ of Parisi's solution.\\

Parisi's solution was later rederived  with an independent method in 1986 by M\'ezard et al \cite{mezard1986sk}, who reobtained  the full-RSB solution  starting from simple physical grounds, and presented it in a more compact form. \\

Finally, the proof of the exactness of Parisi's solution came in 2006 by Talagrand \cite{talagrand2006parisi}, whose results are based on previous works by Guerra  \cite{guerra2003broken},  and who showed with a rigorous formulation that the full-RSB ansatz provides the exact solution of the problem. \\ 

 This ensemble of works clarified the nature of the spin-glass phase in the mean-field case. According to its clear physical interpretation, the RSB mechanism of Parisi's solution became a general framework to deal with systems with a large number of quasi-degenerate states. In particular, in 2002  the RSB mechanism was applied in the domain of constraint satisfaction problems \cite{mezard2002analytic,mezard2002random,monasson1997statistical,biroli2000variational}, showing the existence of a new replica-symmetry broken phase in the satisfiable region which was unknown before then. \\

Despite the striking success in describing mean-field spin glasses, it is not clear whether the RSB scheme is correct also beyond mean field. Amongst the other scenarios describing the low-temperature phase of non-mean-field spin glasses, the droplet picture has been developed in the middle $80$'s by Bray, Moore, Fisher, Huse and McMilllan \cite{mcmillan1984scaling,fisher1986ordered,fisher1987pure,fisher1987absence,fisher1988equilibrium,fisher1988nonequilibrium,bray1986heidelberg}. According to this framework, in the whole low-temperature phase there  is only one ergodic component and its spin reversed counterpart, as in a ferromagnet. Differently from a ferromagnet, in a finite-dimensional spin glass spins arrange in a random way determined by the interplay between quenched disorder and temperature.  \\

On the one hand, there have been several efforts to understand the striking phenomenological features of three-dimensional systems in terms of the RSB \cite{joh1996spin}, the droplet or alternative pictures \cite{jonason1998memory}. Still, none of these was convincing enough for one of these pictures to be widely accepted by the scientific community as the correct framework to describe finite-dimensional systems. \\

On the other hand, there is no analytical framework describing non-mean-field spin glasses. Perturbative expansions around Parisi's solution have been widely investigated by De Dominicis and Kondor \cite{dedominicis1984on,6}, but proved to be difficult and non-predictive.
Similarly, several efforts have been done in the implementation in non-mean-field spin glasses of a perturbative field-theory approach based on the replica method \cite{harris1976critical,kotliar1983one,chen1977mean}, but they  turned out to be non-predictive, because nonperturbative effects are completely untamed. Amongst the possible underlying reasons, there is the fact that such field-theory approaches are all based on a $\phi^3$-theory, whose upper critical dimension is $d_c=6$. Accordingly, a predictive description of physical three-dimensional  systems would require an expansion in $\epsilon = d_c - 3 = 3$, which can be quantitatively predictive only if a huge number of terms of the $\epsilon$-series were known  \cite{zinnjustin}. Finally, high-temperature expansions for the free energy \cite{plefka1982convergence} turn out to be badly behaved in three dimensions \cite{daboul2004test}, and non-predictive. \\

Since analytical approaches do not give a clear answer on such finite-dimensional systems, most of the knowledge comes from MC simulations, which started with the first pioneering works from Ogielsky \cite{ogielsky1985dynamics}, and were then intensively carried on during the $90$'s and $00$'s \cite{bhatt1988numerical,marinari1993on,marinari1998phase,leuzzi1999critical,mari1999ising,palassini1999universal,krzakala2000spin,ballesteros2000critical,katzgraber2001monte,young2004absence,katzgraber2005probing,katzgraber2006universality,jorg2008behavior,contucci2007ultrametricity,jorg2008comment,contucci2008answer,leuzzi2008dilute,hasenbusch2008critical,belletti2009depth,contucci2009structure,katzgraber2009ultrametricity,alvarez2010static,banos2010nature}. 
None of these gave a definitive answer on the structure of the low-temperature phase, and on the correct physical picture describing it. This is because a sampling of the low-temperature phase of a strongly-frustrated system like a non-mean-field spin glass has an exponential complexity in the system size \cite{barahona1982on,welsh1993complexity}. Accordingly, all such numerical simulations are affected by small system sizes, which prevent from discerning which is the correct framework describing the low-temperature phase. 
An example of how finite-size effects played an important role in such analyses is the following. According to the RSB picture, a spin-glass phase transition occurs also in the presence of an external magnetic field \cite{MPV}, while in the droplet picture no transition occurs in such a field \cite{fisher1986ordered}. MC studies \cite{katzgraber2005probing,katzgraber2009study}  of a one-dimensional spin glass with power-law interactions yielded evidence that there is no phase transition beyond mean field in  a magnetic field. Later on, a further MC analysis  \cite{leuzzi2009ising} claimed that the physical observables considered in such a previous work were affected by strong finite-size effects, and yielded evidence of a phase transition in a magnetic field beyond mean field through a new method of data analysis. Interestingly, a recent analytical work \cite{bray2011disappearance} based on a replica analysis suggests that below the upper critical dimension the transition in the presence of an applied magnetic field does disappear, in such a way that there is no RSB in the low-temperature phase \cite{moore2010ordered}.  \\

This  exponential complexity  in probing the structure of the low-temperature phase has played the role of a perpetual hassle in such numerical investigations, and strongly suggests that the final answer towards the understanding of the spin-glass phase in finite dimensions will not rely on numerical methods \cite{huse2011private}. 

\subsubsection*{The Random Energy  Model}

The simplest mean-field model for a structural glass was introduced in 1980 by Derrida \cite{derrida1980random,derrida1981random}, who named it the Random Energy Model (REM).
 In the original paper of 1980, the REM was introduced from a   spin-glass model with quenched disorder, the $p$-spin model. It was shown that in the limit $p\rightarrow \infty$ where  correlations between the energy levels are negligible, the $p$-spin model reduces to the REM:  a model of $N$ spins $S_i = \pm 1$, where the energy $\epsilon[ \vec S]$ of each spin configuration $\vec S$ is a random variable distributed according to a Gaussian distribution with zero mean and variance $1/N$. Accordingly, for every sample of the disorder $\{ \epsilon[ \vec S ] \} _{\vec S}$, the partition function of the REM is given by 
\be
Z = \sum_{\vec S} \textrm{e}^{-\beta \epsilon[\vec S]}. 
\ee

This model became interesting because, despite its striking simplicity, its solution reveals the existence of a phase transition  reproducing all the main physical features of the glass transition observed in laboratory phenomena. Indeed, there exists a finite value $T_c$ of the temperature, such that in the high temperature phase the system is ergodic, and has an exponentially-large number of states available, while in the low-temperature phase the system is stuck in a handful of low-lying energy states. The switchover between these two regimes is signaled by the fact that the entropy is positive for $T>T_c$, while it vanishes for $T<T_c$.  Interestingly, this transition does not fall in any of the universality classes of phase transitions for ferromagnetic systems \cite{zinnjustin}. Indeed, on the one hand the transition is strictly second order, since there is no latent heat. On the other hand, the transition presents the typical freezing features of first-order phase transitions of crystals \cite{0}. \\

Later on, people realized that the phenomenology of the REM is more general, and typical of some spin-glass models with quenched disorder, like the $p$-spin model. Indeed,  the one-step RSB solution scheme of the SK model was found  \cite{crisanti1992spherical} to be exact  for both of the $p$-spin model and the REM \cite{0}, and the resulting solutions show a critical behavior very similar to each other. Accordingly, the REM, the $p$-spin model and other models with quenched disorder are nowadays considered to belong to the same class, the $1$-RSB class \cite{biroli2009random}. \\

The solution of the $p$-spin spherical model reveals that the physics of such $1$-RSB mean-field models is the following \cite{biroli2009random}. There exists a finite temperature $T_d$ such that for $T > T_d$ the system is ergodic, while for $T < T_d$  it is trapped in one amongst exponentially-many metastable states: 
These are the Thouless  Almeida Palmer (TAP) \cite{thouless1977solution} states.
 Since the energy barriers between metastable states are infinite in mean-field models, the system cannot escape from the metastable state it is trapped in. The nature of this transition is purely dynamical, and it  shows up in the divergence of dynamical quantities like the relaxation time $\tau$, while there is no footprint of it in thermodynamic quantities.
 We will denote by $f_\ast(T)$ the free energy of each of these TAP states and by $f_p(T)$ the free energy of the system in its paramagnetic state. Accordingly, the total free energy of the glass below $T_d$ is given by $f_\ast(T) - T \Sigma(T)$.
Since there is no mark of the dynamical transition in thermodynamic quantities, one has that the free energy of the glass below $T_d$ must coincide with $f_p(T)$ 
\be \label{free_energy}
f_p(T) = f_\ast(T) - T \Sigma(T).
\ee
 Below $T_d$, there exists a second finite temperature $T_K < T_d$, such that the complexity vanishes at and below $T_K$: the number of TAP states is no more exponential, and the system is trapped in a bunch of low-lying energy minima: the system undergoes a Kauzmann transition at $T_K$. The nature of this transition is purely static, and shows up in the singularities of thermodynamic quantities such as the entropy. \\

An important physical question is whether this mean-field phenomenology persists beyond mean field.  
In 1989 Kirkpatrick, Thirumalai and Wolynes (KTW) \cite{kirkpatrick1989scaling} proposed a theoretical framework to handle finite-dimensional glass formers, which is known as the Random First Order Transition Theory (RFOT). 
Their basic argument was inspired by the following analogy with ferromagnetic systems. Consider a mean-field ferromagnet in an external magnetic field $h>0$. The free energy has two minima, $f_+$ and $f_-$, with positive and negative magnetization respectively. Being $h>0$, one has $f_- > f_+$. Even though the $+$-state has a lower free energy, it cannot nucleate because the free-energy barriers are infinite in mean field. Differently, in finite dimensions $d$ the free-energy barriers are finite, and the free-energy cost for nucleation of a droplet of positive spins with radius $R$ reads 
\be \label{delta_f_ferro}
\Delta f = C_1 R^{d-1} - (f_- - f_+) C_2 R^d,
\ee
where the first addend is the surface energy cost due to the mismatch between the positive orientation of the spins inside the droplet and the negative orientation of the spins outside the droplet, while the second addend represents the free-energy gain due to nucleation of a droplet of positive spins, and is proportional to the volume of the droplet.  According to the above free-energy balance, there exists a critical value $R_\ast$ such that droplets with $R < R_\ast$ do not nucleate and shrink to zero, while droplets with  $R > R_\ast$ grow indefinitely. 
Inspired by the physics emerging from mean-field models of the $1$-RSB class, KTW applied a similar argument to  glass-forming liquids. Before discussing KTW theory, is important to stress that the dynamical transition at $T_d$ occurring in the mean-field case disappears  in finite dimensions.  This is because the free-energy barriers between metastable states are no more infinite in the thermodynamic limit. Thus, the sharp mean-field dynamical transition is smeared out in finite dimensions, and it is plausible that $T_d$ is replaced by a crossover temperature $T_\ast$, separating a free flow regime for $T>T_\ast$ from an activated dynamics regime for $T<T_\ast$ \cite{biroli2009random}.

According to KTW, for $T<T_\ast$ the system is trapped in a TAP state with free energy $f_\ast$. 
 Following the analogy with the ferromagnetic case, the TAP state is associated with the $-$-state, while the paramagnetic state with the $+$-state. Accordingly, by Eq. (\ref{free_energy}) one has $f_--f_+ = T \Sigma$.   Nucleation of a droplet  of size $R$ of spins in the liquid state into a sea of spins in the TAP state has a free-energy cost
\[
\Delta f = C_1 R^{\theta} - T \Sigma C_2 R^d,
\]
where the exponent $\theta$ is the counterpart of $d-1$ in the ferromagnetic case, Eq. (\ref{delta_f_ferro}). Since the presence of disorder is expected to smear out such a surface effects with respect to the ferromagnetic case, one has $\theta < d-1$. 
Liquid  droplets with radius smaller than $R_\ast \equiv  \left( \frac{ C_1 \theta }{ T \Sigma C_2 d } \right)^{\frac{1}{d-\theta}}$ disappear, while droplets with radius larger that $R_\ast$ extend to infinity. Since there are many spatially localized TAP states, droplets can't extend to infinity as in the mean-field case. The system is rather said to be in a mosaic state, given by liquid droplets that are continually created and destroyed \cite{biroli2009random}. \\

In analogy with the $1$-RSB phenomenology, RFOT theory predicts that  $\Sigma$ vanishes at a finite temperature $T_K < T_\ast$. Below this temperature liquid droplets cannot nucleate anymore, because $R_\ast = \infty$, and the system is said to be in a ideal glassy state, i. e. a collectively-frozen and mechanically-stable low-lying energy state. Sill, the crucial question of the  existence of a Kauzmann transition in real glass-formers is an open issue. It cannot be amended experimentally, because real glasses are frozen in an amorphous configuration below $T_g$, and the entropy measured in laboratory experiments in this temperature range  does not  give an estimate of the number of degenerate metastable states. Accordingly, analytical progress in non-mean-field models of the $1$-RSB class describing the equilibrium properties below $T_g$ would yield a significant advance on this fundamental issue.\vspace{5cm}\\

A clear way to explore critical properties of non-mean-field systems came from the RG theory developed by Wilson in his papers of 1971 \cite{wilson1971renormalization1,wilson1971renormalization2}. The RG theory started from a very simple physical feature observed experimentally in physical systems undergoing a phase transition \cite{wilson1982renormalization}. Consider, for instance, a mixture of water and steam  put under pressure at the boiling temperature.  As the pressure approaches a critical value, steam and water become indistinguishable. In particular, bubbles of steam and water of all length scales, from microscopic ones to macroscopic ones, appear. This empirical observation implies that the system has no characteristic length scale at the critical point. In particular, as the critical point is approached, any typical correlation length of the system must tend  to infinity, in such a way that no finite characteristic length scale is left at the critical point. Accordingly, if we suppose to approach the critical point by a sequence of elementary steps, the physically important length scales must grow at each step. This procedure was implemented in the original work of Wilson, by integrating out all the length scales smaller than a given threshold. As a result, a new system with a larger typical length scale is obtained, and by iterating this procedure many times one obtains a system whose only characteristic length is infinite, and which is said to be critical. \\

The above RG scheme yields a huge simplification of the problem. Indeed, systems having a number of microscopic degrees of freedom which is typically exponential in the number of particles are reduced to a handful of effective long-wavelength degrees of freedom. These are the only physically relevant degrees of freedom in the neighborhood of the critical point, and all the relevant physical information can be extracted from them.  \\

In the first paper of 1971 Wilson's made quantitative the above qualitative picture for the Ising model. Following Kadanoff's picture \cite{kadanoff1966scaling},  short-wavelengths degrees of freedom were integrated out by considering blocks of spins acting as a unit, in such a way that one could treat all the spins in a block as an effective spin. Given the values of the spins in the block, the value of this effective spin could be easily fixed to be $+1$ if the majority of spins in the block are up, and $-1$ otherwise. The resulting approximate RG equations were analyzed in the second paper of 1971 \cite{wilson1971renormalization2}, where Wilson considered a simplified version of the Ising model and showed that this framework could make precise predictions on physical quantities like the critical exponents, which were extracted in perturbation theory. There the author realized that if the dimensionality $d$ of the system was larger than $4$ the resulting physics in the critical regime was the mean-field one, while for $d<4$ non-mean-field effects emerge.
 These RG equations for the three-dimensional Ising model were treated perturbatively in the parameter $\epsilon \equiv 4-d$, measuring the distance from the upper critical dimension $d=4$, in a series of papers in the $70$'s \cite{wilson1972critical,wilson1972feynman}. The validity of this perturbative framework was later confirmed by the reformulation of Wilson's RG equations in the language of field theory. There, the mapping of the Ising model into a $\phi^4$-theory and the solution of the resulting Callan-Symanzik (CS) equations \cite{callan1970broken, symanzik1970small,zinnjustin} for this theory made the RG method theoretically grounded, and the proof of the renormalizability of the $\phi^4$-theory \cite{callan1975methods} to all orders in perturbation theory  served as a further element on behalf of this whole theoretical framework. 
Finally, the picture  was completed some years later by high-order implementations of the $\epsilon$-expansion for the critical exponents \cite{vladimirov1979calculation,chetyrkin1981five,chetyrkin1981errata,chetyrkin1981integration,chetyrkin1983five,kazakov1983method,gorishny1984epsilon,kleinert1991five,kleinert1993five} which were in excellent agreement with experiments \cite{zinnjustin,transitions1980vol} and MC simulations \cite{pawley1984monte,baillie1992monte}.\\

Because of this ensemble of works, the RG served as a fundamental tool in understanding the critical properties of finite-dimensional systems. Hence, it is natural to search for a suitable generalization of Wilson's ideas to describe the critical regime of non-mean-field spin or structural glasses.  The drastic simplification resulting from the reduction of exponentially many  degrees of freedom to a few long-wavelength degrees of freedom would be a breakthrough to tackle the exponential complexity limiting our understanding of the physics of such systems. \\

Still, a construction of  a RG theory for spin or structural glasses is far more difficult than the original one developed for ferromagnetic systems. Indeed, in the ferromagnetic case it is natural to identify the order parameter, the magnetization, and then implement the RG transformation with Kadanoff's majority rule. Conversely, in non-mean-field spin or structural glasses, the order parameter describing the phase transition is fundamentally unknown. \\

For non-mean-field spin glasses, the RSB and droplet picture make two radically different predictions on the behavior of a tentative order parameter in the low-temperature phase.  In the RSB picture the order parameter is the probability distribution of the overlap $P(q)$, being $P(q) = \delta(q)$ in the high-temperature phase and $P(q)$ a smooth function of $q$ in the low-temperature phase \cite{parisi1979infinite}. Such a smooth function reflects  the hierarchical organization of many pure states in the low-temperature phase.  In the droplet picture \cite{fisher1986ordered} $P(q)$ reduces to two delta functions centered on the value of a scalar order parameter, the Edwards-Anderson order parameter $q_{\textrm{EA}}$ \cite{edwards1975theory}. Such an order parameter is nonzero if the local magnetizations are  nonzero, i. e. if the system is frozen in the unique low-lying ergodic component of the configuration space. \\

For structural glasses, after important developments in the understanding of the critical regime came in 2000 \cite{franz2000non}, a significant progress in the identification of the order parameter has been proposed in 2004 \cite{bouchaud2004adam} and  numerically observed in 2008 \cite{biroli2008thermodynamic} by Biroli et al., who suggested that the order parameter is the overlap between two equilibrated spin-configurations with the same boundary conditions: the influence of the boundary conditions propagates deeper and deeper into the bulk as the system is cooled, signaling the emergence of an amorphous order at low temperatures. \\

A justification of the difficulty in the definition of a suitable order parameter for a spin or structural glass has roots in the frustrated nature of the  spin-spin interactions. 
To illustrate this point, let us consider a spin system like the SK where the sign of the couplings $J_{ij}$ are both positive and negative, Eq. (\ref{h_sk}), and try to mimic Wilson's  block-spin transformation \cite{wilson1971renormalization1} for the SK model. Given the values of the spins in a block, Kadanoff's majority rule does not give any useful information on which should the value of the effective spin. Indeed, choosing the effective spin to be $+1$ if most of the spins in the block are up and $-1$ otherwise does not make sense: being the $J_{ij}$s positive or negative with equal probability, the magnetization inside the block is simply zero on average, and does not give any useful information on which value should be assigned to the effective spin.  Again, frustration is the main stumbling block in the theoretical understanding of such systems. \vspace{2cm}\\

In order to overcome this difficulty, we recall that Wilson's approximate RG equations were found to be exact \cite{wilson1982renormalization} on a particular  non-mean-field model for ferromagnetic interactions, where the RG recursion formulas have a strikingly simple and natural form. This is Dyson's Hierarchical Model (DHM), and was introduced by Dyson in 1969 \cite{dyson1969existence}. There, the process of integrating out long-wavelength degrees of freedom emerged naturally in  an exact integral equation for the probability distribution of the magnetization. This equation was the forerunner of Wilson's RG equations. \vspace{2cm}\\

The aim of this thesis is to consider a  suitable generalization of DHM describing non-mean-field  spin or structural glasses, and construct a RG framework for them.
These models will be generally denoted by Hierarchical Models (HM), and will be introduced in Section \ref{hm}. The definition of HM is quite general, and by making some precise choices on the form of the interactions, one can build up a HM capturing the main physical features  a non-mean-field spin or structural glass. Thanks to  their simplicity, HM allow for a simple and clear construction of a RG framework. Our hope is that such a RG framework could shed light on the criticality of the glass transition beyond mean field, and on the identification of the order parameter describing the emergence of an amorphous long-range order, if present. As a long-term future direction, the  RG method on HM could  also be useful to understand the features of the low-temperature phase of such glassy systems.
\\

The thesis is structured as follows. In Chapter \ref{hierarchical_models}  of Part \ref{introduction} we   discuss DHM, and introduce HM for spin or structural glasses. 
In Part \ref{hrem} we study a HM mimicking the physics of a non-mean-field structural glass, the Hierarchical Random Energy Model (HREM), being a hierarchical version of the REM. 
In this Part we show how one can work out a precise solution for thermodynamic quantities of the system, signaling the existence of a Kauzmann phase transition at finite temperature. The HREM constitutes the first non-mean-field model of a structural glass explicitly exhibiting such a freezing transition as predicted by  RFOT. Interestingly, the solution suggests also the existence of a characteristic  length growing as the critical point is approached, in analogy with the predictions of KTW. 
In Part \ref{hea} we study a HM mimicking the physics of a non-mean-field spin glass, the Hierarchical Edwards-Anderson model (HEA), being a hierarchical version of the Edwards-Anderson model. The RG transformation is first implemented with the standard replica field-theory approach, which turns out to be non-predictive because nonperturbative effects are completely untamed. Consequently, a new RG method in real space is developed. This method avoids the cumbersome formalism of the replica approach, and shows the existence of a phase transition, making precise predictions on the critical exponents. The real-space method is also interesting from a purely  methodological viewpoint, because it yields the first suitable generalization of Kadanoff's RG decimation rule for a strongly frustrated system. 
Finally, in Part \ref{conc} we discuss the overall results of this work, by paying particular attention to its implications and future directions in the physical understanding of realistic systems with short-range interactions.

\chapter{Hierarchical models}\label{hierarchical_models}

In this   Chapter we introduce hierarchical models. In Section \ref{dyson} we first introduce the ferromagnetic version of hierarchical models originally introduced by Dyson, and in Section \ref{hm} we extend this definition  to the disordered case, in the perspective to build up a non-mean-field hierarchical  model of a spin or structural glass. 

 \section{Hierarchical models for ferromagnetic systems}\label{dyson}
 
A hierarchical model for ferromagnetic systems has been introduced in the past to describe non-mean-field spin systems \cite{dyson1969existence}, and is  known as Dyson's Hierarchical Model (DHM).   DHM has been of great interest in the past, because Wilson's RG equations \cite{wilson1971renormalization1,wilson1971renormalization2, wilson1974renormalization,wilson1973quantum,wilson1974critical,wilson1972critical} turn out to be exact  in models with power-law ferromagnetic  interactions built  on hierarchical lattices like DHM. Indeed, in this model one can explicitly write an exact RG transformation for the  probability distribution of the magnetization of the system. All the relevant physical information on the paramagnetic, ferromagnetic and critical fixed point, and  the existence of a finite-temperature phase transition are encoded into these RG equations. Moreover, all the physical RG ideas emerge naturally from  these recursion relations, whose solution can be  explicitly built up with the $\epsilon$-expansion technique \cite{cassandro1978critical, collet1978renormalization, collet1977epsilon, collet1977numerical}.\\
 
   DHM  is defined    \cite{dyson1969existence, cassandro1978critical} as a system of $2^{k+1}$  Ising spins $S_1,\ldots,S_{2^{k+1}},\, S_i = \pm 1$, with an energy function which is built up  recursively by coupling two systems of $2^k$ spins
\begin{eqnarray}\label{20}
H^ F_{k+1}\left[S_1,\ldots,S_{2^{k+1}}\right] &= &
 H^ F_{k}\left[S_1,\ldots,S_{2^{k}}\right] +  
 H^ F_{k}\left[S_{2^k+1},\ldots,S_{2^{k+1}}\right]+ \\ \no 
&&-J C_F^ {k+1} \left( \frac{1}{2^{k+1}}\sum_{i=1}^{2^{k+1}}  S_i \right)^2  ,
\end{eqnarray}
where 

\be \label{c_f}
C_F \equiv 2^ {2(1-\sigma_{F})},
\ee
and  $\textrm{F}$ stands for ferromagnetic.  The model is defined for 
\be \label{27}
1/2<\sigma_{F} <1.
\ee
The limits  (\ref{27}) can be derived by observing that for $\sigma_{F}>1$ the interaction energy goes to $0$ for large $k$, and no finite-temperature phase transition occurs, while for $\sigma_{F} <1/2$ the interaction energy grows with $k$ faster than $2^k$, i. e. faster than the system volume, in such a way that the model is thermodynamically unstable. \\

The key issue of DHM is that the recursive nature of the Hamiltonian function encoded in Eq. (\ref{20}) results naturally into an exact RG equation. This equation can be easily derived \ by defining the probability distribution of the magnetization $m$ for a $2^k$-spin DHM, as 
\be \label{def_pcont_dhm}
p_k(m) \equiv \mathcal{C}  \sum_{\vec S} \textrm{e}^{- \beta H_k^F[\vec S]} \delta \left( \frac{1}{2^k} \sum_{i=1}^{2^k} S_i - m \right),
\ee 
where $\delta$ denotes the Dirac delta function, and $\mathcal{C}$ a constant enforcing the normalization condition $\int dm p_k(m)=1$. Starting from Eq. (\ref{20}), one can easily derive a recursion equation relating $p_k$ to $p_{k+1}$. This equation is derived in Section \ref{app_field_dhm_1} of Appendix \ref{app_field_dhm}, and  reads
\be \label{rec_dhm}
p_{k+1}(m)  = \textrm{e}^{\beta J C_F^{k+1} m ^2} \int d \mu   \,   p_{k}(m+\mu ) p_k(m- \mu), 
\ee
where any $m$-independent multiplicative constant has been omitted to simplify the notation. Eq. (\ref{rec_dhm}) relies the probability distribution of a DHM with $2^k$ spins with that of a DHM with $2^{k+1}$ spins. Accordingly, Eq. (\ref{rec_dhm}) is nothing but the flow of the function $p_{k}(m)$ under  reparametrization  $2^k \rightarrow 2^{k+1}$  of the length scale of the system. Historically, Eq. (\ref{rec_dhm}) has been derived by Dyson \cite{dyson1969existence}, and then served as the starting point for the construction of the RG theory for ferromagnetic systems  like the Ising model. Indeed, Wilson's RG recursion formulas for the Ising model \cite{wilson1971renormalization1,wilson1971renormalization2, wilson1974renormalization} are   approximate, while they   turn out to be exact when applied to DHM, because they reduce to Eq. (\ref{rec_dhm}). DHM has thus played a crucial role in the construction of the RG theory for ferromagnetic systems, because in a sense the work of Wilson on finite-dimensional systems has been pursued in the effort to generalize the exact recursion formula (\ref{rec_dhm})  to more realistic systems with no hierarchical structure, like the three-dimensional Ising model. \medskip \\

Equation (\ref{rec_dhm}) has also been an important element in the probabilistic formulation of RG theory, originally foreseen  by Bleher, Sinai \cite{bleher1973investigation} and Baker \cite{baker1972ising}, and later developed by Jona-Lasinio and Cassandro \cite{jona1975renormalization,cassandro1978critical}. Indeed, Eq. (\ref{rec_dhm}) aims to establish the probability distribution of the average of $2^k$ spin variables $\{ S_i \}_i$ for $k \rightarrow \infty$. In the case where the spins are independent and identically distributed (IID), the above analogy becomes transparent, because the answer to the above question is yield by the central limit theorem. Following this connection between RG and probability theory, one can even prove the central limit theorem starting from the RG equations (\ref{rec_dhm}) \cite{cassandro1978critical}.
\\

Equation (\ref{rec_dhm}) has been of interest in the last decades also because it is simple enough to  be solved with high precision, and   the resulting solution gives a clear insight into the critical properties of the system, showing the existence of a phase transition.
The crucial observation  is that Eq. (\ref{rec_dhm}) can be iterated $k\gg1$ times in $2^k$ operations. Indeed, the magnetization $m$ of a $2^k$-spin DHM can take $2^k+1$ possible values $\{ -1, -1 + 2/2^k, \cdots, 0, \cdots, 1- 2/2^k, 1\}$. According to Eq. (\ref{def_pcont_dhm}), the function $p_k(m)$ is nonzero only if $m$ is equal to one of these $2^k+1$ values. It follows that in order to compute $p_{k+1}(m)$, one has to perform a sum in the right-hand side of Eq. (\ref{rec_dhm}), involving $2^k+1$ terms. This implies that the time to calculate $p_{k}(m)$ for $k \gg 1$ is proportional to $2^k$. Thus, the use of the hierarchical structure encoded in Eq.  (\ref{rec_dhm}) yields a significant  improvement in the computation of $p_k(m)$ with respect to  a brute-force evaluation of the sum in the right-hand side of Eq. (\ref{def_pcont_dhm}), which involves $2^{2^k}$ terms. \\

 Let us now discuss the solution of Eq. (\ref{rec_dhm}). For  Eq.  (\ref{rec_dhm}) to be nontrivial for  $k \rightarrow \infty$, one needs to rescale the magnetization variable. Otherwise, the $C_F^{k+1}$-term in the right-hand side of Eq. (\ref{rec_dhm}) would diverge for $k\rightarrow \infty$.
 Setting 
\be \label{redef_dhm}
\mathfrak{p}_{k}(m) \equiv p_k(C_F^{-k/2}m), 
\ee
Eq. (\ref{rec_dhm}) becomes 
\be \label{rec2_dhm}
\mathfrak{p}_{k+1}(m)  = \textrm{e}^{\beta J m ^2} \int d \mu   \,  \mathfrak{ p}_{k}\left(\frac{m+\mu}{C_F^{1/2}} \right) \mathfrak{p}_k \left(\frac{m- \mu}{C_F^{1/2}} \right) . 
\ee

The structure of the fixed points of Eq. (\ref{rec2_dhm}) is discussed  in Section \ref{app_field_dhm_2} of Appendix \ref{app_field_dhm}. In particular, it is shown that
there exists a value $\beta_{c \, F}$ of $\beta$, such that if $\beta <\beta_{c \, F}$   Eq. (\ref{rec2_dhm}) converges to a  high-temperature fixed point, while if $\beta >\beta_{c \, F}$   Eq. (\ref{rec2_dhm}) converges to a  low-temperature fixed point.  Both of these fixed points are stable, and can be qualitatively represented  as basins of attraction in the infinite-dimensional space where $\mathfrak{p}_k(m)$ flows \cite{wilson1974renormalization}. These basins of attraction are separated by an unstable fixed point $\mathfrak{p}_\ast(m)$, which is reached by iterating Eq. (\ref{rec2_dhm}) with $\beta = \beta_{c \, F}$. $\mathfrak{p}_{\ast}(m)$ is called the critical fixed point, and is characterized by the fact that  the convergence of $\mathfrak{p}_k$ to $\mathfrak{p}_\ast$ for $\beta = \beta_{c \, F}$ implies the divergence of the characteristic length scale $\xi_F$ of the system  in the thermodynamic limit $k\rightarrow \infty$. Accordingly, in what follows $\beta_{c \, F}$ will denote  the inverse critical temperature of DHM.  In the neighborhood of the critical temperature the divergence of $\xi_F$ is characterized by a critical exponent $\nu_F$, defined by 
\be \label{def_nu_f}
\xi_F \overset{T \rightarrow T_{c \,F} }{\approx} \frac{A}{(T-T_{c \, F} )^{\nu_F}}, 
\ee
where $A$ is independent of the temperature.  The critical exponent $\nu_F$ is an important physical quantity characterizing criticality, and is  quantitatively predictable from the theory.
 In Section \ref{app_field_dhm_3} of Appendix \ref{app_field_dhm} we show how $\nu_F$ can be computed starting from the RG equation (\ref{rec2_dhm}). This derivation serves as an important example of the techniques that will be employed in generalizations of DHM involving quenched disorder, that will be discussed in the following Sections.\\

The calculation of $\nu_F$ relies on the fact that for $0<\sigma_F \leq 3/4$ the critical fixed point $\mathfrak{p}_\ast(m)$ is a Gaussian function of $m$, while for $3/4<\sigma_F<1$ $\mathfrak{p}_\ast(m)$ is not Gaussian, as illustrated in Section  \ref{app_field_dhm_2}.  We recall \cite{collet1978renormalization, collet1977epsilon, cassandro1978critical, wilson1974renormalization, zinnjustin, zinn2007phase} that  a Gaussian $\mathfrak{p}_\ast(m)$  corresponds to a mean-field regime of the model. The expression mean field is due to the following. Consider for instance the thermal average at the critical point of a physical observable 
$O(1/2^k \sum_{i=1}^{2^k}S_i)$, depending on  the spins $\vec S$ through the magnetization of the system. This can be expressed as an average of $O(m)$ with weight $p_{\ast}(m)$, where $p_{\ast}(m) = \mathfrak{p}_\ast(C_F^{k/2}m)$
\be \label{152}
\mathbb{E}_{\vec S}^ F[O] = \int dm \, p_{\ast}(m) O(m),
\ee
where $\mathbb{E}^F_{\vec{S}}$  stands for the thermal average  
\[
\mathbb{E}^F_{\vec{S}}[ O] \equiv \frac{\sum_{\vec S} e^ { -\beta \, H_k^F[\vec S] } O(1/2^k \sum_{i=1}^{2^k}S_i)  }{  \sum_{\vec S} e^ { -\beta \, H_k^F[\vec S] } }. 
\] 
In the mean-field approximation one evaluates integrals like that in  the right-hand side of Eq. (\ref{152}) with the saddle-point approximation \cite{zinnjustin, huang, NishimoriBook01}. If $0<\sigma_F \leq 3/4$, $\mathfrak{p}_\ast(m)$ is Gaussian, and so is $p_\ast (m)$, in such a way that the saddle-point approximation is exact, i. e. the mean-field approximation is correct.  On the contrary, for $3/4<\sigma_F  \leq 1$, $\mathfrak{p}_\ast(m)$ is not Gaussian, and the system has a non-mean-field behavior.  In particular, fluctuations around the mean-field saddle point in the right-hand side of Eq. (\ref{152}) are not negligible. According to this discussion, we call $\sigma_F= 3/4$ the upper critical dimension \cite{huang, zinnjustin, zinn2007phase, itzykson1991statistical} of DHM. \\

In the mean-field region $0<\sigma_F \leq 3/4$, $\nu_F$ can be computed exactly, and is given by Eqs. (\ref{nu_f}), (\ref{150}). In the non-mean-field region $\nu_F$ can be calculated by supposing that  the physical picture  emerging for $0<\sigma_F \leq 3/4$ is slightly modified in the non-mean-field region. As discussed in Section \ref{app_field_dhm_2}, this assumption is equivalent to saying that  corrections to the mean-field estimate of integrals like (\ref{152}) are small, i. e. they can be handled perturbatively. Whether this assumption is correct or not can be checked  a posteriori, by expanding physical quantities like $\nu_F$ in powers of $\epsilon_F=\sigma_F-3/4$,  and investigating the convergence properties of the expansion.  If the $\epsilon_F$-expansion is found to be convergent or resummable \cite{zinnjustin}, the original assumption is confirmed to be valid. If it is not, the non-mean-field physics is presumably radically different from that arising in the mean-field region, and cannot be handled perturbatively. As an example, the result to  $O(\epsilon_F)$  for $\nu_F$ is given by Eqs. (\ref{nu_f}), (\ref{151}).   In \cite{collet1977numerical,collet1978renormalization}, the $\epsilon_F$-expansion has been performed to high orders, and found to be nonconvergent. Even though, the authors showed  that the  application of a resummation method originally presented in \cite{loeffel1975transformation} yields a convergent series for $\nu_F$, which is in quantitative agreement with the values of the exponent obtained by Bleher \cite{bleher1975critical}. Finally, we mention that the results from the $\epsilon_F$-expansion of DHM have been found to be in excellent agreement with those obtained with the high-temperature expansion, which has been studied by Y. Meurice et al. \cite{meurice1997oscillatory}. \\

Since DHM allows for a relatively simple implementation of the RG equations for a non-mean-field ferromagnet, it is natural to ask oneself whether there exists a suitable generalization of DHM that can describe a non-mean-field spin or  structural glass. This generalization will be exposed in the following Section.

\section{Hierarchical models for spin and structural glasses}\label{hm}
In the effort to clarify the non-mean-field scenario of both spin glasses and structural glasses, it is  useful to consider a suitable generalization of DHM to the disordered case. Concerning this, it is important to observe that the extension of DHM  to the random case has been performed only for some particular models.\\

Firstly, models with local interactions on hierarchical lattices built on diamond plaques   \cite{berker1979renormalisation},  have been widely studied  in their spin-glass version, and lead  to weakly frustrated systems even in their mean-field limit  \cite{gardner1984spin}. Notwithstanding this, such models yield a very useful and interesting playground to show how to implement the RG ideas in  disordered hierarchical lattices, and in particular on the construction  of  a suitable decimation rule for a frustrated system. \\

 Secondly, a RG analysis for random weakly frustrated models on Dyson's hierarchical lattice has been done in the past by A. Theumann  \cite{theumann1980critical,theumann1980ferromagnetic}, and the structure of the physical and unphysical infrared (IR) fixed points has been obtained  with the $\epsilon$-expansion technique. Unfortunately, in these models spins belonging to the same hierarchical block interact with each other with the same   \cite{theumann1980critical} random coupling, in such a way that frustration turns out to be relatively weak and they are not a good representative for  realistic strongly frustrated systems. This is because these models are obtained from DHM by replacing the coupling $J$  in Eq. (\ref{20}) with a random variable $J_k$. Thus, the interaction energy between spins $S_1, \ldots, S_{2^k}$ is fixed, and purely ferromagnetic or antiferromagnetic, depending on the sign of $J_k$.  Differently, in strongly frustrated systems like the SK model, the coupling $J_{ij}$ between any spin pair $S_i, S_j$ is never fixed to be ferromagnetic or antiferromagnetic, because its sign is randomly drawn  for any $i$ and $j$. \\

Thirdly, disordered spin models on Dyson's hierarchical lattice have been studied by A. Naimzhanov \cite{naimzhanov1982hierarchical,naimzhanov1983hierarchical}, who showed that the probability distribution of the magnetization converges to a Gaussian distribution in the infinite-size limit. Also in this case, the interaction between spins $S_1, \ldots, S_{2^k}$ is fixed to  be  ferromagnetic or antiferromagnetic, depending on the sign of a random energy $\varepsilon_k$ which is equal to $\pm 1$ with equal probability. \medskip\\

Here we present a different generalization of DHM to a disordered and strongly frustrated case, first introduced in \cite{franz2009overlap}, and simply call these models   \textit{hierarchical models} (HM). Indeed, the definition (\ref{20}) holding in the ferromagnetic case can be easily generalized as follows. We define  a  HM    as  a system of $2^{k+1}$  spins $S_1,\ldots,S_{2^{k+1}},\, S_i = \pm 1$, with an energy function defined recursively by coupling two systems, say system $1$ and system $2$, of $2^k$ Ising spins
\begin{eqnarray}\label{1}
H_{k+1}\left[S_1,\ldots,S_{2^{k+1}}\right] &=&
 H_{k}^{1}\left[S_1,\ldots,S_{2^{k}}\right]  + H_{k}^{2}\left[S_{2^k+1},\ldots,S_{2^{k+1}}\right]+ \\ \no 
&&+ \epsilon_{k+1}\left[S_1,\ldots,S_{2^{k+1}}\right].
\end{eqnarray}
The energies $H_k^ {1}, H_k^ 2$ are to be considered as the energy  of system $1$ and system $2$ respectively, while $\epsilon_k$ is the coupling energy between system $1$ and  system $2$. Differently from the ferromagnetic case, here the coupling energy $\epsilon_{k+1}\left[S_1,\ldots,S_{2^{k+1}}\right]$ of any spin configuration $S_1,\cdots, S_{2^ {k+1}}$ is a random variable, which is chosen to have zero mean for convenience. \\

 Since the interaction energy $\epsilon_{k+1}$ couples $2^ {k+1}$ spins, and since its order of magnitude is give by its variance,  one must have 
\be \label{171}
 \mathbb{E}_\epsilon [\epsilon_{k+1}^2 ] < 2^{ k+1},
 \ee
where  $\mathbb E_{\epsilon}$ stands for the expectation value with respect to all the coupling energies $\epsilon_k$ of the model.
Eq. (\ref{171}) states that the interaction energy between $2^{k+1}$ spins is sub extensive with respect to the system volume $2^{k+1}$, and ensures \cite{MPV, NishimoriBook01} that HM  are  non-mean-field models. The mean-field limit will be constantly recovered in the following chapters as the limit where $ \mathbb{E}_\epsilon [\epsilon_{k+1}^2 ] $ becomes of the same order of magnitude as the volume $2^{k+1}$. \\

\begin{centering}
  \begin{figure}[htb] 
\centering
\includegraphics[width=10cm]{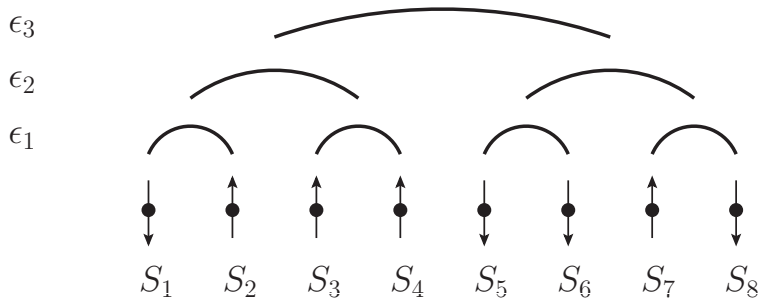}
\caption{A $2^3$-spin hierarchical model obtained by iterating Eq. (\ref{1}) until $k=3$. The arcs coupling pairs of spins represent the energies $\epsilon_1$ at the first hierarchical level $k=1$.  Those coupling quartets of spins represent  $\epsilon_2$  at the second hierarchical level $k=2$.  Those coupling octets of spins represent the energies $\epsilon_3$ at the third hierarchical level $k=3$.}
\label{fig1}
\end{figure}
  \end{centering}

  As we will show in the following,  the form (\ref{1}) of the Hamiltonian  corresponds to dividing the system in hierarchical embedded blocks of size $2^k$, so that the interaction between two spins depends on the distance of the blocks to which they belong   \cite{franz2009overlap,castellana2010renormalization, castellana2011renormalization}, as shown in Fig \ref{fig1}.\medskip\\

 The random energies $\epsilon_k$ of HM can be suitably chosen to mimic the interactions of a strongly frustrated structural glass (Part \ref{hrem}), or of a spin glass (Part \ref{hea}), in the perspective to give some insight into the non-mean-field behavior and criticality of both of these models. In this thesis such features will  be investigated by means of  RG techniques.  Indeed, as for DHM, the recursive nature of the definition (\ref{1}) suggests that HM are particularly suitable for an explicit implementation of the RG transformation. As a matter of fact, the definition (\ref{1}) is indeed  a RG flow transformation from the length scale $2^ k$ to the length scale $2^ {k+1}$. As we will show explicitly in Part \ref{hea}, one can  analyze the  fixed points of such an RG flow, in order to establish if a phase transition occurs, and  investigate the critical properties of the system. 
 
  It is important to observe that without the hierarchical structure this would be extremely difficult. This is mainly because of the intrinsic and deep difficulty in identifying the correct order parameter discussed in Section \ref{historical_outline}, and thus write an RG equation for a function (or functional) of it without making use of the replica method \cite{6,MPV} which, up to the present day, could not be used to make predictions for the non-mean-field systems under consideration in this thesis.   \\

  After introducing HM in their very general form, we now make a precise choice for the random energies $\epsilon_k$ in order to build up a hierarchical model for a structural glass, the Hierarchical Random Energy Model, and discuss its solution. 
  
  \part{The Hierarchical Random Energy Model}\label{hrem}

 As discussed in Section \ref{historical_outline}, the REM is a mean-field spin model mimicking the phenomenology of a supercooled liquid. Given the general definition of HM, it is  easy to make a particular choice for the random energies $\epsilon_k$ in (\ref{1}), to build up a non-mean-field version of the REM, i. e. a HM being a candidate for describing the phenomenology of a supercooled liquid beyond mean field. Indeed, we choose the energies $\epsilon_k$  to be independent variables distributed according to a  Gaussian distribution with zero mean and variance proportional to $C^{2k}$
 \be \label{42}
\mathbb{E}_\epsilon [ \epsilon_k^2] \sim C^ {2k},
 \ee
where we set
\be \label{def_c}
C^2 =2^{1-\sigma}.
\ee

 For  $\sigma<0$ the thermodynamic limit $k \rightarrow \infty$  is ill-defined, because the interaction energy $\mathbb{E}_\epsilon[ \epsilon_k ^2]$ grows faster than the volume $2^k$. For $\sigma>1$, $\mathbb{E}_\epsilon[ \epsilon_k ^2]$ goes to $0$ as $k \rightarrow \infty$, implying that there is no phase transition at finite temperature. Hence, the interesting region that we will consider in the following is 
 \be \label{21}
 0<\sigma < 1,
 \ee 
which is the equivalent of Eq. (\ref{27}) for DHM. 
As we will discuss in the following, this HM reproduces  the REM in the mean-field case $\sigma=0$, and will thus  be called the \textit{Hierarchical Random Energy Model} (HREM) \cite{castellana2010hierarchical, castellana2011real}. 
According to the general classification of models with quenched disorder given in Section \ref{historical_outline}, the HREM has to be considered as a model mimicking  a structural glass.\\
 
Before discussing the solution of the HREM, it is important to focus our attention on some important features of the model that make it interesting in the perspective of investigating the non-mean-field regime of a structural glass. \\

 Firstly, the hierarchical structure of the HREM allows an almost explicit solution with two independent and relatively simple methods. \\
The first method will be described very shortly here (a complete discussion can be found in \cite{castellana2010hierarchical,castellana3}) and relies on the fact that the recursive nature of Eq. (\ref{1}) implies a recursion relation for the function  $\mathcal{N}_k(E)$, defined as the number  of states with energy $E$  at the $k$-the step of the recursion.  By solving this recursion equation for large $k$, one can compute the entropy of the system 
\be 
s(E) \equiv \frac{1}{2^ k} \log \left[ \mathcal{N}_k(E) \right],
\ee 
 and thus investigate its equilibrium properties. The computation time needed to  implement this recursion at the $k$-th step is proportional to a power of $2^ k$, and represents a neat improvement on the exact computation of the partition function, involving a time proportional to $2^ {2^k}$. This recursive  method is also significantly better than estimating thermodynamic quantities with MC simulations, because the latter are affected by a severe increase of the thermalization time when approaching the critical point, as discussed in Section \ref{historical_outline}. \\
 The second method investigates the thermodynamic properties of the HREM by a perturbative expansion in the parameter $C$, physically representing the coupling constant between spins. As a matter of fact, the relatively simple structure of the model allows for a fully automated expansion in $C$ of the equilibrium thermodynamic quantities, which exhibits a neat and clear convergence when increasing the perturbative order as discussed in Chapter \ref{pert_hrem}.  \\
 It follows that the HREM is a model that hopefully encodes the non-mean-field features of a structural glass, and that is solvable with relatively simple and reliable methods, such as the recursion equation for $\mathcal{N}(E)$ and the perturbative  expansion in $C$. In particular, as we will show in Chapter \ref{pert_hrem}, with such methods one can identify  the existence of a phase transition in the HREM,  and then analyze its physical features.  \\ 
 
\begin{centering}
  \begin{figure}
\centering
\includegraphics[width=10cm]{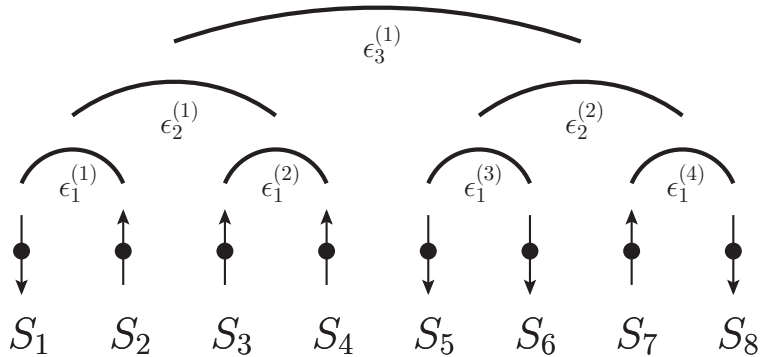}
\caption{A $2^3$-spin HREM and its interaction structure. Each arc represents an interaction energy $\epsilon_j^{(i)}$ amongst spins falling below the arc.}
\label{fig20}
\end{figure}
  \end{centering}

 Secondly, it turns out that  the $2^{2^k}$ energy levels $\{ H_k[\vec S]\}_{\vec S}$ of the HREM are not independent variables as in the REM \cite{derrida1980random}, because here they are correlated to each other. Indeed, by iterating $k$ times Eq. (\ref{1}), one obtains explicitly the Hamiltonian for a HREM with $2^ k$ spins
\be  \label{24}
H_k[\vec S] = \sum_{j=0}^ k  \sum_{i=1}^ {2^{k-j}} \epsilon^ {(i)}_j[\vec S^ {(j,i)} ],
\ee
 where $\vec S \equiv \{ S_1, \cdots, S_{2^ k} \}$, while  $\vec S^ {(j,i)} \equiv  \{ S_{2^ j(i-1)+1}, \cdots,  S_{2^j i}\}$ are the spins in the $i$-th embedded block at the $j$-th hierarchical level, and  $\epsilon^ {(i)}_j$ is the interaction energy $\epsilon_j$ (see Eq. (\ref{1})) of the $i$-th hierarchical embedded block. The interaction energies $\epsilon_j^{(i)}$ of Eq. (\ref{24}) are depicted in Fig \ref{fig20} for a HREM with $2^3$ spins.

 According to Eq. (\ref{24}), the energy levels are clearly correlated to each other.  As we will show in Chapter \ref{pert_hrem}, this fact implies that some critical features  of the HREM turn out to be quite different from to those of the REM. In particular, we will show by an explicit calculation how a naive estimate of the critical temperature based on the hypothesis that the energy levels are uncorrelated fails miserably, proving the relevance of energy correlations in the critical regime.\\
 
  Thirdly, the existence of the hierarchical structure depicted in Fig. \ref{1} allows for the introduction of a notion of distance between spins in the HREM, whereas in the REM there is no notion of distance, because  mean-field models have  no spatial geometry \cite{MPV}. As we will show in Chapter \ref{corr_hrem}, such a length scale can be introduced in the HREM by defining a suitable correlation function, and extracting the characteristic length scale associated with its exponential decay at large distances.  It is then interesting to ask oneself whether such a length diverges at the critical point as in ferromagnetic systems \cite{huang,landau1980statistical, zinnjustin, zinn2007phase, wilson1974renormalization,wilson1983renormalization, parisi1998statistical}. This point will be investigated in Chapter \ref{corr_hrem}, by means of the perturbative expansion method.  \\
  
  We will now present the perturbative computation of the equilibrium properties of the HREM, and discuss the results on the critical behavior of the model \cite{castellana2010hierarchical}. 
  
  \chapter{Perturbative computation of the free energy }\label{pert_hrem}
  
Given a sample of the random energies $\{ \epsilon \} \equiv \{ \epsilon_j^ {(i)} \}_{j,i}$, the free energy of a HREM with $2^ k$ spins is defined as \cite{0,MPV}
\be  \label{25}
f[T, \{ \epsilon \}] \equiv - \frac{1}{\beta 2^ k} \log \left[ Z\left[ T, \{ \epsilon \} \right] \right], 
\ee
  where
\be  \label{46}
Z\left[ T, \{ \epsilon \} \right] \equiv \sum_{\vec S} \exp\left(  - \beta H_k[\vec S] \right) ,
\ee  
$\beta \equiv 1/T$ is the inverse temperature, and   $H_k[ \vec S]$ is given by Eq. (\ref{24}). To simplify the notation,  in the following we omit the volume label $k$ in the free energy $f$ and in the partition function $Z$ unless necessary.

  The free energy (\ref{25}) of a \textit{typical} sample $\{ \epsilon \}$ can be computed by hypothesizing that the self-averaging property holds. According to this property, holding  in  the thermodynamic limit of a broad class of disordered systems with quenched disorder \cite{MPV, castellani2005spin},   the free energy computed on a fixed and typical sample of the disorder is equal to the average value of the free energy over the disorder.  Here we hypothesize that this property holds, so that in the thermodynamic limit $k\rightarrow \infty$ we compute $f[ T, \{\epsilon \}] $ on a typical sample $\{\epsilon\}$ as the average of Eq. (\ref{25})  over the random energies 
  \be  \label{26}
 \lim_{k \rightarrow \infty} f[ T, \{\epsilon \}]  =  \lim_{k \rightarrow \infty} \mathbb{E}_{\epsilon}  \left[ f[ T, \{\epsilon \}]  \right]. 
  \ee  \\

The advantage of using the self-averaging property is that the right-hand side of Eq. (\ref{26}) is easier to compute than the left-hand side by using the replica trick \cite{MPV,0}
\be  \label{28}
 \mathbb{E}_{\epsilon}  \left[ f[ T, \{\epsilon \}]  \right] = - \frac{1}{\beta 2^ k}\lim_{n\rightarrow 0} \frac{\mathbb{E}_{ \epsilon } \left[ Z[ T, \{ \epsilon \}] ^ n \right]-1}{n}. 
\ee  
According to the general prescriptions of the replica trick \cite{MPV, parisi1980order, parisi1979infinite, castellani2005spin}, the argument of the limit in Eq. (\ref{28}) is here computed for integer $n$, and an analytic  function of $n$ is obtained. The left-hand side of (\ref{28})  is then  computed by continuing such a function to real $n$, and taking its $n\rightarrow 0$ limit.\\
 As observed in Section \ref{historical_outline}, the use of the replica trick in mean-field models can be   non-rigorous, because of the assumption that one can exchange the thermodynamic limit and the $n\rightarrow 0$ limit \cite{MPV, 0, parisi1980order, castellani2005spin}. It is important to observe that this issue does not occur in this case. Indeed,  by using Eqs. (\ref{25}), (\ref{26}) and (\ref{28}), one has
 \be \label{30}
\lim_{k \rightarrow  \infty} f[ T, \{\epsilon \}]  = \lim_{k \rightarrow \infty}    \lim_{n\rightarrow 0}   \frac{1-\mathbb{E}_{ \epsilon } \left[ Z[ T, \{ \epsilon \}] ^ n \right]}{n \beta 2^ k} . 
 \ee
  In order to compute Eq. (\ref{30}) in   mean-field models, one hypothesizes that  one can  first compute the right-hand side of Eq. (\ref{29}) in the thermodynamic limit $ k \rightarrow \infty$ by using the saddle-point approximation, and then take $n \rightarrow 0$, by exchanging the limits. Being the HREM a non-mean-field model, the saddle-point approximation is wrong even in the thermodynamic limit, so that the right-hand side of Eq. (\ref{30}) cannot be computed by taking its saddle point, and we do not need to exchange the limits. Hence, the subtleties resulting from the exchange of the limits do not occur in this case. In other words, here the replica trick is simply a convenient way to perform the computation of the quenched free energy, and a direct inspection of Eq. (\ref{30})   in perturbation theory shows that one can do the  computation without replicas, and obtain the same result as that obtained with the replica trick to any order in $C$. We observe that this fact is true also in  the mean-field theory of spin glasses, where the full-RSB solution \cite{parisi1979infinite} can  be rederived \cite{mezard1986sk} without making use of the replica method. \\

 Let us now focus on the explicit computation of the right-hand side of Eq. (\ref{30})  for integer $n$ and on the $n \rightarrow 0$-limit.  One has
 \be \label{29}
 \mathbb{E}_{ \epsilon } \left[ Z[ T, \{ \epsilon \}] ^ n \right] =\sum_{\{ \vec{S}_{a}\}_{a=1,\cdots, n}}\exp\left(\frac{\beta^{2}}{4}\sum_{j=0}^{k}C^{2j}\sum_{i=1}^{2^{k-j}}\sum_{a,b=1}^{n}\delta_{ \vec{S}_{a}^{(j,i)} , \vec{S}_{b}^{(j,i)}}\right),
\ee
where $\vec{S}_1, \cdots, \vec{S}_n$   denote   the spin configurations of the $n$ replicas of the system \cite{parisi1980order,parisi1983order,castellani2005spin,NishimoriBook01}. We then expand Eq. (\ref{29}) in power of $C^2$, and take the $n\rightarrow 0, k\rightarrow \infty$-limits. It is important to observe that this $C^2$-expansion is equivalent to a high-temperature expansion. Indeed, in Eq. (\ref{29}) any power  $C^ {2j}$ of the coupling constant is multiplied by a factor $\beta^2$, so that   the smallness of $C^ 2$ is equivalent to the smallness of the inverse temperature $\beta$.\\
 By Eq. (\ref{30}), the  expansion of Eq. (\ref{29})  in powers of $C^2$ results into an expansion for $f[ T,  \{ \epsilon \} ]$, that can be written as 
\be  \label{31}
f[ T,  \{ \epsilon \} ]  = \sum_{i=0}^ {\infty} C^{2i} \phi_i(T), 
\ee
where for simplicity we omit the $k \rightarrow \infty$-limit, and the dependence of $f$ on $\{ \epsilon \}$ has disappeared because of the self-averaging property (\ref{26}). The coefficients $\phi_i(T)$ can be explicitly calculated for large $i$ by means of a symbolic manipulation program \cite{wolfram1996mathematica}, handling the tensorial operations on the replica indices \cite{castellana2010hierarchical, castellana3}. This computation is carried on for integer $n$ and an analytic function of $n$ is obtained, so that the limit $n\rightarrow 0$ can be safely taken. In Appendix \ref{app0} we give an  example of how these computations are performed, by doing the explicit calculation of the coefficient $\phi_0$.\\ In the following, the expansion (\ref{31}) will be worked out at a fixed order $l$, under the underlying assumption that the resulting free energy 
\be  \label{32}
f_l( T) \equiv   \sum_{i=0}^ {l} C^{2i} \phi _i(T)
\ee
approximates the exact free energy (\ref{31}) as $l$ is large
\[
f_l(T) \overset{l \rightarrow \infty} {\rightarrow} f_{\infty}(T) = f[T, \{ \epsilon \} ]. 
\]
\medskip\\

Before discussing the result of this computation for $0<\sigma<1$, it is interesting to test perturbation theory in the region $\sigma<0$  for the following reason. As stated in Section \ref{hm}, for $\sigma<0$ the thermodynamic limit of the model is ill-defined. This is because the interaction energy $\epsilon_k$ defined in Eq. (\ref{1}) grows with $k$ faster than the volume $2^ k$ according to Eq. (\ref{42}). Notwithstanding this, having the HREM  $2^k$ spins, one can redefine the inverse temperature 
\be \label{39}
\beta \rightarrow  2^ {k \sigma/2} \beta,
\ee
 in such a way that the variance of $\epsilon_k$ defined in Eq. (\ref{42}) becomes 
\be \label{34}
\mathbb{E}_\epsilon [ \epsilon_k^2 ] \rightarrow 2^ k.
\ee 
 The thermodynamic limit is now well-defined, because the coupling energy scales as the volume, and the model is a purely mean-field one. A direct numerical inspection of the expansion (\ref{32}) after such a redefinition of $\beta$ for $\sigma<0$ shows that as $l$ is increased the free energy of the HREM converges to that of a REM \cite{derrida1980random, derrida1981random} with critical temperature 
\be  \label{33}
T_{c\,U}^ {\sigma<0} \equiv \frac{1}{2 \sqrt{\log 2 (1-2^ \sigma)}}. 
\ee
 The label  U in Eq. (\ref{33}) stands for uncorrelated, because the value (\ref{33}) of the critical temperature can be easily worked out by hypothesizing that the energy levels are uncorrelated as in the REM. Indeed, the fact that the free energy  (\ref{32})  converges to that of the REM for $\sigma<0$ tells us that in this region correlations are irrelevant, and the model reduces to a purely mean-field one with the same features as the REM. This is  what we expected from the fact that the energy scales as the system volume (Eq. (\ref{34})), and serves as an important test of the perturbative expansion (\ref{31}).\medskip \\
 
\begin{centering}
  \begin{figure}[htb] 
\centering
\includegraphics[width=10cm]{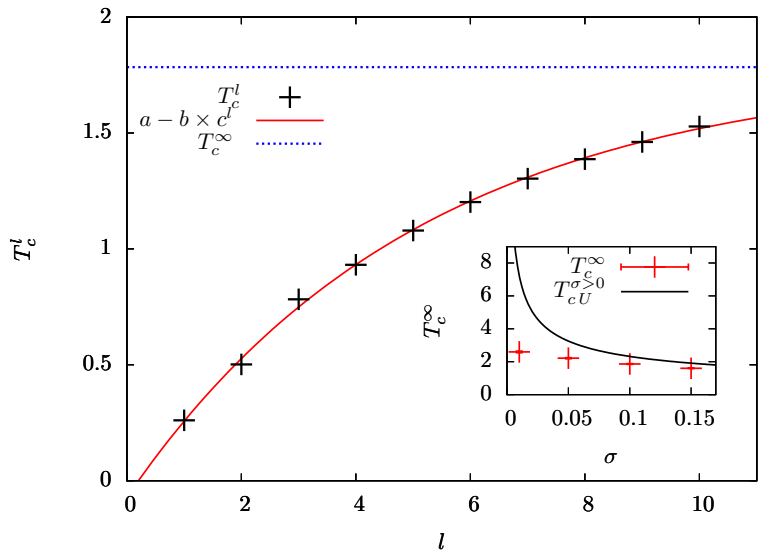}
\caption{Critical temperature $T_c^l$ (black points), its fitting function $a-b \times c^ l$ (red solid curve) and asymptotic value $T_c^ {\infty}=1.861\pm0.021$ (blue dashed line) as a function of $l$ for $\sigma = 0.1$. $T_c^ {\infty} = a$ has been determined as a fit parameter. Inset: critical temperature $T_c^ {\infty}$ (red points) in the region $0<\sigma<0.15$ where the first $10$ orders of  the perturbative expansion show a clear convergence, and critical temperature $T_{c\,U}^{\sigma >0}$ (black curve), as a function of $\sigma$. The error bars on $T_c^ {\infty}$ are an estimate of the error resulting from the fit on the parameter $a$. 
 $T_c^ {\infty}$  is clearly non-consistent with $T_{c\, U}^ {\sigma>0}$, showing that correlations between energy levels are important. 
The $\sigma \rightarrow 0^+$-limit of $T_c^\infty$ does not coincide with the $\sigma \rightarrow 0^-$-limit of $T_{c\, U}^{ \sigma <0}$ because of the abrupt change (\ref{39}) in the normalization of the temperature when switching from $\sigma >0$ to $\sigma <0$. 
}
\label{fig2}
\end{figure}
\end{centering}

 We now focus on the region $0<\sigma<1$. From a direct analysis of the data for the free energy $f_l(T)$, it turns out that there exists an $l$-dependent critical temperature $T_c^ l$, defined in such a way that the entropy at the $l$-th order in $C^2$ vanishes at $T=T_c^l$
 \be \label{36}
 s_l(T_c^ l) \equiv - \left. \der{ f_l(T)}{ T}\right|_{T=T_c^ l } =0. 
 \ee
 As discussed in Section \ref{historical_outline}, in the REM the fact that the entropy vanishes at a given temperature signals a Kauzmann phase transition. Hence,  by definition $T_c^ l$ can be considered as the $l$-th order critical temperature of the system. Since perturbation theory is approximate, and there is no guarantee that a perturbative expansion converges at a critical point \cite{huang, zinnjustin, zinn2007phase, parisi1998statistical}, it is important to check the behavior of $T_c^ l$ as $l$ is increased. In Fig. \ref{fig2}, $T_c^ l$ as a function of $l$ is depicted for $\sigma = 0.1$. Even for $l\leq 10$, a clear convergence is observed, and the resulting `exact'  critical temperature $T_c^ {\infty}$  is easily determined by fitting $T_c^ l$ vs. $l$ with a function of the form $a-b\times c^ l$, with $c<1$, and setting $T_c^ {\infty} = a$. In this way, $T_c^ {\infty}$ as a function of  $\sigma$ is determined in the region $0\leq\sigma\leq 0.15$, where  $T_c^ l$ vs. $l$ for $l\leq 10$ exhibits a clear convergence as a function of $l$, and the extrapolation for $l\rightarrow \infty$ is meaningful.\\
 
   According to Eq. (\ref{36}),  \mnote{The HREM has a finite temperature phase transition \` a la Kauzmann.} the  entropy of the HREM
\be \label{37}
s(T) \equiv - \der{f_\infty(T)}{T}
\ee
    vanishes for $T=T_c^ {\infty}$. This allows a straightforward interpretation of the phase transition occurring at $T=T_c^ {\infty}$, resembling to that occurring in the REM \cite{derrida1980random}: for $T>T_c^ {\infty}$ the entropy is positive, and the system explores an exponentially large number of states in the configuration space, while for $T< T_c^ {\infty}$ the system is trapped in a handful of low-lying energy states. We have thus shown that the HREM  undergoes a phase transition \` a la Kauzmann at a finite temperature $T_c^ {\infty}$, whose features are similar to that of the phase transition of the REM and, more generally, of mean-field structural glasses \cite{biroli2009random}.
 \\

  In the inset of Fig. \ref{fig2},  $T_c^ {\infty}$ as a function of $\sigma$ is depicted, and $T_c^{\infty}$ turns out to be a decreasing function of $\sigma$. This fact is physically meaningful, because according to Eq. (\ref{def_c}), the larger $\sigma$ the smaller the coupling $C$ between spins, and so the smaller the temperature $T_c^ {\infty}$ such that for $T<T_c^ {\infty}$ all the spins are frozen in a low-lying energy state. \\
  As in the $\sigma<0$-case,  we can hypothesize that the energy levels act as uncorrelated random variables, in such a way that the HREM behaves as a REM. In this case, the critical temperature can be computed exactly, and is given by 
 \be  \label{33}
T_{c\,U}^ {\sigma>0} \equiv \frac{1}{2 \sqrt{\log 2 (1-2^ {-\sigma})}}. 
\ee
Differently from the $\sigma<0$-case, here the decorrelation hypothesis turns out to be wrong. Indeed, by looking at the inset of Fig \ref{fig2}, $T_c^ {\infty}$ does not coincide with $T_{c\, U}^ {\sigma>0}$. This fact is a clear evidence that correlations between the energy levels  play a crucial role in the region $\sigma>0$, and cannot be neglected. \\
 
\begin{centering}
  \begin{figure}[htb] 
\centering
\includegraphics[width=10cm]{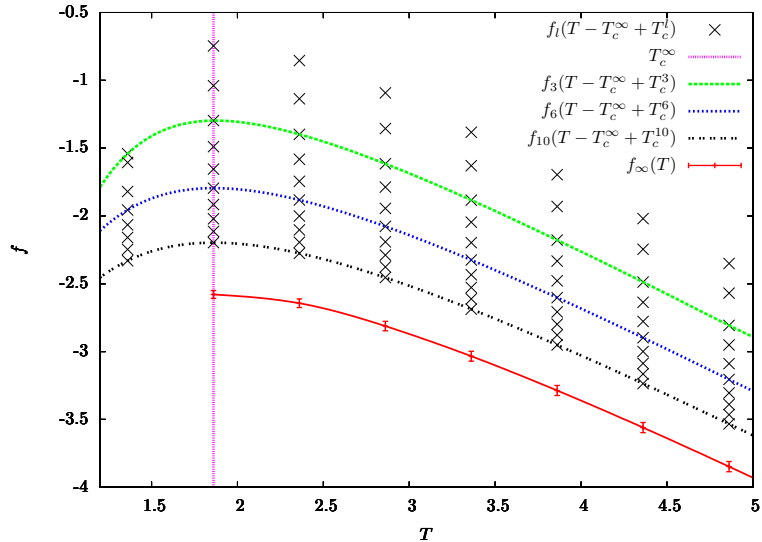}
\caption{Free energy $f$ as a function of temperature $T$ for $\sigma=0.1$. We depict $f_l(T-T_c^ {\infty} + T_c^ l)$  for $l=1,2,\cdots,10$ (gray points), $f_3(T-T_c^ {\infty} + T_c^ 3)$   (green dashed curve),   $f_6(T-T_c^ {\infty} + T_c^ 6)$   (blue dashed curve), $f_{10}(T-T_c^ {\infty} + T_c^ {10})$   (black dashed curve), and the extrapolated free energy $f_\infty(T)$ as a function of $T$ (red points and solid curve). We also depict the critical temperature $T_c^{\infty}$ (violet dashed line). For any fixed $T$, $f_\infty(T)$ has been obtained by fitting the sequence $f_l(T-T_c^ {\infty} + T_c^ l)$ vs. $l$ with a function of the form $a-b\times c^ l$, with $c<1$, and setting $f_\infty(T)=a$. To compute $f_\infty(T)$, we used the sequence $f_l(T-T_c^ {\infty} + T_c^ l)$ vs. $l$ instead of $f_l(T)$ vs. $l$ because the former has the same limit as the latter for $l\rightarrow \infty$, and exhibits a better convergence for the accessible values of $l\leq 10$. The  error bars on $f_\infty(T)$ are given by an estimate of the fit error on the parameter $a$.}
\label{fig3}
\end{figure}
\end{centering}

 In Fig \ref{fig3} the free energy $f_l(T)$ as a function of the temperature $T$ for $\sigma = 0.1$ and different values of $l\leq 10$ is depicted. $f_l(T)$ is found to converge to a finite value $f_{\infty}(T)$ for $T>T_c^ {\infty}$, while for $T<T_c^ {\infty}$ the lower the temperature the worse the convergence of the sequence $f_l(T)$ vs. $l$.  Hence, when descending into the low-temperature phase from $T>T_c^ {\infty}$, a breakdown of perturbation theory occurs, signaling the possibility of a nonanalyticity of the free energy at the critical point, resembling to the nonanalytical behavior of physical quantities occurring in second-order phase transitions for ferromagnetic systems \cite{huang, landau1980statistical, zinnjustin, zinn2007phase, parisi1998statistical}. \bigskip\\
 
According to the above discussion, the perturbative expansion (\ref{31}) yields  a reliable method to estimate physical quantities in the high-temperature phase $T>T_c^ {\infty}$. Notwithstanding this, no conclusions can be drawn on the behavior of the free energy in the low-temperature phase with this perturbative framework. In particular, this method gives no insight into the structure of the states of the system in the low-temperature phase. An interesting approach yielding a tentative solution in the low-temperature phase can be worked out by hypothesizing that the $n$ replicas $\vec S_1, \cdots, \vec S_n$  in Eq. (\ref{29}) 
are grouped into $n/x$ groups, where each group is composed by $x$ replicas \cite{MPV,0}. For any two replicas
$a, b$ in the same group one has $\vec S_a = \vec S_b$.
We can look at the small $C^2$-expansion  (\ref{31}) in the particular case where the replicas are grouped as described above. We call the free energy  to the  $l$-th order obtained with this ansatz $f_l^ {\textrm{RSB}}(T,x)$.  $\textrm{RSB}$  stands for replica-symmetry-breaking, and has the same physical interpretation as the ordinary RSB mechanism described in Section \ref{historical_outline} for structural glasses: as in the REM, a replica-symmetry-broken structure in the low-temperature phase implies that the system is no more ergodic, because it is trapped in a handful of low-lying energy states \cite{biroli2009random, MPV,NishimoriBook01,derrida1980random,derrida1981random, parisi1980order,parisi1983order,0}.   By performing the computation explicitly, it is easy to find out that $f_l^{ \textrm{RSB}}(T,x) =  f_l(T/x)$. According to the general prescriptions of the replica approach \cite{MPV,0, parisi1980order, parisi1979infinite, parisi1983order}, as we take the $n\rightarrow 0$-limit the parameter $x$, originally defined as an integer number, has to be treated as a real number lying in the interval $[0,1]$. Hence, the maximization of $f_l^ {\textrm{RSB}}(T,x)$ with respect to $x$ gives $x = 1$ for $T \geq T_c^ l$, and $x = T/T_c^l$ for $T < T_c^ l$. It follows that according to this this RSB ansatz the exact free energy reads
\bea \label{38}
f^ {\textrm{RSB}}_\infty(T) & =  &  \left\{ 
\begin{array}{cc}
f_\infty(T_c^ \infty) & T< T_c^ \infty \\
f_\infty(T) & T \geq T_c ^ \infty 
\end{array}
\right . .
\eea
The form (\ref{38}) of the RSB free energy is the  same as that of the REM \cite{MPV,derrida1980random,derrida1981random}, and predicts that in the low-temperature phase the HREM has a one-step RSB, reflecting ergodicity breaking. On the one hand,  this RSB ansatz predicts a free energy $f_\infty(T)$ which is exact for $T \geq T_c^ {\infty}$, because it coincides with the free energy computed with perturbation theory without making use of any ansatz. On the other hand, there is no guarantee that $f^ {\textrm{RSB}}$ is  exact in the low-temperature phase. In particular, the $n$ replicas could be grouped in a more complicated pattern than the RSB one described above, and this configuration could yield a free energy that is larger than $f^ {\textrm{RSB}}$ for $T<T_c^ \infty$. Since in the replica method the exact free energy is not the minimum, but the maximum of the free energy as a function of the order parameter configurations \cite{0,MPV}, such a more complicated pattern would yield the exact free energy of the system. 
The investigation of the existence of such an optimal pattern is an extremely interesting question that could be subject of future work, and give some insight into  the low-temperature phase of the HREM, and more generally into the low-temperature features of non-mean-field structural glasses.\medskip\\

Once the existence of a phase transition has been established,  we ask ourselves what are its physical features. In particular, an interesting question is whether, as in second-order phase transitions \cite{huang, wilson1974renormalization, wilson1983renormalization, zinnjustin, zinn2007phase}, the system has no characteristic scale length at the critical point. Indeed, answering   this question for the HREM is particularly interesting, because an analysis of the characteristic length scales of the system in the critical region could give some insight into the construction of a RG theory for non-mean-field structural glasses. 

\chapter{Spatial correlations of the model}\label{corr_hrem}

\begin{centering}
  \begin{figure}[htb] 
\centering
\includegraphics[width=9cm]{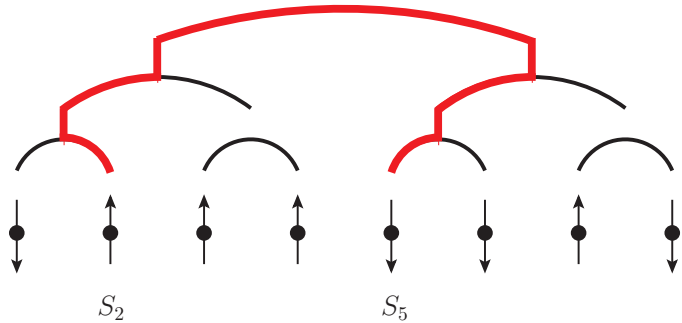}
\caption{Ultrametric distance between spins $S_2$ and $S_5$ in a HREM with $k=3$. In order to find a  root shared by  $S_2$ and $S_5$, one has to go  $2$ levels up  in the binary tree. Hence, the ultrametric distance between $S_2$ and $S_5$ is $m=2$.}
\label{fig4}
\end{figure}
\end{centering}

Being a non-mean-field model, the HREM allows for the definition of a distance between spins. This definition is yield naturally by the hierarchical structure of the couplings shown in Fig. \ref{fig1}. Indeed, given two spin sites $i$ and $j$, one can define  their ultrametric distance $m$ as the number of levels one has to get up in the binary tree starting from the leaves, until one finds a root that is shared by $i$ and $j$. This geometrical construction of the ultrametric distance is depicted in Fig. \ref{fig4} for a HREM with $k=3$. One can thus define the distance between $i$ and $j$ as 
\be
\parallel i-j \parallel \equiv 2^ m.
\ee

Once the notion of distance has been clarified, we want to know if the system has a characteristics length defined in terms of this distance, and what is the behavior of this length in the critical region. In order to do so \cite{castellana3}, we  define a correlation function whose exponential decay at large distances  yields a characteristic length scale $\xi$ of the system.  This  correlation  function is defined as 
\be \label{100_a}
Y(2^m,T)\equiv  \mathbb{E}_\epsilon \left[ \mathbb{E}_{\vec{S}_1,\vec{S}_2} \left[ \prod_{i=1}^ {2^m}  \delta_{S_{1,i}, S_{2,i}} \right] \right]\overset{m\rightarrow\infty}{\equiv} \exp\left[ {-\frac{2^m}{\xi(T)}} \right] ,
\ee
where $\mathbb{E}_{\vec{S}}$  stands for the thermal average  
\be \label{102}
\mathbb{E}_{\vec{S}}[\mathcal{O}[\vec{S}]] \equiv \frac{\sum_{\vec S} e^ { -\beta \, H_k[\vec S] }\mathcal{O}[\vec S]}{Z[T,\{ \epsilon \}]},
\ee
 and $\delta_{S_i,S_j}$ denotes the Kronecker delta function. The correlation function (\ref{100_a}) has the following physical meaning. Given two spin configurations $\vec S_1,\, \vec S_2$, $Y(2^ m,T)$ physically represents the mean overlap between  $\vec S_1$ and  $\vec S_2$ on the sites $1,\cdots, 2^ m$ of the lattice.\medskip\\
 
Before studying the behavior of $\xi$ in the region $0 < \sigma < 1$, we compute $\xi$ in the region $\sigma<0$ where the model is purely mean field. As discussed in Chapter \ref{pert_hrem}, for the thermodynamic limit to be well-defined for $\sigma <0$, one has to rescale the temperature according to Eq. (\ref{39}). By plugging Eq. (\ref{39}) into Eq. (\ref{100_a}) and taking $\sigma<0$, one easily obtains the correlation function in the mean-field case
\bea \label{111}
Y(2^ m, 2^ {-k \sigma /2}T) & \overset{k\rightarrow\infty}{=} & \frac{\sum_{\vec S_1, \vec S_2 } \prod_{i=1}^ {2^m}  \delta_{S_{1,i}, S_{2,i}}}{2^ {2\, 2^{ k}}}\\ \no 
& = & \exp(-2^ m \log 2).
\eea
Comparing Eq. (\ref{111}) to the definition of $\xi(T)$ in Eq. (\ref{100_a}), we obtain the mean-field value of the correlation length 
\be  \label{112b}
\xi_{MF}(T) = \frac{1}{\log 2}. 
\ee
Eq. (\ref{112b}) is consistent with the fact that in the mean-field case there must be no notion of  physical distance between spins \cite{NishimoriBook01}, and so the system has no  physical length scale signaling the range of  spatial correlations between spins.  \medskip\\

 This picture should radically change for $0< \sigma < 1$, where a physical spatial structure and distance does exist.  In  Chapter \ref{pert_hrem} we showed that  the HREM has a phase transition at $T_c^ \infty$. According to the above physical meaning  of the correlation function (\ref{100_a}), one expects long-range  spatial correlations to occur at $T_c^ \infty$, because for $T\rightarrow T_c^ {\infty}$ both  $\vec S_1$ and  $\vec S_2$ should stay trapped in the same handful of low-lying energy states, and exhibit a high degree of overlap with each other. Hence, $Y(2^ m,T)$ should tend to $1$, in such a way that $\xi$ diverges. \\
    
 In the following we  compute $\xi$ for $0<\sigma <1$ in the same perturbative framework as in Chapter \ref{pert_hrem}, to investigate the existence of such a long-range spatial correlations at the critical point. Firstly, Eq.  (\ref{100_a}) can  be rewritten with the replica trick
\bea \label{104}
 Y(2^m,T) & = & \mathbb{E}_\epsilon  \left[ \frac{\sum_{\vec {S}_1, \vec {S}_2} e^ { -\beta ( H_k[\vec {S}_1] + H_k[\vec {S}_2] ) }  \prod_{i=1}^ {2^m}  \delta_{{S_{1,i}, S_{2,i}}} }{{Z[T,\{ \epsilon\} ]}^2} \right]   \\ \no 
 & = & \mathbb{E}_\epsilon  \left[ \lim_{n\rightarrow 0} \sum_{\vec{S}_1,\cdots,\vec{S}_n} e^ { -\beta \sum_{a=1}^ {n}\, H_k[\vec{S}_a] }  \prod_{i=1}^ {2^m}  \delta_{{S_{1,i}, S_{2,i}}} \right] \\ \no
  &=& \lim_{n\rightarrow 0} \sum_{\vec{S}_1,\cdots,\vec{S}_n} \exp  \left(  \frac{\beta^2}{4} \sum_{j=0}^ k C^{2j} \sum_{i=1}^ {2^ {k-j}} \sum_{a,b=1}^ {n} \delta_{\vec{S}^{(j,i)}_a, \vec{S}^{(j,i)}_b} \right) \prod_{i=1}^ {2^m}  \delta_{{S_{1,i}, S_{2,i}}}.
\eea
The last line of Eq. (\ref{104}) is very similar to (\ref{29}), used in Chapter \ref{pert_hrem} to compute the free energy of the HREM. Hence, the very same techniques used to compute $f$ in perturbation theory can be employed here to calculate the correlation function $Y$. In particular, one can  expand the correlation function (\ref{104})  in the coupling constant  $C^ 2$
 \be 
  Y(2^m,T) = \sum_{i=0}^ {\infty} C^{2 i}\Upsilon_{m,i}(T),
\ee
and  explicitly evaluate the coefficients $\Upsilon_{m,i}$ by a symbolic manipulation program \cite{wolfram1996mathematica} until the order $i=9$.  In Appendix \ref{app1} we present the steps of the computation  of  $\Upsilon_{m,0}(T)$, to give some insight into the main techniques employed in the calculation to high orders.  \medskip\\ 

For any fixed $m$ and $T$, the exact value of  $Y(2^m,T)$ has been computed by extrapolating the sequence 
\be  \label{40}
Y_l(2^m,T) \equiv \sum_{i=0}^ {l} C^ {2i} \Upsilon_{m,i}(T) 
\ee
 to $l\rightarrow \infty$, with the underlying assumption that for large $l$ Eq. (\ref{40}) converges to the exact value of the correlation function 
\[
Y_l(2^ m,T) \overset{l \rightarrow \infty} {\rightarrow} Y_{\infty}(2^ m,T) = Y(2^ m,T). 
\]
 
\begin{figure} 
\centering  
\includegraphics[scale=0.8]{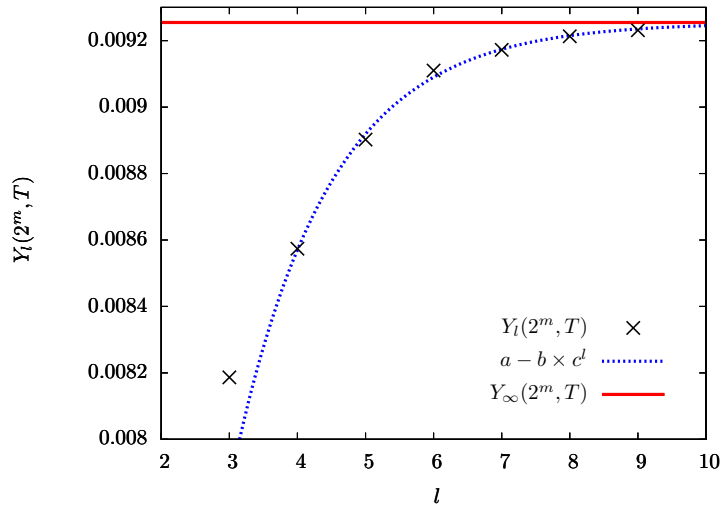}
\caption{$Y_l(2^ m,T)$ (black points), its  fitting function   $a -b\times  c^l$ (blue dashed curve) as a function of $l$, and $Y_\infty(2^ m,T) = a$ (red solid line), determined as a fit parameter. Here $\sigma = 0.1, m=3$ and $T=3.5$. }   
\label{fig5}  
\end{figure}

  The sequence $Y_l(2^ m,T)$ as a function of $m$ for fixed $\sigma, l$ and $T$ is shown in Fig. \ref{fig5} for $m=3, T=3.5$. Even though $Y_l$ is nicely convergent even to relatively low orders for the  values of $m$ and $T$ considered in Fig. \ref{fig5}, an explicit analysis of  $Y_l(2^ m,T)$ for different values of $m$ shows that the larger $m$, the larger the number of orders needed to see a nice  convergence  with respect to $l$. This fact can be easily understood by recalling that the $C^2$-expansion is equivalent to a high-temperature expansion (see Chapter \ref{pert_hrem}). It is a general feature of high-temperature expansions \cite{zinn1979analysis,campostrini1999improved,sykes1972high,daboul2004test} that with a finite number of orders of the $\beta$-series,  one cannot describe arbitrarily large length scales. Hence, with a finite number of orders ($9$ in our case) for $Y_l(2^m,T)$, one cannot describe the correlations $Y_l(2^m,T)$ for too large $m$. 
  \\
  Another important fact is that, for any fixed $m$ the convergence of $Y_l(2^ m,T)$ gets worse as the temperature $T$ is decreased, because more terms in the $\beta$-expansion, and so in the $C^ 2$ expansion, are needed.\\
  Practically speaking, these limitations of the perturbative expansion made us  take $m\leq 3$ and $T>T_0$, where $T_0$ is a $\sigma$-dependent value of the temperature signaling a breakdown of perturbation theory. As we will discuss in the following, notwithstanding the  very small values of $m$ here available, it has been possible to compute the correlation length $\xi(T)$ defined in Eq. (\ref{100_a}) in a wide interval of temperatures.\medskip\\
 
  The correlation length $\xi(T)$ has been computed for every temperature $T$ by fitting the data for $Y_\infty(2^ m,T)$ vs. $m$, according to the definition of $\xi(T)$ given in Eq. (\ref{100_a}).
  
\begin{figure} 
\centering  
\includegraphics[width=10cm]{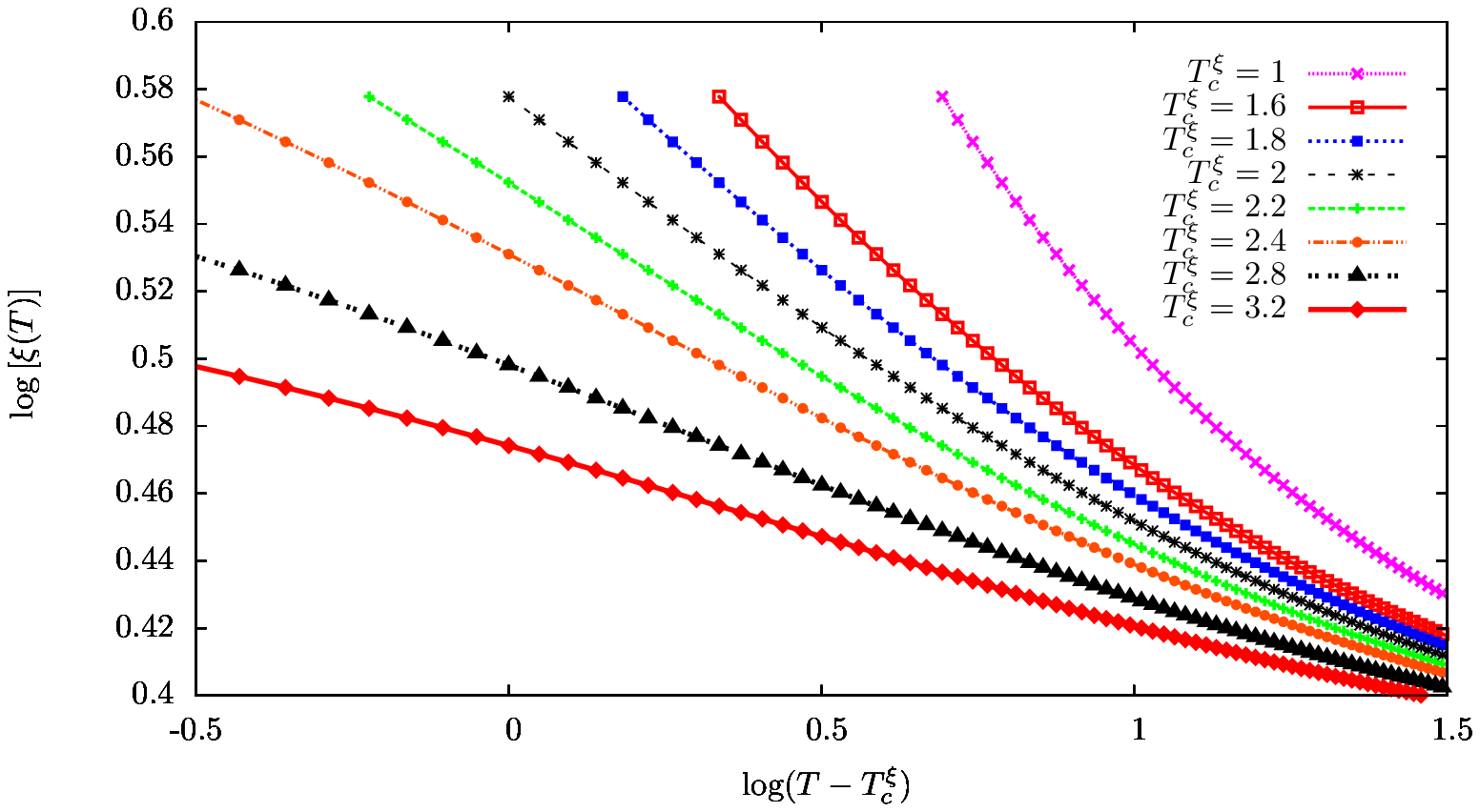}
\vspace{1cm}\\
\includegraphics[width=10cm]{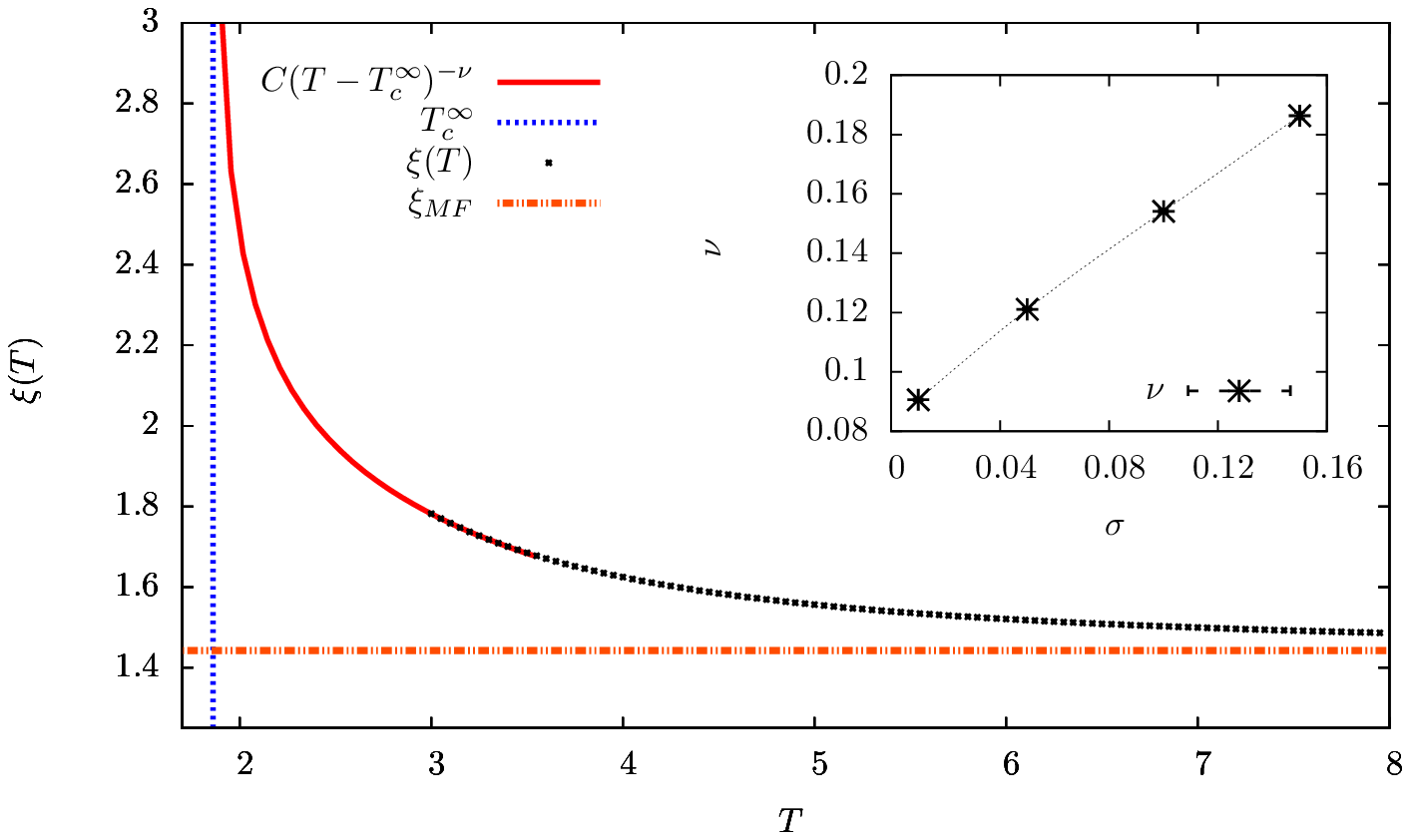}
\caption{Top: $\log\left[ \xi(T) \right]$ vs. $\log(T-T_c^ \xi)$ for different tentative values of $1 < T_c^ \xi < 3.2$ and $\sigma=0.1$. The value of $T_c^ \xi$ such that $\xi(T)$ best fits with the hypothesis (\ref{107}) is such that $\log\left[ \xi(T) \right]$ vs. $\log(T-T_c^ \xi)$ best fits with a straight line. This value lies in the interval $[1.8,2]$, and is thus consistent with the Kauzmann transition  temperature $T_c^ \infty=1.861\pm0.021$ of the model.
Bottom: $\xi(T)$ (black points), its fitting function (\ref{107}) with $T_c^ \xi = T_c^{\infty}$ (solid red curve),  $\xi_{MF}$ (orange dashed line) as a function of $T$ for $\sigma=0.1$, and  the Kauzmann transition temperature $T_c^{\infty}$ (blue dashed line). In the region $0<\sigma < 1$, the correlation length $\xi(T)$ at low temperature  is significantly larger than the mean-field value $\xi_{MF}$, because of the existence of physical spatial correlations between spins. Inset:  critical exponent $\nu$ defined by Eq. (\ref{107}) as a function of $\sigma$. $\nu$ has been computed by setting $T_c^  \xi = T_c^ \infty$ and fitting $\xi(T)$ with Eq. (\ref{107}). Error bars on $\nu$ are an estimate of the uncertainty resulting from the determination of $\nu$ as a fit parameter. }
\label{fig6_7}  
\end{figure}

Once $\xi(T)$ is known, we investigate its behavior at low temperatures. As stated above, one cannot take too low values of $T$ because of the non-convergence of the perturbative expansion. Notwithstanding this,  it is still possible to approach enough the critical point $T_c^ \infty$ and investigate the existence of long-range spatial correlations.   In particular, we   test the validity of the hypothesis of a diverging $\xi(T)$ for $T \rightarrow T_c^ \infty$. In order to do so, we check whether the data for $\xi(T)$ is consistent with a power-law divergence at some temperature $T_c^ \xi$ 
 \be  \label{107}
 \xi(T)  \overset{T \rightarrow T_c^ \xi }{\approx \, \, \,} \frac{C}{(T-T_c^ \xi)^ \nu} . 
 \ee 
 The validity of the hypothesis (\ref{107}) has been tested in the following way. We suppose that Eq. (\ref{107}) holds, and determine the value of $T_c^ \xi$ such that the data for $\xi(T)$ best fits with Eq. (\ref{107}). We fit the data for  $\log\left[\xi(T)\right]$ vs. $\log(T-T_c^ \xi)$ for different values of $T_c^ \xi$. The value $T_c^ \xi$ such that  $\log\left[\xi(T)\right]$ vs. $\log(T-T_c^ \xi)$ best fits  with a straight line, is such that  the data for $\xi(T)$ is consistent with a power-law divergence at $T_c^ \xi$, according to  (\ref{107}).  The top panel of Fig. \ref{fig6_7}  shows that for $\sigma=0.1$ the optimal value of $T_ c^ \xi$ is compatible with the critical temperature  $T_c^ \infty$ for $\sigma=0.1$ obtained in Chapter \ref{pert_hrem}.\\
 
  The data \mnote{The data for the correlation length of the HREM is consistent with a power-law divergence at the Kauzmann transition temperature.} for $\xi(T)$ is thus consistent with a diverging correlation length at the Kauzmann transition temperature $T_c^ {\infty}$. In the bottom panel of Fig. \ref{fig6_7}, $\xi(T)$ as a function of $T$ for $\sigma=0.1$ is depicted, together with its fitting function (\ref{107}) with  $T_c^ \xi=T_c^ {\infty}$. $\xi(T)$ increases as the temperature is decreased, and its shape  is compatible with a power-law divergence at the Kauzmann transition temperature $T_c^ {\infty}$. \vspace{2cm}\\

Since this work establishes the existence of a thermodynamic phase transition and the possibility of a diverging correlation length in the HREM, it also shows the way forward to the study of finite-dimensional spin or structural-glass models. \\
For instance, the REM has been found  \cite{dynamical_rem,bouchaud2002lecture}  to have  a dynamical phase transition at finite temperature if a particular dynamics is chosen: it would be interesting to study these  dynamical properties in the HREM, by introducing some suitable dynamics of the spins, and by investigating the existence of a dynamical phase transition, and of a diverging dynamical correlation length. \\
Another  interesting future direction would be to generalize the techniques used in the solution of the REM to more realistic non-mean-field spin or structural-glass models, like  $p$-spin models \cite{crisanti1992spherical} built on a hierarchical lattice. Indeed, even though the HREM serves as a model representing a non-mean-field structural glass, its structure is still far from being realistic: strictly speaking, the spins $\vec S$ in the HREM are not  physical degrees of freedom, but serve as mere labels for the energy variables $H_k[\vec S]$ as in the REM \cite{derrida1980random,  0, NishimoriBook01}. Moreover, a criticism one could address to the solution techniques developed in Chapters \ref{pert_hrem} and \ref{corr_hrem} is that these do not give any insight into the construction of a suitable RG protocol. In particular, a decimation rule on spins is still lacking. \\

In the following Part we  introduce a HM of a spin glass, in the effort to address these two points: in such a HM spins are real physical degrees of freedom, in such a way that a decimation  rule on spins  naturally emerges,  and one can explicitly solve the resulting RG equations.

  \part{The Hierarchical Edwards-Anderson Model}\label{hea}

As discussed in Section \ref{historical_outline}, the SK model is a mean-field spin glass, whose critical properties have been clarified in Parisi's solution \cite{parisi1979infinite, parisi1980order, parisi1983order,MPV, 0, NishimoriBook01}. Whether the physical features of this solution persist also in a non-mean-field version of the SK, or the non-mean-field case is described by a radically different scheme \cite{mcmillan1984scaling,fisher1986ordered,fisher1987pure,fisher1987absence,fisher1988equilibrium,fisher1988nonequilibrium,bray1986heidelberg,moore1998evidence}, is one of the most hotly debated topics in statistical physics. Being this a very difficult question, it is interesting to consider non-mean-field versions of the SK that presumably capture all the non-mean-field physics, and that are simple enough to be solved quite easily, by means of an explicit implementation of the RG transformation. \\

Starting from  the general definition (\ref{1}) of Hierarchical models given in Section \ref{hm}, here we explicitly chose the random energies $\epsilon_k$ to build up a non-mean-field version of the SK model having these features: the Hierarchical Edwards-Anderson model (HEA) \cite{franz2009overlap,castellana2010renormalization,castellana2011renormalization,castellana2011real}. The HEA is defined by choosing the random energy $\epsilon_k$ in the following way \cite{franz2009overlap}
\be \label{43}
\epsilon_{k+1}[S_1, \ldots, S_{2^ {k+1}} ] \equiv - \left( \frac{C^2}{2} \right)^ {k+1} \sum_{i<j=1}^ {2^ {k+1}} J_{12,ij} S_i S_j,
\ee
where $C$ is still defined by Eq. (\ref{def_c}), and $J_{12, ij}$ are Gaussian random variables with zero mean and unit variance. \\

Starting from Eq. (\ref{43}) and from the definition of the $J_{12}$s, it is easy to show that $\epsilon_{k+1}$ has zero mean, and that its variance satisfies
\be \label{175}
\mathbb{E}_\epsilon[\epsilon_k^2  ] \overset{k \rightarrow \infty}{\sim} 2^{2k(1-\sigma)}. 
\ee
It follows that for $\sigma<1/2$ the interaction energy (\ref{175}) grows faster than the volume $2^k$, and the thermodynamic limit $k \rightarrow \infty$ is not defined. 
The purely mean-field case, i. e. the case where the interaction energy scales with $k$  as the volume $2^k$, is recovered for $\sigma=1/2$. 
Moreover, for $\sigma>1$ the interaction energy (\ref{175}) goes to $0$ as $k \rightarrow \infty$, in such a way that no phase transition can occur. Hence, in the following we will take
\be \label{sigma_hea}
1/2<\sigma <1 ,
\ee 
which is the equivalent of Eq. (\ref{27}) for DHM. \\

Physically speaking, the interaction energy (\ref{43}) of the HEA introduces two-spin interactions, while the interaction energy of the HREM  defined in Part \ref{hrem}  has  $2^{k+1}$-spin interactions. Another fundamental difference between the HEA and the HREM is that, according to the general classification of models with quenched disorder given in Section \ref{historical_outline}, the HEA has to be considered as a model for a spin glass, while the HREM as a model mimicking the physics of a structural glass \cite{biroli2009random}.  Compared to the HREM, the HEA is  a more realistic model, because according to the definition (\ref{43}), here the spins of the system are  physical degrees of freedom, and  not mere labels for the energy variables as in the REM and in the HREM. As we will see in the following, the HEA also allows for an explicit construction of a suitable decimation rule on spins, and so of a RG transformation.\\

An equivalent definition of the HEA can be given without using the recursion relation (\ref{1}). Indeed, one can recover Eq. (\ref{1}) by defining the  HEA as a system of $2^ k$  spins $S_i=  \pm 1,\, 0 \leq i \leq 2^ k -1 $, with  Hamiltonian
\be \label{214}
H_k[\vec S] = -\sum_{i,j=0}^ {2^ k-1}J_{ij}S_i S_j,
\ee 
where the $J_{ij}$s are Gaussian random variables with zero mean and  variance $\varsigma^2_{ij}$.   $\varsigma^2_{ij}$ is given by the following expression: 
consider the binary representation of the points $i,j$
\[
 i  = \sum_{a=0}^ {k-1} c_a 2^ {k-1-a},\,  j  = \sum_{a=0}^ {k-1} d_a 2^ {k-1-a}. 
\]
If  only the last $m$ digits $\{ c_{k-m}, \cdots,c_{k-1} \}$ of the binary representation of $i$ are different from the last $m$ digits $\{ d_{k-m}, \cdots,d_{k-1} \}$ of the binary representation of $j$, one has 
\be  \label{44}
\varsigma^2_{ij}= 2^{-2\sigma m}.
\ee
 The  definition (\ref{214}) is equivalent to the definition given by (\ref{1}) and (\ref{43}).\\

 The form (\ref{214}) of the Hamiltonian  can be obtained by dividing the system in hierarchical embedded blocks of size $2^m$, as shown in Fig. \ref{fig1}. More precisely, the integer $m$  can be considered as the ultrametric distance between spins $S_i$ and $S_j$ defined in Fig. \ref{fig4} for the HREM. \\

The HEA is a hierarchical counterpart of the one-dimensional spin glass with power-law interactions (PLSG) \cite{kotliar1983one}, which has received attention recently \cite{katzgraber2008computer,katzgraber2003monte,katzgraber2005probing,leuzzi1999critical,leuzzi2008dilute}. The only difference between this PLSG  and the HEA is the form of $\varsigma^2_{ij}$. In the PLSG, Eq. (\ref{44}) is replaced by $\varsigma^2_{ij}= \lvert i - j  \rvert^ {-2 \sigma}$, where $\lvert i - j  \rvert$ denotes the ordinary absolute value of $i-j$. 
This form of the interaction structure, even though apparently simpler than that of the HEA, makes the implementation of a RG transformation extremely harder to pursue practically. Indeed, the form (\ref{44}) of the interactions of the HEA keeps  track of the hierarchical structure of the model, and so of a symmetry that is absent in the PLSG. Differently from the PLSG, thanks to this symmetry the  RG equations of the HEA allow for a direct solution that can be in principle computed with extremely high precision \cite{castellana2010renormalization,castellana2011renormalization}, as we will show in Chapter \ref{replica_approach}. \\

We now proceed by exposing the techniques developed to solve the HEA \cite{castellana2010renormalization,castellana2011renormalization,castellana2011real}. An important observation is that an explicit evaluation of the partition function for large $k$ is practically impossible, because this would involve $2^ k$ spins, and so a sum of $2^ {2^k}$ terms. Moreover, as discussed in Section \ref{historical_outline}, there is no guarantee \cite{huse2011private} that even the most refined MC techniques \cite{parallel_tempering1, parallel_tempering2,hasenbusch2008critical, banos2010nature, alvarez2010static, belletti2009depth} work properly at low temperatures for reasonably large system sizes, because of the existence of many metastable minima in the energy landscape. \\
An interesting method to overcome these difficulties is to use the hierarchical structure of the model. Indeed, the recursion equation (\ref{1}) stemming from this structure results into some RG equations, whose thermodynamic limit $k \rightarrow \infty$ can be studied with some suitable approximation schemes. In Chapter \ref{replica_approach}, we derive these RG equations with the replica approach, and analyze their fixed points with standard field-theory techniques. We show that notwithstanding their simplicity, the perturbative solution of these RG equations results into a  perturbative series which is probably non-convergent. Hence, an alternative real-space approach which does not rely on the replica method is developed in Chapter \ref{real_space}. In this latter approach, the hierarchical structure of the model is again used to write some RG equations that can be solved numerically with high precision. The replica RG approach and the real-space approach are then compared, by considering their predictions on the critical exponents of the model.

\chapter{The RG in the replica approach}\label{replica_approach}

In this Chapter we derive and solve the RG equations for the HEA model  with the replica approach \cite{castellana2010renormalization,castellana2011renormalization}. These RG equations can be derived with two different methods. The first, exposed in Section \ref{a_la_wilson}, derives the RG equations by using directly the hierarchical structure of the model. We will call this approach method \` a la Wilson, because it implements a coarse-graining RG step relating a $2^ k$-spin HEA to a $2^ {k+1}$-spin HEA, and this yields a RG equation similar to the RG equations originally obtained by Dyson \cite{dyson1969existence, collet1978renormalization, collet1977epsilon, cassandro1978critical} for DHM, and so analogous to Wilson's RG equations  \cite{wilson1982renormalization,wilson1971renormalization1,wilson1971renormalization2,wilson1974renormalization}. The second approach, exposed in Section \ref{field_theoretical}, reformulates the problem in terms of a $\phi^ 3$-field theory \cite{zinnjustin}, and the resulting RG equations are nothing but the Callan-Symanzig equations \cite{zinnjustin, zinn2007phase} for such a field theory. The two formulations are tested to be equivalent by an explicit computation of the critical exponents. 

\section{The RG method \` a la Wilson}\label{a_la_wilson}

Let us consider the partition function $Z[ T, \{ \epsilon \} ]$ of a HEA with $2^k$ spins, which is defined by Eq. (\ref{46}). According to the  general features of the replica approach \cite{MPV,0, NishimoriBook01, franz2009overlap}, the physics of the model is encoded into the $n\rightarrow 0$ limit of the $n$-times replicated partition function
\be \label{48}
\mathbb{E}_\epsilon[ Z[ T, \{ \epsilon \} ] ^ n] = \mathbb{E}_\epsilon \left[ \sum_{\{\vec{S}_a\}_{a=1,\ldots,n}} \exp\left( - \beta \sum_{a=1}^ n H_k[\vec{S}_a ] \right)   \right] ,
\ee
 where $\mathbb{E}_\epsilon$ denotes the expectation value with respect to the random distribution of the energies $\epsilon_k$, i. e. with respect to all the random couplings $J_{12, \,ij}$ of Eq. (\ref{43}), and $\vec{S}_a$ is the spin configuration of the $a$-th replica of the system. One can then consider  the $n \times n$ matrix $Q_{ab}$ \cite{MPV,0, NishimoriBook01} physically representing the overlap between replicas $a$ and $b$, which is defined as 
 \beas \label{overlap}
 Q_{ab}  &\equiv &  \frac{1}{2^ k} \sum_{i=1}^ {2^ k} S_{a,i} S_{b,i}\, \, \forall a \neq b,\no \\
    Q_{aa} & \equiv  & 0\, \forall a. 
\eeas
An interesting quantity which is derived from (\ref{48}) is the probability distribution of the overlap over the quenched disorder $\{ \epsilon \}$
 \be  \label{z}
 \textrm{P}_{k}[Q] \equiv  \mathbb{E}_{\epsilon} \left[ \sum_{\{\vec{S}_a\}_{a=1, \ldots, n}} \exp\left( -\beta \sum_{a=1}^ n H_{k}[\vec{S}_a ] \right)  \prod_{a<b=1}^ n \delta\left(Q_{ab} - \frac{1}{2^ k}\sum_{i=1}^ {2^{k}} S_{ a,i} S_{b,i} \right)\right],
 \ee
 where the volume dependence  has been explicitly restored with the label $k$ in $\textrm{P}_k$, and $\delta$ denotes the Dirac delta function.  The quantity $\textrm{P}_k[Q]$ is  interesting   because when one iterates the  recursion equation (\ref{1}), and a $2^ {k+1}$-spin HEA is built up, the resulting $\textrm{P}_{k+1}$ can be related to $\textrm{P}_k$ by a simple recursion equation, which is 
\bea \label{recz}
\textrm{P}_{k+1}[Q] & = & \exp \left(  \frac{\beta^2 C^{4(k+1)} }{4} \text{Tr}[Q^2] \right)  \int [d Q_1 dQ_2] 
\textrm{P}_{k}[Q_1]\textrm{P}_{k}[Q_2]  \times \\ \no 
&& \times   \prod_{a<b=1}^ n \delta\left(Q_{ab} - \frac{Q_{1,ab}+ Q_{2, ab}  }{2} \right),
\eea
where $\text{Tr}$ denotes the trace over the replica indices $a,b,\ldots$, and the integral over the matrix $Q$ is denoted by $\int [ d Q ] \equiv \int \prod_{a<b=1}^ n dQ_{ab}$. Eq. (\ref{recz}) is equivalent to  the recursion equation  (\ref{rec_dhm}) for DHM \cite{dyson1969existence},  and it yields the flow of the function $\textrm{P}_k$ obtained  by coupling two systems with volume $2^ {k}$ to obtain  a system with volume $2^ {k+1}$. It follows that Eq. (\ref{recz}) can be considered as the flow of $\textrm{P}_k$ under the reparametrization $2^ k \rightarrow 2 \times 2^ k$  of the length scale  \cite{wilson1971renormalization1,wilson1971renormalization2,wilson1974renormalization, zinnjustin, zinn2007phase,itzykson1991statistical, parisi1998statistical, huang}. According to these considerations,  Eq. (\ref{recz}) is a RG equation. \\

The very same techniques developed in Section \ref{dyson} and in Appendix \ref{app_field_dhm} to solve Eq. (\ref{rec_dhm}) for DHM can be used to solve Eq. (\ref{recz}). Notwithstanding this, the solution of Eq. (\ref{recz})  turns out to be much more cumbersome than that of Eq. (\ref{rec_dhm}), because  the former is a flow equation for a function $\textrm{P}_k[Q]$  of a matrix $Q_{ab}$, while  the latter is a flow equation for a function $p_k(m)$ of a number $m$.  In what  follows we will show the main steps of the solution of Eq. (\ref{recz}). \\

First of all, we seek for a solution of  Eq. (\ref{recz}) for $k\rightarrow \infty$, in order to investigate the critical properties of the HEA in the thermodynamic limit. To this end, let us rescale the variable $Q$ in Eq. (\ref{recz}), by setting 
\be \label{50}
 \mathscr{P}_k[Q] \equiv \textrm{P}_k[C^ {-2k}Q],
 \ee
 in such a way that Eq. (\ref{recz}) becomes 
 \be\label{51}
\mathcal{\mathscr{P}}_k[Q]=\exp\left({\frac{\beta^2}{4}  \text{Tr} [ Q^2 ]} \right)  \int \left[ d P \right] \mathcal{\mathscr{P}}_{k-1}\left[ \frac{Q+P}{C^ {2}} \right] \mathscr{P}_{k-1}\left[ \frac{Q-P}{C^ {2}} \right].
\ee 
 Similarly to  Eq. (\ref{redef_dhm}), Eq. (\ref{50}) is  the correct rescaling for Eq. (\ref{recz}) to converge to a nontrivial fixed point for $k\rightarrow \infty$, because  according to Eqs. (\ref{def_c}), (\ref{sigma_hea}) one has $C>1$, and  the $ C^{4(k+1)}$-term in the right-hand side of Eq. (\ref{recz}) is an increasing function of $k$ allowing for no nontrivial fixed point  for $k \rightarrow \infty$. It follows that physically speaking, this rescaling aims to look at the RG equations (\ref{recz}) on the scales that are relevant for large $k$, by means of a `zoom' on the function $\textrm{P}_k[Q]$, which is encoded in the definition  (\ref{50}). An analogous rescaling will be presented in the real-space approach discussed in Chapter \ref{real_space}.

Eq. (\ref{51}) can now be solved by making an ansatz for   $\mathscr{P}_{k}[Q]$, following the same lines as in the solution of  DHM illustrated in  Appendix \ref{app_field_dhm}. The simplest form one can guess for $\mathscr{P} _k[Q]$ is the  Gaussian 
\be \label{gauss}
\mathscr{P}_k[Q]=\exp \left(- r_k \text{Tr}[Q^2]\right). 
\ee
This form corresponds to a mean-field solution  \cite{franz2009overlap}. By plugging Eq. (\ref{gauss}) into Eq. (\ref{51}), one finds an evolution equation relating $r_{k}$ to $r_{k-1}$
\be 
r_{k}=\frac{2 r_{k-1}}{C^ 4}- \frac{\beta^2}{4}. 
\ee \bigskip\\

Even though the mean-field solution (\ref{gauss}) is a fixed point of the RG equation (\ref{51}), there is no guarantee that more complex and physically meaningful fixed points do not exist. By hypothesizing that $\mathscr{P}_k[Q]$ can be expanded in powers of $Q$, non-gaussian fixed points can be explicitly built up in a perturbative framework, by following the same lines as in the Ising model \cite{wilson1974renormalization} and in  DHM \cite{collet1978renormalization, collet1977epsilon, cassandro1978critical}. Indeed, we can add  non-Gaussian terms in Eq.  (\ref{gauss}),  proportional to  higher powers of $Q$, and consistent with the symmetry properties of the model. For instance, in principle there would be several possible replica invariants proportional to $Q^3$, but it is possible to show that the only invariant that is consistent with the original symmetries of the Hamiltonian $H_k$ is $\text{Tr}[Q^3]$. 
It follows  that the simplest non-mean-field ansatz for $\mathscr{P}_k[Q]$ reads
\begin{equation}\label{55}
\mathscr{P}_k[Q]=\exp\left[ - \left(r_k  \text{Tr} [ Q^2 ] + \frac{w_k}{3}  \text{Tr}[Q^3]  \right) \right].
\end{equation}
This non-Gaussian ansatz can  be handled by supposing that the coefficient  $w_k$,  representing the deviations from the Gaussian solution, is small for every $k$. It is important to point out that this is an hypothesis which is  equivalent to assuming that the non-mean-field regime of this model can be described in terms of a perturbation of the mean-field regime \cite{6}. As discussed in Section \ref{historical_outline}, there is no general agreement on the fact that the non-mean-field behavior of a spin glass can be described in terms of a slight modification of the mean-field picture \cite{mcmillan1984scaling,fisher1986ordered,fisher1987pure,fisher1987absence,fisher1988equilibrium,fisher1988nonequilibrium,bray1986heidelberg}. Hence, one should keep in mind that this assumption is far from being trivial and  surely innocuous.\\
 As shown in Appendix \ref{app_rec}, if one plugs the ansatz (\ref{55}) into the RG equation (\ref{51}) and expands up to $O(w_k^3)$, one finds a recursion relation for the vector $(r_k,w_k)$, which is expressed as a function of $(r_{k-1}, w_{k-1})$
 \be \label{56}
\left\{
\begin{array}{lcl}
r_{k}&= &\frac{2 r_{k-1}}{C^ 4} -\frac{\beta^ 2}{4}-  \frac{n-2}{4 } \left( \frac{w_{k-1}}{2 C^ {2} r_{k-1}} \right)^2 + O(w_{k-1}^4),\\ 
w_{k}&= &\frac{ 2 w_{k-1} }{C^ {6}}+ \frac{n-2}{2}\left(\frac{w_{k-1}}{2 C^ {2} r_{k-1}} \right)^3+O(w_{k-1}^ 4).
\end{array}
\right . 
\ee

Eqs. (\ref{56}) are analogous to Wilson's RG equations for the Ising model. Indeed, the mass $r_k$ and the coupling constant $u_k$  of the $\phi^ 4$-theory describing the Ising model \cite{zinnjustin} satisfy a recursion equation very similar to Eq. (\ref{56}) \cite{wilson1972critical, wilson1974renormalization}. \\

In general Eqs. (\ref{56}) do not have a finite fixed point $r_k = r_{k-1} \equiv  r_\ast, w_k = w_{k-1}= w_\ast$ for every value of $\beta$. Concerning this, let us \textit{suppose} that there exists a finite and nonzero inverse temperature $\beta_c$ such that Eqs. (\ref{56}) have a finite fixed point. By definition, $\beta_c$ physically represents the inverse temperature such that the system is invariant under the RG step $k \rightarrow k+1$, i. e. under reparametrization of the length scale. Hence, at $\beta_c$ the system has no characteristic length scale, i. e. it is critical \cite{wilson1982renormalization}. We  call $\beta_c$  the inverse critical temperature of the HEA, because it separates the high and low-temperature phases $\beta < \beta_c, \beta > \beta_c$ where the system is not invariant under reparametrization of lengths.   

We now set $\beta= \beta_c$ and  sketch qualitatively the flow of the coefficient $w_k$ towards its fixed-point value $w_\ast$.  Let us consider first the case   $2/C^ 6<1$,  i. e. $\epsilon <0$, with
\be  \label{epsilon}
\epsilon \equiv \sigma - 2/3. 
\ee
In this case $w_k$ is decreased as $k\rightarrow k+1$, and tends to zero, in such a way that $\mathscr{P}_k[Q]$ tends to a Gaussian solution for large $k$. \\ 
The situation is different when  $\epsilon>0$.  In order to better understand the case $\epsilon >0$, let us rewrite  the recursion equation for $w_k$ as
\be  \label{57}
w_{k}-w_{k-1}= \left(\frac{ 2  }{C^ {6}}-1\right)  w_{k-1}+ \frac{n-2}{2}\left(\frac{w_{k-1}}{2 C^ {2} r_{k-1}} \right)^3+O(w_{k-1}^ 4),
\ee
and suppose for simplicity that $w_{k-1}$ is positive. Being $2/C^ 6>1$, the first addend in the right-hand side of Eq. (\ref{57}) is positive, while the second addend   is negative  in the physical limit $n \rightarrow 0$. Hence, the first addend aims to increase $w_k$, while the second  aims to decrease it. As we will show in the following, these two effects compensate each other, in such a way that $w_k$ tends to a finite and nonzero fixed point for $k \rightarrow \infty$. \\ 

 Following the very same lines as Wilson, we call $\mathscr{P}_\ast[Q], \textrm{P}_\ast[Q]$ the fixed-point probability distributions of the overlap, obtained by setting $r_k  = r_\ast, w_k = w_\ast$ in $\mathscr{P}_k[Q], \textrm{P}_k[Q]$  respectively. The  equations for $n =0$ yield
 \be \label{58}
w_\ast^2   = \left\{
\begin{array}{ll}
0 & \epsilon \leq 0 \\
 48 \log 2 \left( \frac{\beta^2/4}{2^ {1/3}-1} \right)^ 3 \epsilon + O(\epsilon^{3/2}) & \epsilon >0. 
\end{array}
\right . 
\ee
As anticipated above, for $\epsilon>0$, $w_k$ tends to a finite and nonzero value, which is  found to be proportional to $\epsilon$ in Eq. (\ref{58}). \\
 
 We recall \cite{collet1978renormalization, collet1977epsilon, cassandro1978critical, wilson1974renormalization, zinnjustin, zinn2007phase} that  a Gaussian $\mathscr{P}_\ast[Q]$  corresponds to a mean-field regime of the model.
Indeed, in the mean-field approximation   one evaluates with the saddle-point method the functional integral yielding the replicated partition function \cite{MPV,NishimoriBook01, 0, parisi1979infinite, parisi1980order}
\be \label{59}
\mathbb{E}_\epsilon[ Z\left[ T, \{ \epsilon \} \right]^ n ] = \int [d Q ] \textrm{P}_\ast [Q], 
\ee
where Eq. (\ref{z}) and Eq. (\ref{48}) have been used. If $\sigma \leq 2/3$, i. e.  $\epsilon <0$, $\mathscr{P}_\ast[Q]$ is Gaussian, and so is $\textrm{P}_\ast [Q]$, in such a way that the saddle-point approximation is exact in the right-hand side of Eq. (\ref{59}), i. e. the mean-field approximation is correct.  On the contrary, for $2/3<\sigma  \leq 1$, $\mathscr{P}_\ast[Q]$ is not Gaussian, and the system has a non-mean-field behavior.  In particular, fluctuations around the mean-field saddle point in the right-hand side of Eq. (\ref{59}) cannot be neglected. Hence, we call $\sigma= 2/3$ the upper critical dimension \cite{huang, zinnjustin, zinn2007phase, itzykson1991statistical} of the HEA. \\
 In  the computation of a given  physical quantity $\mathcal{O}(\sigma)$ for $\sigma>2/3$,   fluctuations show up in the guise of some corrections proportional to $\epsilon$ to the mean-field value of this quantity.  It is  important to emphasize that these $\epsilon$-corrections are not merely a numerical improvement on the predictions for the observable, but somehow encode the strength of the corrections to the mean-field physics. For instance, if  $\mathcal{O}(\sigma)$  was  expanded in powers of $\epsilon$ around $\sigma=2/3$, and if this expansion could be resummed and  made convergent, this would mean that the non-mean-field physics of the system could be considered as a `small correction' to the mean-field physics.
 On the contrary, if such a non-mean-field physics were substantially different, the latter statement would be incorrect, and this fact would dramatically show up in a  divergent and  non-resummable $\epsilon$-series for $\mathcal{O}(\sigma)$.   \\
 
These observations can be  directly illustrated by considering as observable $\mathcal{O}$ the  critical exponent $\nu$, related to the divergence of the correlation length $\xi$
\be \label{xi_hea}
\xi \overset{T \rightarrow T_c}{\sim} (T-T_c)^ {-\nu}.  
\ee
$\nu$ can be computed \cite{wilson1974renormalization} by linearizing the transformation (\ref{56}) in the neighborhood of $r_\ast, w_\ast$. Such a linearization is performed by considering the $2\times 2$-matrix 
\[
\mathscr{M}_{ij} \equiv \left. \frac{\partial  (r_{k+1}, w_{k+1}) }   {\partial (r_{k}, w_{k})  } \right|_{ r_k  = r_\ast, w_k = w_\ast}. 
\]
It can be shown that $\nu$ is related to the largest eigenvalue $\Lambda$ of $\mathscr{M}$ by the relation 
\be \label{90}
\nu = \frac{\log 2}{\log \Lambda}. 
\ee
A straightforward calculation yields $\nu$ at order $\epsilon$, for $n=0$
\be \label{nu}
\nu = \left\{
\begin{array}{ll}
\frac{1}{2 \sigma-1} & \sigma \leq 2/3\\
3 + 36 \epsilon + O(\epsilon^2) & \sigma > 2/3.  
\end{array}
\right. 
\ee

\begin{figure}
\centering
\includegraphics[width=10cm]{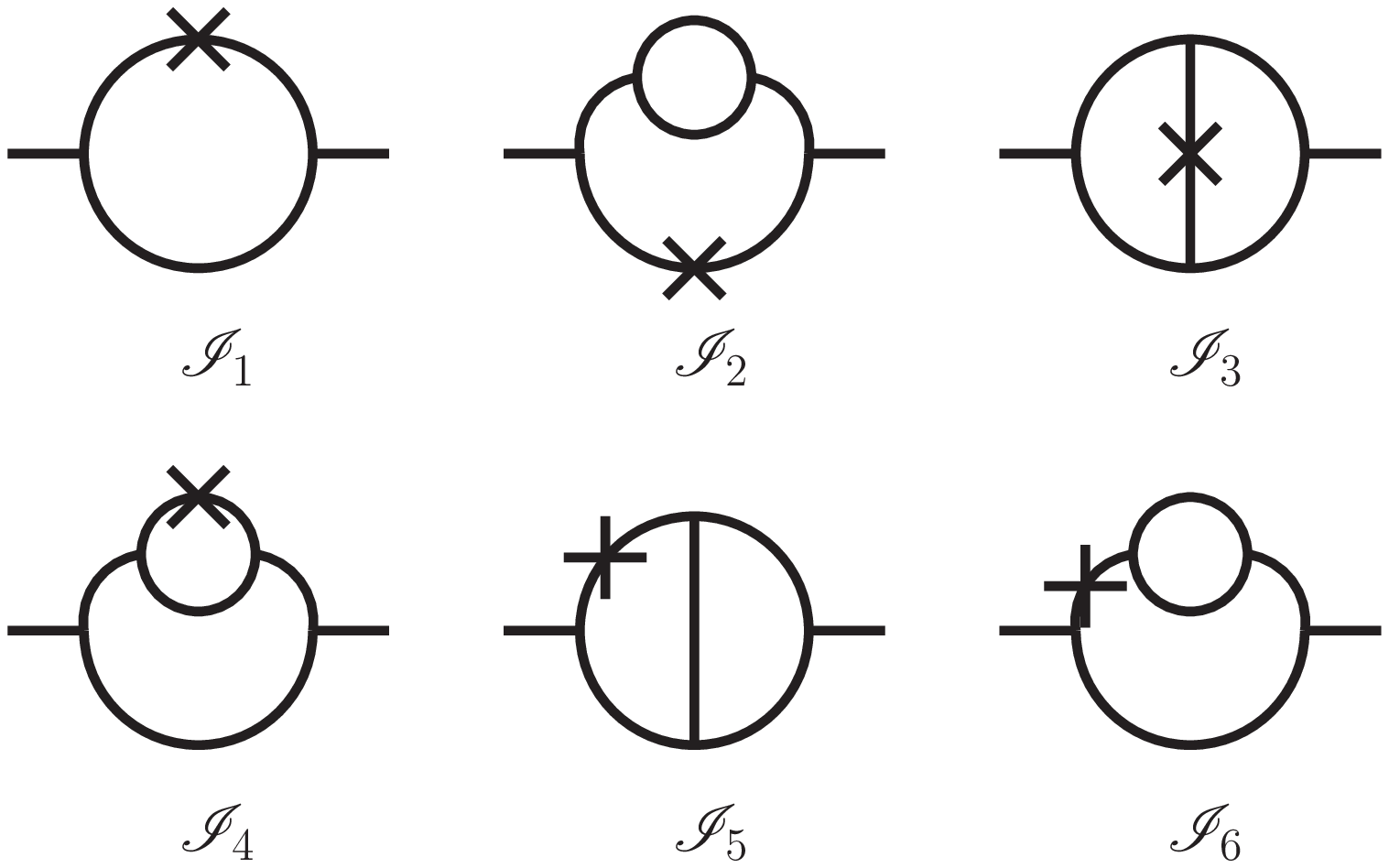}
\vspace*{1cm}\\
\includegraphics[width=10cm]{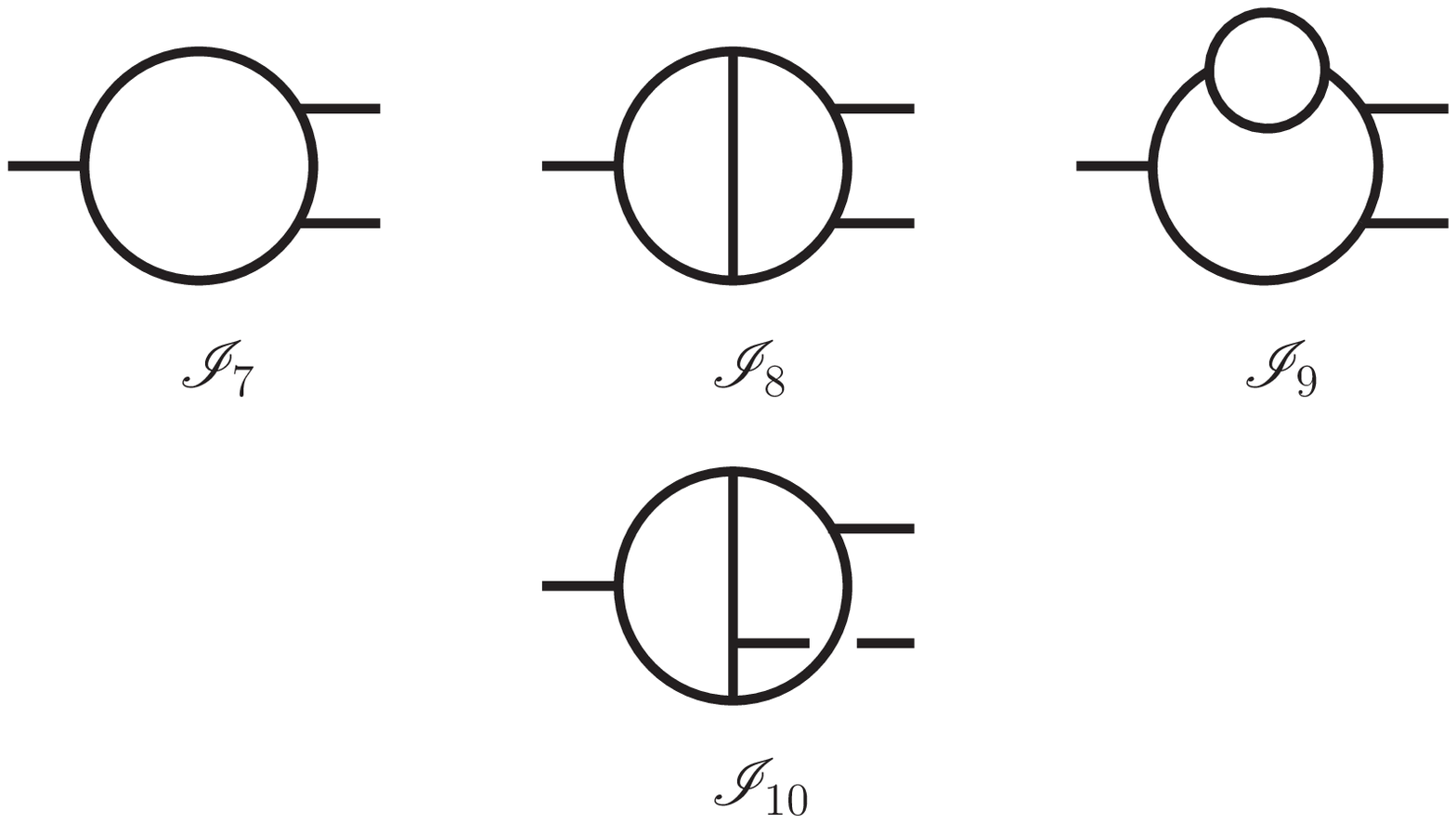}
\caption{One and two-loop $1$PI Feynman diagrams deriving from the $\textrm{Tr}[Q^3]$-interaction term, which contribute to the computation of the critical exponent $\nu$ in the method \`a la Wilson presented in Section \ref{a_la_wilson}, and in the field-theory method presented in Section \ref{field_theoretical}. 
In the method \`a la Wilson,  diagrams $\mathscr{I}_1, \cdots, \mathscr{I}_6$ with two external lines have to be considered without  crosses on the internal lines, and represent the diagrams contributing to the $\textrm{Tr}[Q^2]$-term in $\mathscr{P}_k[Q]$. In the same method, it can be shown that diagrams $\mathscr{I}_7, \cdots, \mathscr{I}_{10}$ with three external lines contribute to the $Q^3,Q^4, Q^5$-terms in $\mathscr{P}_k[Q]$.
In the field-theory method, diagrams $\mathscr{I}_1, \cdots, \mathscr{I}_6$ and  $\mathscr{I}_7, \cdots, \mathscr{I}_{10}$ contribute to the $1$PI two-point and three-point correlation functions respectively,  and crosses stand for $\textrm{Tr}[Q^2]$-insertions \cite{zinnjustin}. 
This graphical identification of the same diagrams in the two methods is an important test of their mutual consistency.}
\label{fig8}
\end{figure}

A detailed analysis of the computation of Appendix \ref{app_p} reveals that in this $O(w_k^3)$-calculation  resulting in the $O(\epsilon)$-estimate of $\nu$, one  considers all the one-particle irreducible ($1$PI) \cite{zinnjustin, weinberg2005quantum1, weinberg2005quantum2} one-loop Feynman diagrams generated by the $Q^3$-vertex in Eq. (\ref{55}). These are diagrams  $\mathscr{I}_1, \mathscr{I}_7$ in Fig. \ref{fig8}.  Similarly,  in the computation at order $w_k^5$, resulting in the $O(\epsilon^2)$-estimate  of $\nu$,  one considers two-loop $1$PI Feynman diagrams $\mathscr{I}_2, \cdots, \mathscr{I}_6$ and $\mathscr{I}_8, \cdots, \mathscr{I}_{10}$ in Fig. \ref{fig8}, and so on.  \\
%
%

As can be seen by Eq. (\ref{nu}), the coefficient of $\epsilon$ in Eq. (\ref{nu}) is quite large, an it is plausible that the full $\epsilon$-series of $\nu$ does not converge. According to the above discussion, the convergence or resummability of the series would give some deep insight into how strongly the mean-field physical picture should be modified in the non-mean-field region $\sigma>2/3$. In particular, a non-resummable series would suggest that the non-mean-field physics cannot be described in terms of a perturbation of the mean-field one, and would have a strong impact on the general problem of spin glasses in finite dimensions. It is thus interesting to investigate the properties of the $\nu$-series by directly computing higher orders in $\epsilon$, and checking the convergence or the resummability of the series. \\

The computation of the $\epsilon$-expansion at high orders can be performed with a computer program. This is a particular feature of the HEA, because thanks to the hierarchical structure of the system, the RG equations (\ref{51}) have a simple form compared to the RG equations of the PLSG \cite{kotliar1983one} or of the EA model \cite{harris1976critical}. Indeed, the latter are nothing but the Callan-Symanzik equations \cite{callan1970broken, symanzik1970small,zinnjustin} for a $Q^3$-field theory, and their solution in perturbation theory requires an explicit enumeration of all the Feynman diagrams deriving from the $Q^3$-vertices, and the computation of their IR-divergent part. This enumeration is extremely hard to implement in a computer program, and has never been automated even in the simplest case of the Ising model  \cite{guida2006towards}. On the contrary, we believe that a  symbolic manipulation program could handle Eqs. (\ref{51}) and automate the computation to higher orders in $w_k$, with no need to evaluate the IR-divergent part of Feynman integrals, which is not needed in the present approach \`a la Wilson.\\
 In order to perform this automation, we have evaluated by hand the first few terms of the series, which has been computed to $O(\epsilon^2)$ with  the method \`a la Wilson, and with a quite independent field-theory method exposed in Section \ref{field_theoretical}. Both methods give the same two-loop result for $\nu$, which will serve as a  severe test for a future automation of the $\epsilon$-expansion to high orders.  This automation is very difficult from a  purely technical viewpoint, and is beyond the scope of this thesis.   \medskip\\

Here we sketch the main steps of the two-loop computation of $\nu$, more details are given in \cite{castellana2011renormalization}.  We showed that if we plug the ansatz (\ref{55}) into the right-hand side of Eq. (\ref{51}) and  systematically neglect terms of order higher than $w_k^3$, we get a $\mathscr{P}_{k+1}[Q]$ which is still of the form (\ref{55}), i. e. the RG equations are closed. This is not true if we keep also terms of order higher than $w_k^3$. Indeed, it is easy to show that in this case terms of order $Q^4$ arise in the right-hand side of Eq. (\ref{51}), and these terms are not present in the original ansatz. This fact implies that to $O(w_k^4)$, $\mathscr{P}_{k}[Q]$ must contain also $Q^4$-terms, and that these terms must be proportional to $w_k^4$. By plugging the fourth-degree polynomial $\mathscr{P}_k[Q]$ in Eq. (\ref{51}) and expanding the right-hand side up to $O(w_k^5)$, $Q^5$-terms are generated. It follows that  at $O(w_k^5)$, $\mathscr{P}_{k}[Q]$ must contain also $Q^5$-terms, and that these terms must be proportional to $w_k^5$. If this perturbative framework is consistent, by iterating such a procedure to higher orders we reconstruct the exact function $\mathscr{P}_k[Q]$. In particular, at the $j$-th step of this procedure we generate $n_j$ monomials of order $Q^j$, and call these monomials $\{  I_l^ {(j)}[Q] \}_{l=1,\ldots,n_j }$. Hence, if this procedure is iterated until step  number $j=p$, $\mathscr{P}_k[Q]$  can be written in the compact form
\begin{equation}
\mathscr{P}_k[Q]=\exp \left\{ - \left[c^ {(2)}_{1,\, k} I^ {(2)}_1[Q] + \sum_{j=3}^{p} \frac{1}{j} \sum_{l=1}^ {n_j} c^{(j)}_{l,\, k} I_l^ {(j)}[Q]  \right]\right\},
\end{equation}
where  
\bea \no 
I^{(2)}[Q] \equiv \textrm{Tr}[Q^2],   \,I^{(3)}[Q] \equiv \textrm{Tr}[Q^3],\\ \no 
n_3\equiv  1 , \, c^{(2)}_{1,\,k} \equiv r_k, \,  c^{(3)}_{1,\,k} \equiv w_k. 
\eea
 \\
In the present work such a procedure has been pushed up to order $w_k^5$, and $\mathscr{P}_{k}[Q]$ has been computed as a fifth-degree polynomial in $Q$.  In particular, one generates $n_4=4$ invariants $I^{(4)}[Q]$ of fourth degree and $n_5=4$ invariants $I^{(5)}[Q]$ of fifth degree in $Q$.   The explicit expression for all the monomials $I^{(j)}_l[Q]$ at this order  is given in  Table \ref{tab1} of Appendix \ref{app_p}. The set of two RG equations (\ref{56}) for the two-dimensional vector $(r_k,w_k) = (c^{(2)}_{1,\,k} , c^{(3)}_{1,\,k})$ obtained in the one-loop calculation here becomes a set of ten equations for the vector $c^{(2)}_{1,\,k}, c^{(3)}_{1,\,k}, c^{(4)}_{1,\,k} , \cdots, c^{(4)}_{4,\,k}, c^{(5)}_{1,\,k} , \cdots, c^{(5)}_{4,\,k}$, Eqs. (\ref{80})-(\ref{89}). By linearizing these equations at the critical fixed point one can compute the $10 \times 10$-matrix $\mathscr{M}$, and extract $\Lambda$. By Eq. (\ref{90}), one can then compute the exponent $\nu$ at two loops for $\epsilon>0$ and $n=0$
\be \label{nu_2}
\nu =  3 + 36 \epsilon +  \big[ 432 - 27 \big(50 + 55 \cdot 2^{1/3} + 53 \cdot  2^{2/3}\big) \log 2 \big] \epsilon^2 + O ( \epsilon^3 ).
\ee \\

The  \mnote{The replica $\epsilon$-expansion for the critical exponents of the HEA is presumably badly behaved, and non-predictive. } coefficient of $\epsilon^2$ in Eq. (\ref{nu_2}) is about $-5.1\times 10^3$: the first two orders of the $\epsilon$-expansion show that this is probably badly-behaved and out of control. In particular, it is impossible to make any prediction on $\nu$ with the first two orders of the series.  Differently, the $\epsilon$-expansion for the critical exponents of the Ising model (consider for instance the exponent $\gamma$ in \cite{zinnjustin}) is nonconvergent, but it settles to a reasonable value as the  order  is increased from zero up to at least three, and then it  deviates from such a value when higher orders are considered (see \cite{vladimirov1979calculation,chetyrkin1981five,chetyrkin1981errata,chetyrkin1981integration, chetyrkin1983five,kazakov1983method,gorishny1984epsilon,kleinert1991five,kleinert1993five} for  five-loop computations of the exponents). Though, in that case the expansion can be resummed and made finite, giving a result for the exponents which is in excellent agreement with experiments \cite{transitions1980vol} and MC simulations \cite{pawley1984monte,baillie1992monte}. \medskip\\

As anticipated above, it is interesting to reproduce the two-loop result (\ref{nu_2}) with an independent method. Indeed, in the effort to build up a fully automated $\epsilon$-expansion it is important to check that the prediction (\ref{nu_2}) is not only correct, but also well-defined, i. e. it does not depend on the RG scheme used in the calculation. For instance, in this approach \`a la Wilson the IR limit of the theory is taken by requiring invariance under the transformation $k \rightarrow k+1$, which doubles the system volume at each step. On the contrary, in Section \ref{field_theoretical}  we perform the IR limit by considering a real parameter $\lambda$ physically representing the typical energy scale of the system, and by sending it  smoothly to zero. As we will show in the next Section, these two independent ways of taking the IR limit yield the same two-loop result for $\nu$.

\section{The RG method in the field-theory approach}\label{field_theoretical}

The replica formulation for the HEA allows for a quite general treatment of the IR behavior of the system, based on the path-integral formulation. This formulation can be developed along the lines of the path-integral formulation of the Ising model \cite{zinnjustin, zinn2007phase, parisi1998statistical, weinberg2005quantum2}.  In the latter the partition and correlation functions are represented in terms of a path integral over a field $\phi$, weighted with a $\phi^4$-action. Setting $\varepsilon \equiv 4-d$ (for the Ising model  we use a different font for $\varepsilon$, to avoid confusion with the $\epsilon$ of the HEA defined in Eq. (\ref{epsilon})),  one finds that in the physically relevant case $\varepsilon >0$ this $\phi^4$-field theory presents IR divergences occurring when the temperature $T$ approaches its critical value. These divergences are removed by means of the observation that one can construct an auxiliary field theory, the renormalized one, which makes the same physical predictions as the original one and has no IR divergences.  Indeed,  it  has been  shown  that  these divergences can be  removed at any order in perturbation theory \cite{callan1975methods}, and reabsorbed in the renormalization constants.  Once the renormalized theory has been built up, one can extract the critical exponents in perturbation theory from the $\varepsilon$-expansion of the renormalization constants.\\

Here we show how the result (\ref{nu_2}) for $\nu$ can be reproduced along these lines, more details can be found in \cite{castellana2011renormalization}, while an extensive treatment of the renormalization group theory used here is given in \cite{zinnjustin}. \\

First, this computation is better performed by taking a definition of the HEA which is slightly different from that of Eqs. (\ref{1}), (\ref{43}), and that has the same critical exponent $\nu$. First, let us relabel the spins $S_1, \cdots, S_{2^k}$ as $S_0, \cdots, S_{2^k-1}$. We redefine  the interaction term  in Eq. (\ref{43}) as   
\begin{equation}\label{redefinition}
\epsilon_{k+1}[S_0, \ldots, S_{2^ {k+1}-1} ]  \rightarrow - \left( \frac{C^2}{2} \right) ^{k+1}  \sum_{i=0}^{2^{k}-1} \sum_{j=2^{k}}^{2^{k+1}-1} J_{12, ij} S_i  S_j. 
\end{equation}
The redefinition (\ref{redefinition}) has the following physical meaning. In the original definition (\ref{43}), one couples two systems, say system $1$ and system $2$, with $2^k$ spins each, and obtains a $2^{k+1}$-spin system. The interaction energy between $1$ and $2$ is given by couplings between spins belonging to $1$ (given by the terms in the sum in Eq. (\ref{43}) with $1\leq i,j \leq 2^k$),  couplings between spins belonging to $2$ (given by the terms in the sum in Eq. (\ref{43}) with $2^k+1\leq i,j \leq 2^{k+1}$), and couplings between  $1$ and $2$ (given by the terms in the sum in Eq. (\ref{43}) with $1\leq i \leq 2^{k}, \, 2^k+1\leq j \leq 2^{k+1}$). In the redefinition (\ref{redefinition}), only the latter couplings are kept, and neither couplings within system $1$ nor $2$ appear in the definition. \\

The equivalence between the two definitions can be shown as follows. If one considers two spins $S_i, S_j$ and their interaction energy, it is easy to show \cite{franz2009overlap} that in the model defined by Eq. (\ref{redefinition})  the variance of such an interaction energy scales with the ultrametric distance  between $S_i$ and $S_j$ in the same way as  in the model defined by  Eq. (\ref{43}), and that the two variances  differ only in a constant multiplicative factor. It follows that the long-wavelength features of the two models  are the same. According to general universality arguments, both of the models must have the same critical exponents, because these depend only on the long-wavelength features of the system, like the way interactions  between spins decay at large distances  \cite{wilson1974renormalization,wilson1982renormalization}.
Notwithstanding this fact, non-universal quantities are generally different in the two models. For instance, it is well known that if one multiplies the interaction strength between spins by a constant factor, one changes the microscopic energy scale of the system. According to dimensional analysis \cite{zinnjustin}, the critical temperature must be proportional to this energy scale, and is thus multiplied by the same factor. Hence, the critical temperature of the model defined by Eq. (\ref{43}) is different from that of the model defined by Eq. (\ref{redefinition}). 
This can be verified by considering how  the recursion relation (\ref{51}) is modified when one  applies the redefinition (\ref{redefinition}). Indeed, if one starts from Eq.  (\ref{redefinition}) and goes through the steps of the derivation of Eq. (\ref{51}), one finds a recursion equation that differs from Eq. (\ref{51}) by a factor multiplying $\beta$. \\
The reason why the definition (\ref{redefinition}) is more suitable for this field-theory approach will be clarified below. \\

The path-integral formulation of the HEA can now be introduced by observing that one can write the replicated partition function (\ref{48}) in terms of a functional integral over a \textit{local} overlap field $Q_{i,\, ab}\equiv S_i^a S_i^b, \, 0 \leq i \leq 2^k-1$
\be \label{97}
 \mathbb{E}_\epsilon[ Z[ T, \{ \epsilon \} ] ^ n]  =  \int  \mathscr{D} Q \,  \textrm{e}^{-\textrm{S}[Q]},
\ee
where $\int \mathscr{D}Q \equiv \int \prod_{i=0}^{2^k-1}\prod_{a<b=1}^n d Q_{i,\, ab}$ stands for the functional integral over the field $Q_{i,\, ab}$. The action $\textrm{S}[Q]$ can be worked out by supposing that there exists a critical temperature $T_c$ such that the  characteristic length of the system diverges as $T$ approaches $T_c$. We stress that this hypothesis has been made also in the RG approach \`a la Wilson, where we assumed the existence of a $T_c$ such that the RG equations (\ref{56}) have a nontrivial fixed point, i. e. a fixed point such that the system has no finite characteristic length. \\

 By taking  $T \approx T_c$, one can select the IR-dominant terms in $\textrm{S}[Q]$, and obtain
\be \label{92}
\textrm{S}[Q] =\frac{1}{2} \sum_{i,j=0}^{2^k-1} \Delta'_{i,j} \text{Tr}\big[Q_i  Q_j \big]  +\frac{g}{3 !} \sum_{i=0}^{2^k-1}  \text{Tr}[Q_i^3].
\ee
In Eq. (\ref{92}) the  propagator $\Delta'_{i,j}$  depends on $i,j$  through the difference $\mathcal{I}(i)-\mathcal{I}(j)$, where for any $0 \leq i \leq 2^k-1$ the function
$\mathcal{I}(i)$ is defined  in terms of the expression in base $2$ of $i$ as
\begin{equation}
i = \sum_{j=0}^{k-1} a_{j} 2^j,\, \mathcal{I}(i) \equiv \sum_{j=0}^{k-1} a_{k-1-j} 2^j. 
\end{equation}
 According to the above definition of $\mathcal {I}(i)$, the quadratic term of Eq. (\ref{92}) is not invariant under spatial translations $i \rightarrow i + l$. 
Accordingly \cite{weinberg2005quantum1, zinnjustin}, the Fourier transform of the propagator $\Delta'_{i,j}$ does not depend only on the momentum $p$ associated with the variable $i-j$, but it generally depend on both of the momenta $p,q$ associated with the variables $i,j$  respectively. With this complicated form of the propagator, any explicit computation of the loop integrals, which is necessary for the computation of the critical exponents, is extremely  difficult to perform.  This problem can be overcome with a simple relabeling of the sites of the lattice. Indeed, if one sets
\[
\mathcal{I}(i) \rightarrow i, \, \forall i=0, \ldots, 2^k-1,
\] 
it is easy to show that Eq. (\ref{97}) still holds, with an  action $\textrm{S}[Q]$ which now reads
\be \label{93}
\textrm{S}[Q] \rightarrow \frac{1}{2} \sum_{i,j=0}^{2^k-1}  \Delta_{i,j} \text{Tr}\big[Q_i  Q_j \big]  +\frac{g}{3 !} \sum_{i=0}^{2^k-1}  \text{Tr}[Q_i^3],
\ee
where $\Delta_{i,j} \equiv \Delta'_{\mathcal{I}{^-1}(i), \mathcal{I}^{-1}(j)}$. 
Since  $\Delta'_{i,j}$ depends on $i,j$  through the difference $\mathcal{I}(i)-\mathcal{I}(j)$,  $\Delta_{i,j} $ depends on $i,j$ through the difference $i-j$. It follows that $\textrm{S}[Q]$ is now translationally invariant, and the ordinary Fourier transform  techniques \cite{taibleson1976fourier, meurice1995high}  can be used. In particular, the Fourier transform $\Delta(p)$ of $\Delta_{i,j}$ depends only on the dyadic norm  $|p|_2$ (see \cite{parisi2000p} for a precise definition of the dyadic norm) of the momentum $ p$ relative to the variable $i-j$, and can be written as
\be  \label{prop}
\Delta(p) = |p|_2^{2 \sigma-1}+ m ,
\ee 
where the mass $m$ is proportional to $T - T_c$, and  has dimensions $[m]=2\sigma -1$. \smallskip\\

The action defined by  Eq. (\ref{93}) yields a $\text{Tr}[Q^3]$-field theory, which is known to describe the spin-glass transition in both short-range   \cite{chen1977mean} and long-range   \cite{chang1984spin, kotliar1983one} spin glasses. Notwithstanding this, an interesting and novel feature of the propagator  (\ref{prop}) is that it depends on the momentum $p$ through its dyadic norm $|p|_2$. This fact is rather interesting, because it implies a direct analogy with  the original derivation of the RG equations for the Ising model in finite dimensions,  in particular with the Polyakov derivation  \cite{wilson1974renormalization, polyakov1975interaction}. Indeed, the basic approximation scheme in the Polyakov  derivation consists in introducing an ultrametric structure in momentum space, such that the momentum space is divided into shells and the sum of two
momenta in a given shell cannot give a momentum of a higher momentum
scale cell. This feature is analogous to a  general property of the dyadic norm, such that  if $p_1,\, p_2$ are two integers in $0, \ldots, 2^k-1$, their dyadic norms satisfy  \cite{parisi2000p}
$|p_1+p_2|_2 \leq \max \left( |p_1|_2,|p_2|_2\right)$. This fact implies that if $p_1, p_2$ are inside a shell of radius $\max \left( |p_1|_2,|p_2|_2\right)$, the momentum $p_1+p_2$ is still in that shell. Thus, the dyadic structure of Eq. (\ref{prop})  emerges naturally in more general contexts where there is no hierarchical structure, such as finite-dimensional systems with short-range interactions. \\

A perturbative expansion in $g$ of the two and three-point $1$PI correlation functions deriving from the action (\ref{93}) reveals that if $\epsilon<0$ the field theory (\ref{93}) is well-defined and finite, while if $\epsilon>0$ IR divergences occur when $m \rightarrow 0$. According to a simple dimensional argument, if $\epsilon<0$ the critical exponent $\nu$ is given by the first line of Eq. (\ref{nu}). On the contrary, if $\epsilon>0$ a more elaborated treatment is needed to deal with IR divergences and compute $\nu$. \\

For $\epsilon>0$,  IR divergences can be eliminated by defining a renormalized field theory having a renormalized mass and coupling constant $m_r,g_r$, which are defined in terms of the bare mass and coupling constant 
\bea \label{parameter_m}
 m&=& m_r+\delta m ,\\
 g&=&m_r^{\frac{3 \epsilon}{2 \sigma -1 }}g_r Z_g, \label{parameter_g}
 \eea
where $\delta m$ is the mass shift due to renormalization, and $Z_g$ is the renormalization constant of the coupling $g$.  According to general results on models with long-range interactions, the field $Q_{i,\, ab}$ is not renormalized, i. e.  its renormalization constant  $Z_Q$ is equal to one.  On the contrary, one has to introduce a renormalization constant $Z_{Q^2}$ enforcing the renormalization of the $\textrm{Tr}[Q_{i}^2]$-field.
 According to the minimal subtraction scheme, the renormalization constants $\delta m, Z_g, Z_{Q^2}$ are chosen in such a way that they subtract the divergences occurring in the two and three-point  $1$PI  correlation functions of the renormalized theory. In the IR limit $m_r\rightarrow 0$, these divergences appear in the shape of poles in $\epsilon$. Since here $\delta m,Z_g,Z_{Q^2}$ are expanded in powers of the renormalized coupling $g_r$,  and since these IR divergences are subtracted order by order in $g_r$, the constants $\delta m, Z_g,Z_{Q^2}$ are  given by a series in $g_r$ whose coefficients contain poles in $\epsilon$. From a detailed analysis it turns out that these series contain only even powers of $g_r$,  that only $Z_g$ and $Z_{Q^2}$ are needed to compute $\nu$, and that $a$-loops Feynman diagrams contribute to order $g_r^{2a}$ in $Z_g,Z_{Q^2}$. In Appendix \ref{app_field_theor} we sketch the main steps of the one-loop computation of $Z_g,Z_{Q^2}$, which are given in Eqs. (\ref{109}), (\ref{108}).  \\

By following the very same techniques as those exposed in Appendix \ref{app_field_theor}, we computed $Z_g,Z_{Q^2}$ at two loops.  For $n=0$ one has
\bea
Z_g &=&1+\frac{g_r^2}{48 \epsilon \log 2}+g_r^4 \left[ \frac{1}{1536 \epsilon^2 (\log 2)^2}+\frac{5+2 \cdot 2^{2/3}}{512 \epsilon \log 2}\right]+O(g_r^6), \label{zg}\\ \no 
Z_{Q^2}&=&1 + \frac{g_r^2}{24 \epsilon \log 2}+ g_r^4 \left[ \frac{1}{576 \epsilon^2 (\log 2)^2} - 5\frac{ (1 + 11 \cdot 2^{1/3} + 7 \cdot 2^{2/3})}{ 2304 \epsilon  \log 2}\right] +\\ 
&& +O(g_r^ 6). \label{zq2}
\eea
One can also show that  $\delta m = O(g_r^ 4)$.\\

Eqs. (\ref{zg}), (\ref{zq2}) explicitly construct the renormalized theory, which is free of IR divergences. In this theory, we can safely perform the IR limit, and in particular compute physical quantities in this limit. In order to do this, we  introduce a function $g(\lambda)$, physically representing the effective coupling constant of the model  at the energy scale  $\lambda$. $g(\lambda)$ can be computed from the Callan-Symanzik  equations \cite{callan1970broken,symanzik1970small}, 
as the solution of the differential equation 
\be \label{113}
\beta(g(\lambda)) = \lambda \frac{d g(\lambda)}{d \lambda},
\ee
where the $\beta$-function  is defined as
\be 
\beta(g_r) \equiv \left. \mu \frac{\partial g_r}{\partial \mu} \right | _{g,m},
\ee 
and  $\mu \equiv m_r ^{\frac{1}{2\sigma-1}}$. Eq. (\ref{113}) states that $\beta(g_r)$ governs the flow of the effective coupling $g(\lambda)$ under changes in the energy scale $\lambda$ of the system. $\beta(g_r)$ can be explicitly computed in terms of the renormalization constant $Z_g$, Eq. (\ref{zg})
\be \label{beta}
\beta(g_r)= -3 \epsilon g_r + \frac{g_r^ 3}{8 \log 2} + 3 \frac{5+2 \cdot 2^{2/3}}{128 \log 2} g_r^ 5 + O(g_r^ 7).
\ee

The effective coupling $g(\lambda)$ in the IR limit is obtained by letting the energy scale $\lambda$ go to zero, and is given by  $g_r^\ast \equiv g(\lambda=0)$. By definition, $g_r^\ast$ is a fixed point of the flow equation (\ref{113}), and is obtained perturbatively in the shape of a series in $\epsilon$, as the solution of the  fixed-point equation $\beta(g_r^\ast)=0$. Moreover, one can show   from Eq.  (\ref{beta}) that the IR fixed point $g_r^\ast=0$ is stable only for $\epsilon<0$, while for $\epsilon>0$ a nontrivial fixed point $g_r^{*}\neq 0$ of order $\sqrt{\epsilon}$ arises. The same fixed-point structure arises in Wilson's method, Eq. (\ref{58}).  Notwithstanding this,  in  Wilson's method the IR limit is reached by doing a set of discrete steps $k\rightarrow k+1$ each of which doubles the volume of the system, while in this approach this limit is reached by letting the energy scale $\lambda$ go  to zero smoothly.\\

Once the effective coupling in the IR limit is known,  the scaling relations yield the critical exponent $\nu$  in terms of $g_r^\ast$ and  of the renormalization constant $Z_{Q^2}$ 
\be \label{115}
\nu= \frac{1}{\eta_2(g_r^{*}) + 2\sigma -1},
\ee
where
\begin{equation} \label{eta}
\eta_2(g_r)  \equiv \mu  \left. \frac{\partial \log Z_{Q^2}}{\partial \mu } \right|_{g,\, m}.
\end{equation}
By plugging the two-loop result (\ref{zq2}) for  $Z_{Q^2}$ into Eqs. (\ref{eta}) and evaluating $\eta(g_r)$ for $g_r= g_r^\ast$, we can extract $\nu$ for $n=0$ from Eq. (\ref{115}). The result is exactly the same as that of with Wilson's method, Eq. (\ref{nu_2}). \bigskip\\

The fact that the method \`a la Wilson and the field-theory method yield the same two-loop  prediction for $\nu$  shows that the IR limit of the HEA is well-defined, because it does not depend on the RG framework used to reach it: even though the two methods have a few underlying common features, they yield the same result for the universal quantities  of the system, which are encoded into the coefficients of the $\epsilon$-expansion. We want to stress that these universal quantities  stay the same when changing the RG approach and  redefining  the microscopical details of the model, Eq. (\ref{redefinition}). Accordingly, this picture suggests that the ordinary RG ideas for the Ising model work consistently also in this disordered case:  the HEA has a characteristic length diverging at the critical point, and the universal physical features  in the critical region are governed by  long-wavelength degrees of freedom.  \vspace{2cm}\\

Notwithstanding the positiveness of this result, the $\epsilon$-expansion is still non-predictive, because the first few terms of the series (\ref{nu_2}) have a nonconvergent behavior.
On the one hand, one could test empirically the reliability of the present perturbative approach, and so the convergence properties of the perturbative series, by solving Eq. (\ref{51}) numerically for integer $n>0$ and by comparing the result to the first three orders of the perturbative expansion.  On the other hand, it would be much more difficult  to test whether the series could be made resummable by some suitable techniques, by explicitly computing high orders of the expansion. To this end, our two-loop result constitutes a starting point and a severe test of this high-order computation.  An eventual evidence of the resummability of the series would suggest that the non-mean-field behavior of this model can be considered as a perturbation of the mean-field one. On the contrary, a failure of the resummation techniques would imply that the non-mean-field behavior is radically different from the mean-field one.

Another weak spot of this replica RG approach is that the method does not identify the correct spin-decimation rule in a non-mean-field strongly frustrated case, which, as discussed in Section \ref{historical_outline},  is one of the fundamental questions and difficulties in the construction of a RG theory for finite-dimensional spin glasses. 
\medskip\\

Both of these weak spots of the replica RG approach made us seek for an alternative methodology which, based on a transparent spin-decimation rule, could overcome over the difficulties of the replica method, and make quantitative predictions for the critical exponents. This methodology will be illustrated in Chapter \ref{real_space}, and does not rely on the replica  method, but on a real-space RG criterion.

\chapter{The  RG approach in real space}\label{real_space}

As discussed in Section \ref{historical_outline}, a fundamental ingredient for constructing  of a RG theory is the introduction of a decimation rule. Through decimation, one practically implements a coarse-graining process that  changes the length scale with which one looks at the physics of the system. For ferromagnetic systems, a suitable spin decimation rule has been  originally introduced by Kadanoff \cite{kadanoff1966scaling},  and relies on the construction of block spins. As discussed in Section \ref{historical_outline}, Kadanoff's decimation rule does not work in a disordered system like the HEA, because the average over disorder of the   magnetization inside a block of spins is trivially zero. In this Chapter we propose a real-space decimation rule for the HEA which overcomes this problem, and which is not directly based on the block-spin construction.  This method will be first applied to DHM in Section \ref{real_dyson}, and then generalized to the HEA in Section \ref{real_hea}.  

\section{The RG approach in real space for Dyson's Hierarchical Model} \label{real_dyson}

Let us consider a DHM, defined by Eqs. (\ref{20}), (\ref{c_f}). The real-space RG method is built up by initially iterating exactly the recursion equation (\ref{20}) for $k=k_0$ steps, 
assuming for simplicity that $H_0^F[S]=0$. 
 In this way, a DHM with $2^ {k_0}$ spins $S_1, \cdots, S_{2^{k_0}}$  and Hamiltonian $H_{k_0}^F[S_1, \cdots, S_{2^{k_0}}]$ is obtained exactly. Practically speaking, this means that in the following we compute exactly the physical observables of this $2^{k_0}$-spin DHM. 
 For instance, if $k_0=2$ we have a $4$-spin DHM whose Hamiltonian is
\bea \label{h_3} \no 
H_2^F[S_1, \cdots, S_4] & = & -  \Bigg\{  C_F J  \left[ \left( \frac{S_1+S_2}{2} \right) ^2 + \left( \frac{S_3+S_4}{2} \right) ^2 \right]  + \\ 
&&+  C_F^2 J \left(\frac{S_1 + S_2 + S_3 + S_4}{4} \right)^2\Bigg\}.
\eea
We recall for future purpose that in the Hamiltonian (\ref{h_3}), and similarly for arbitrary  values of $k_0$, when one goes one hierarchical level up,  the couplings are multiplied by a factor $C_F$. In particular, the first addend in braces in Eq. (\ref{h_3}) physically represents the couplings at the first hierarchical level, while the second addend represents the couplings at the second hierarchical level. \\

We now want to build up a $2^{k_0+1}$-spin DHM starting from such a $2^{k_0}$-spin DHM. As shown in Section \ref{dyson}, DHM is a special case where this procedure can be iterated $k$ times in $2^{k}$ operations, by using the hierarchical structure of the system resulting in the recurrence equation (\ref{rec_dhm}). On the contrary, if the recurrence equations (\ref{rec_dhm}) did not hold, in order to build up a $2^k$-spin DHM one should compute exactly the partition function, which involves $2^{2^k}$ operations.
To our knowledge, for  the HEA model there is no known recursion equation analogous to (\ref{rec_dhm}). Indeed, the only recursion equation that one can derive for the HEA is Eq. (\ref{recz}), which relies on the replica approach.
To derive a recurrence equation for a function or functional of a suitably defined order parameter without relying on the replica approach is  very difficult. The origin of this difficulty is nothing but the problem of how to identify of a suitable order parameter and  a function of it, which should replace the magnetization $m$ and its probability $p_k(m)$ in the recursion equation (\ref{rec_dhm}) of DHM. Since we have not been able to derive such a recurrence equation without relying on the replica approach, the construction of a $2^k$-spin HEA model still requires a computational effort of $2^{2^k}$. It is hence clear that an approximation scheme is needed to reach the thermodynamic limit $k \rightarrow \infty$ for the HEA. 
We now illustrate this approximation scheme for DHM first, and then generalized it to the HEA in Section \ref{real_hea}. \\

\begin{centering}
  \begin{figure} 
\centering
\includegraphics[width=9cm]{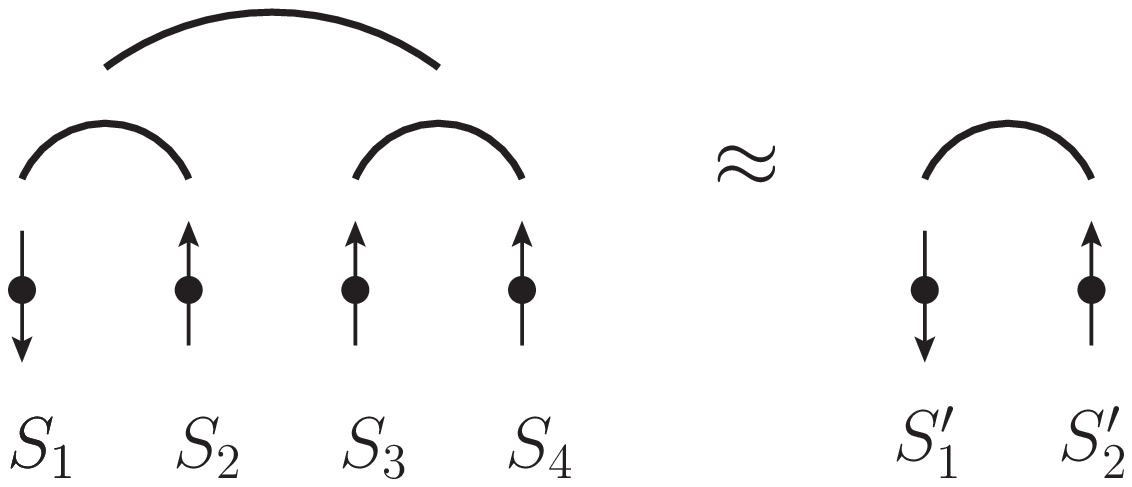}
\caption{Approximation of the real-space approach for $k_0=2$. 
In the implementation of the real-space approach to DHM exposed in Section \ref{real_dyson}, a  $2^2$-spin DHM with spins $S_1,\cdots, S_4$ and Hamiltonian (\ref{h_3}) is approximated by a $2$-spin DHM with spins $S'_1,S'_2$ and Hamiltonian (\ref{h_2}).\\
In the implementation of the real-space approach to the HEA exposed in Section \ref{real_hea}, a  $2^2$-spin HEA with spins $S_1,\cdots, S_4$ and Hamiltonian (\ref{156}) is approximated by a $2$-spin HEA with spins $S'_1,S'_2$ and Hamiltonian (\ref{300}).
}
\label{fig16}
\end{figure}
  \end{centering}

Once a $2^{k_0}$-spin DHM has been built exactly, we consider $2^{k_0-1}$-spin DHM, where $J$ is replaced by another coupling $J'$.  More precisely, such a $2^{k_0-1}$-spin DHM is defined by iterating $k_0-1$ times Eq. (\ref{20}) with $J \rightarrow J'$, and for the sake of clarity its spins will be denoted by $S'_1, \cdots, S'_{2^{k_0-1}}$, and its Hamiltonian by ${H'}^F_{k_0-1}[S'_1, \cdots, S'_{2^{k_0-1}}]$.   For $k_0=2$ the Hamiltonian of this DHM reads 
\be \label{h_2} 
{H'}_1^F[S'_1, S'_2]  = -    C_F J'   \left( \frac{S'_1+S'_2}{2} \right) ^2.
\ee  
Given $J$, the coupling $J'$ is chosen in such a way that the $2^{k_0-1}$-spin DHM represents as well as possible the $2^{k_0}$-spin DHM, as qualitatively depicted  in Fig. \ref{fig16}. The precise meaning of this representation will be illustrated shortly. 
  \\

According to the iterative construction of Eq. (\ref{20}), a new DHM is  then constructed by taking two copies of the $2^ {k_0-1}$-spin DHM. Say that the first copy has spins $S'_1, \cdots, S'_{2^{k_0-1}}$  and Hamiltonian ${H'}^F_{k_0-1}[S'_1, \cdots, S'_{2^{k_0-1}}]$, while the second copy has spins $S'_{2^{k_0-1}+1}, \cdots, S'_{2^{k_0}}$  and Hamiltonian ${H'}^F_{k_0-1}[S'_{2^{k_0-1}+1}, \cdots, S'_{2^{k_0}}]$. We make these two copies  interact and form a $2^{k_0}$-spin DHM with Hamiltonian
\be \label{comp} 
{H'}^F_{k_0-1}[S'_1, \cdots, S'_{2^{k_0-1}}]  
 + {H'}^F_{k_0-1}[S'_{2^{k_0-1}+1}, \cdots, S'_{2^{k_0}}]
 -J' C_F^ {k_0} \left( \frac{1}{2^{k_0}}\sum_{i=1}^{2^{k_0}}  S'_i \right)^2  . 
\ee

Since each of the DHMs that we make interact represents a $2^{k_0}$-spin DHM, the model defined by Eq. (\ref{comp}) represents a $2^{k_0+1}$-spin DHM. Once again, this DHM is then  approximated by a $2^ {k_0-1}$-spin DHM with coupling, say, $J''$, and two copies of such a $2^ {k_0-1}$-spin DHM are then taken and  coupled again, to obtain a system representing a $2^ {k_0+2}$-spin DHM.   Such a recursive construction is iterated $k$ times, and a system representing  a $2^{k_0+k}$-spin DHM is obtained. Setting $J_0\equiv J, J_1 \equiv J', J_2 \equiv  J'', \cdots$, this procedure establishes a relation between $J_{k}$ and $J_{k+1}$. Since at each step $k$ of this procedure we double the system size, this flow physically represents the RG flow of the coupling $J_{k}$ under reparametrization of the unit length $2^k \rightarrow 2^{k+1}$. \\

Let us now describe how a $2^{k_0}$-spin system has been approximated by a  $2^ {k_0-1}$-spin system.  Consider a physical observable $O^ F_{k_0}(\beta J )$ of the $2^ {k_0}$-spin DHM, whose  spins  are $S_1, \cdots, S_{2^{k_0}}$, and whose Hamiltonian is $H_{k_0}^F[S_1, \cdots, S_{2^{k_0}}]$. Consider also an observable $O^ F_{k_0-1}(\beta  J'  )$ of the $2^ {k_0-1}$-spin DHM, whose  spins are $S'_1,\cdots, S'_{2^{k_0-1}}$,  and whose Hamiltonian is ${H'}_{k_0-1}^F[S'_1, \cdots, S'_{2^{k_0-1}}]$. 
  The normalized magnetizations  on the left and right half of the $2^ {k_0}$-spin system are 
   \bea \label{m} \no 
  m_{ L}& \equiv& \frac { \frac{1}{2^ {k_0-1}} \sum_{i=1}^ {2^ {k_0-1}} S_i }{\sqrt{\mathbb{E}_{\vec{S}}  \left[ \left( \frac{1}{2^ {k_0-1}} \sum_{i=1}^ {2^ {k_0-1}} S_i    \right) ^2  \right] }},\\
  m_{ R} &\equiv  & \frac { \frac{1}{2^ {k_0-1}} \sum_{i=2^{k_0-1}+1}^ {2^ {k_0}} S_i }{\sqrt{\mathbb{E}_{\vec{S}}  \left[ \left( \frac{1}{2^ {k_0-1}} \sum_{i=2^ {k_0-1}+1}^ {2^ {k_0}} S_i    \right) ^2  \right] }}
   \eea
respectively,   where $\mathbb{E}_{\vec{S}}$ stands for the thermal average at fixed temperature $T$, performed with the Boltzmann weight $\exp(-\beta H_{k_0}^F)$.  
   Similarly, the normalized magnetizations  on the left and right half of the $2^ {k_0-1}$-spin system are 
   \bea \label{m'} \no 
  m'_{ L} &\equiv &  \frac { \frac{1}{2^ {k_0-2}} \sum_{i=1}^ {2^ {k_0-2}} S'_i }{\sqrt{\mathbb{E}_{\vec{S'}}  \left[ \left( \frac{1}{2^ {k_0-2}} \sum_{i=1}^ {2^ {k_0-2}} S'_i    \right) ^2  \right] }}, \\
m'_{ R} &\equiv&    \frac { \frac{1}{2^ {k_0-2}} \sum_{i=2^{k_0-2}+1}^ {2^ {k_0-1}} S'_i }{\sqrt{\mathbb{E}_{\vec{S'}}  \left[ \left( \frac{1}{2^ {k_0-2}} \sum_{i=2^ {k_0-2}+1}^ {2^ {k_0-1}} S'_i    \right) ^2  \right] }}
   \eea
respectively,  where $\mathbb{E}_{\vec{S'}}$ stands for the thermal average with the Boltzmann weight $\exp(-\beta {H'}_{k_0-1}^F)$. \\

According to Kadanoff's block-spin rule, in order that  the $2^{k_0-1}$-spin DHM might be a good approximation of the $2^{k_0}$-spin DHM, one should map the block of  spins in the left half of the $2^{k_0}$-spin DHM into the block of  spins in the left half of the $2^{k_0-1}$-spin DHM, and so for the right half. Accordingly, one should find a method which quantitatively implements the qualitative equalities 
\be \label{kadanoff} 
m_L = m'_L, \, m_R = m'_R. 
\ee
To this end,  we   choose the following  observables
\bea \label{obs_ferr} \no
    O^ F_{k_0}(\beta J ) & \equiv & \mathbb{E}_{\vec{S}} \left[  m_L m_R  \right]  ,  \\
    O^ F_{k_0-1}(\beta J') & \equiv & \mathbb{E}_{\vec{S}' } \left[  m'_L m'_R \right].
\eea
According to Eqs. (\ref{kadanoff}),  Kadanoff's block-spin rule described in Eq. (\ref{kadanoff}) can be practically implemented by imposing the constraint
\be \label{rg_equation_ferro}
O^ F_{k_0}( \beta J ) =   O^ F_{k_0-1}(\beta J').
\ee
For any fixed $J$, Eq. (\ref{rg_equation_ferro}) is the  equation determining $J'$ as a function of $J$, as the value of the coupling of the $2^{k_0-1}$-spin DHM such that this is the best-possible approximation of the $2^{k_0}$-spin DHM. According to the above discussion, Eq. (\ref{rg_equation_ferro}) is the RG equation relating the coupling $J$ at the scale $2^k$ to the coupling $J'$ at the scale $2^{k+1}$. \\

The RG Eq. (\ref{rg_equation_ferro}) is not exact, because it relies on the fact that a $2^{k_0}$-spin DHM is approximated by a $2^{k_0-1}$-spin DHM. Even though, such an approximation  becomes asymptotically exact for large $k_0$, as we will explicitly show in the following. Another important issue of this RG scheme is that there is a considerable amount of freedom in the choice of the observables $O^ F_{k_0}, O^ F_{k_0-1}$, and that the  RG equations (\ref{rg_equation_ferro}) depend on this choice. This is the reason why in what  follows the whole method will  be systematically tested a posteriori, by comparing its predictions to the predictions obtained heretofore with other methods, if these exist. As we will show shortly, the encouraging outcome of this comparison makes us guess that if $k_0$ is large enough, the results of this RG approach do not depend on the choice of the observables, if this is reasonable. \\

Quite large values of $k_0$ can be achieved by using the hierarchical structure of the system. Indeed, thanks to this structure the thermal averages appearing in Eqs. (\ref{obs_ferr}), which would involve $2^{2^{k_0}}$ terms in a brute-force computation, can be computed in  $2^{k_0}$ operations, as shown in Appendix \ref{app_dhm}. \\

Back to the predictions of Eq. (\ref{rg_equation_ferro}),  one can show that for any $k_0$ they reproduce the interval (\ref{27}). Indeed, for $\sigma_F > 1$ Eq. (\ref{rg_equation_ferro}) gives $J'<J\, \forall J,\beta $, in such a way that the coupling $J$ goes to $0$ when the RG transformation is iterated many times, and no phase transition occurs. 
On the contrary, for $\sigma_F <1/2$ one has $J' > J \, \forall J,\beta$, in such a way that the model is thermodynamically unstable. The fact that the interval (\ref{27}) is reproduced is  a first test of the correctness of the real-space approach.  
For $1/2 <\sigma_F < 1$ there is a finite inverse temperature $\beta_{c\, F}^{RS}$ such that for $\beta < \beta_{c\, F}^{RS}$ one has $J' < J$, while for $\beta>\beta_{c\, F}^{RS}$ one has $J'>J$, where the label RS stands for real space. For $\beta = \beta_{c\, F}^{RS}$, $J'=J$, i. e.  the system is invariant under reparametrization of the length scale $2^k \rightarrow 2^{k+1}$. Hence $\beta_{c\, F}^{RS}$ is the critical temperature of the model \cite{wilson1982renormalization, wilson1974renormalization}. \\

\begin{figure}
\begin{centering}
\includegraphics[width=9cm]{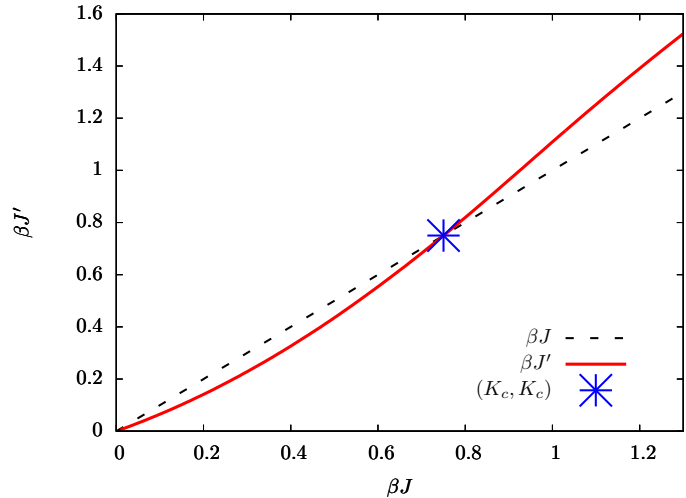}
\caption{$\beta J'$ as a function of $\beta J$ for $C_F=1.2,\, k_0=8$, and point $(K_c,K_c)$ where the two curves intersect. 
According to the discussion in the text, $\beta \lesseqgtr \beta_{c\, F}^{RS}$ implies that  $J' \lesseqgtr J$. 
 }
\label{fig9}
\end{centering}
\end{figure}

The RG transformation (\ref{rg_equation_ferro})  is illustrated in Fig \ref{fig9}, where $\beta J'$ is depicted as a function of $\beta J$ for a given $C_F$ and $k_0$-value. Interesting properties about universality emerge from this plot. Indeed, the curve $\beta J'$ as a function of $\beta J$ intersects the straight line $\beta J$ for a unique value of the  coupling $\beta J \equiv K_c$.  Now let us iterate the RG transformation several times, starting with a given $J= J_0$, then determining $J'=J_1, J''=J_2$, and so on. At the first step of the iteration, $J_1=J_0$ if and only if $\beta J_0 = K_c$. It follows that $\beta_{c\, F}^{RS}=K_c/J_0$. Similarly, at the next steps $K_c/J_0$ is the only value of the inverse temperature such that $J_k=J_0\, \forall k$. Since $K_c$ is defined as the solution of the equations $O^ F_{k_0}(K_c ) =   O^ F_{k_0-1}(K_c)$, it does not depend on the initial condition $J_0$, and  thus it is universal.
As an analogy, $K_c$ corresponds to the dimensionless nearest-neighbor critical coupling of the Ising model, which has been extensively measured in three dimensions by means of Monte Carlo Renormalization Group (MCRG) calculations \cite{baillie1992monte,pawley1984monte}. 
On the other hand, dimensional quantities like $\beta_{c\, F}^{RS}=K_c/J_0$ are not universal. Indeed, $\beta_{c\, F}^{RS}=K_c/J_0$  depends on the coupling $J_0$ at the initial step of the iteration, i. e. at microscopic length scales. This is in agreement with the very general RG picture of ferromagnetic systems like the Ising model \cite{wilson1974renormalization}, where the critical temperature is not universal because it depends on the microscopic properties of the lattice. 
\\

\begin{figure}
\begin{centering}
\includegraphics[width=13cm]{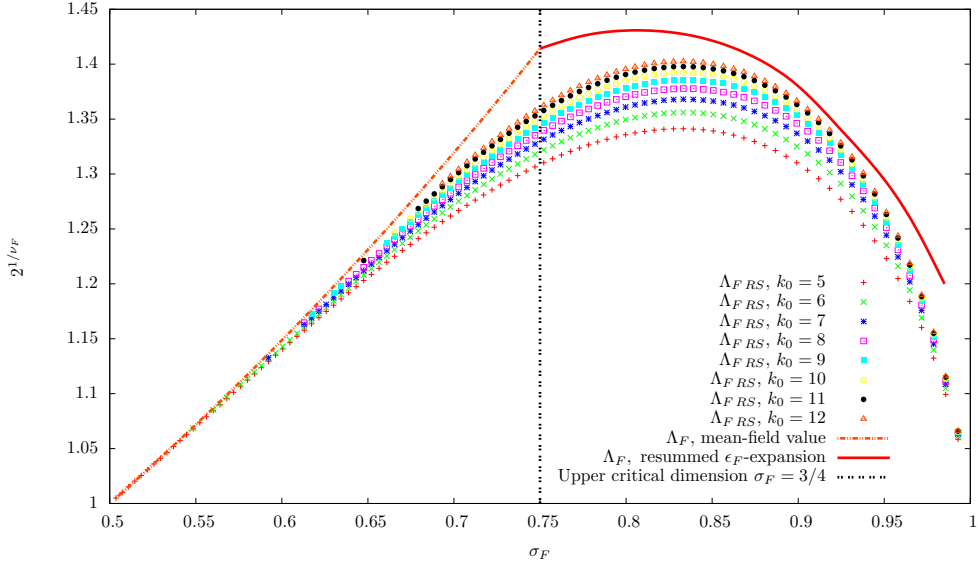}
\caption{$\Lambda_{F\, RS}$ and $\Lambda_F$ as a function of $\sigma_F $ for $1/2<\sigma_F<1$. The black dashed line represents the upper   critical dimension  $\sigma_F=3/4$ discussed in Section \ref{dyson} and Appendix \ref{app_field_dhm}. The points represent $\Lambda_{F\, RS}$ for $5\leq k_0\leq 12$. The orange dashed curve and the red solid curve represent  $\Lambda_F$ obtained with the field-theory method in the mean-field region $1/2<\sigma_F \leq 3/4$ and in the non-mean-field region $3/4<\sigma_F < 1$ respectively. 
The orange dashed curve represent the exact mean-field value of $\Lambda_F$ given by Eq. (\ref{150}), and the red continuous curve represents $\Lambda_F$ computed with the resummed $\epsilon_F=\sigma_F-3/4$-expansion \cite{collet1977numerical} (see Eq. (\ref{151}) for the first order of the $\epsilon_F$-expansion).
}
\label{fig10}
\end{centering}
\end{figure}

An important universal quantity is the critical exponent $\nu_F$, defined in terms of the correlation length by Eq. (\ref{def_nu_f}).  According to the general RG theory in the neighborhood of the critical fixed point $K_c$, $\nu_F$ is given by
\be \label{nu_lambda_f}
\nu_F = \frac{ \log 2} {\log \Lambda_{F\, RS}},
\ee
where $\Lambda_{F\, RS}$ is here defined as 
\be \label{lambda_ferro}
\Lambda_{F\, RS} \equiv \left . \der{ \beta J'}{\beta J}  \right|_{\beta J = K_c} . 
\ee
Since in this case the RG transformation involves only one variable $J$, $\Lambda_{F\, RS}$ is simply the largest eigenvalue of the $1 \times 1$ matrix linearizing the transformation in the neighborhood of the critical fixed point. A more complex case where the RG transformation involves an infinite number of variables will be discussed in Section \ref{real_hea}. 
In Fig. \ref{fig10} we depict $\Lambda_{F\, RS}$ from Eq. (\ref{lambda_ferro})  together with the  values of $\Lambda_F$ (Eq. (\ref{nu_f}))  presented in \cite{collet1977numerical}  resulting from the field-theory approach of  Section \ref{dyson} and Appendix \ref{app_field_dhm}, as a function of $1/2<\sigma_F<1$.  The field-theory method makes the exact prediction (\ref{150}) for $\Lambda_F$ in the region $1/2<\sigma_F<3/4$ where the mean-field approximation is exact, while it estimates  $\Lambda_F$ in the non-mean-field region $3/4<\sigma_F<1$ by means of a resummed $\epsilon_F=\sigma_F-3/4$-expansion. The first order of this expansion is given by Eq. (\ref{151}). \\

As $k_0$ is increased, \mnote{The real-space RG approach makes a prediction for the critical exponents of Dyson's Hierarchical Model which is in good agreement with that obtained with other  methods.}
$\Lambda_{F\, RS}$ computed with the real-space method  approaches the field-theory value $\Lambda_F$, confirming the validity of the real-space RG approach. Notice that  according to Eqs. (\ref{150}) and (\ref{151}), the derivative with respect  to $\sigma_F$ of $\Lambda_F$ is discontinuous at $\sigma_F=3/4$. On the contrary, $\Lambda_{F\, RS}$ is a smooth function of $\sigma_F$. This discrepancy is presumably due to the fact that $k_0$ is not large enough, and should disappear for larger $k_0$, because the real-space approach is exact for $k_0 \rightarrow \infty$.  \bigskip \\

Since the real-space approach  for DHM reproduces the interval (\ref{27}) and, for large $k_0$, also the critical exponents obtained with other methods, it is natural to generalize it to the HEA model. Accordingly, its predictions will be compared to those obtained with the replica approach in the mean-field region $1/2<\sigma \leq 2/3$ where the latter is predictive, yielding a precise consistency test of the two approaches. On the contrary, in the non-mean-field region $2/3<\sigma<1$ the replica approach is not predictive, and thus a quantitative comparison of the values of the critical exponents would be meaningless. Even though, an interesting question is whether the real-space approach is predictive in this non-mean-field region. These points will be addressed in the following Section.

\section{The RG approach in real space for the  Hierarchical Edwards-Anderson model} \label{real_hea}

It turns out that the definition of the HEA which is suitable to implement the real-space approach is the one where the interaction energies are redefined by Eq. (\ref{redefinition}). Accordingly, in what follows the HEA will be defined by Eqs. (\ref{1}), (\ref{redefinition}), and  we set for simplicity $H_0[S]=0$. 

\subsection{Simplest approximation of the real-space method}\label{k_0_equal_2}

Let us now illustrate how to implement in the spin-glass case  the real-space  approach presented in Section \ref{real_dyson} for $k_0=2$.
The reader should follow our derivation in close analogy with the one exposed in Section \ref{real_dyson} for DHM. 
 A  HEA model   with $2^ {2}$ spins $S_1,\cdots, S_4$ is built up exactly by means of the recursion equation (\ref{1}). Setting $\mathcal{J}_{ij} \equiv C^2/2 J_{ij}$,  the Hamiltonian of this model is 
\bea \no \label{156}
H_2[S_1, \cdots, S_4 ]& = &-  \Bigg \{ \left [ \mathcal{J}_{12} S_1 S_2 +  \mathcal{J}_{34} S_3 S_4  \right] +  \frac{C^2}{2} [ \mathcal{J}_{13}S_1S_3 +  \mathcal{J}_{14}S_1S_4+\\
&& +  \mathcal{J}_{23}S_2S_3 +  \mathcal{J}_{24}S_2S_4  ] \Bigg \}.
\eea
By definition, the couplings  $\{ \mathcal{J}_{ij} \}_{ij}$ are IID random variables, and the probability distribution of each of them will be denoted by $p(\mathcal{J})$. \\

Thus, we consider a $2$-spin HEA model, defined by iterating once Eq. (\ref{1}). For the sake of clarity its spins will be denoted by $S'_1, S'_{2}$, and its Hamiltonian reads
\be \label{300} 
{H'}_1[S'_1, S'_2]  = -   \mathcal{ J}'_{12}   S'_1S'_2. 
\ee  

For each realization of the couplings $\{ \mathcal{J}_{ij} \}_{ij}$, we choose $\mathcal{J}'_{12}$ as a function of $\{ \mathcal{J}_{ij} \}_{ij}$  in such a way that the $2$-spin HEA model yields the best-possible approximation of the $2^2$-spin HEA. 
This procedure is analogous to that exposed  in Section \ref{real_dyson} for DHM, and is qualitatively depicted in Fig. \ref{fig16}. 
By choosing $\mathcal{J}'_{12}$ in  this way, the distribution $p(\mathcal{J})$ induces a distribution of $\mathcal{J}'_{12}$, that we will denote by $p'(\mathcal{J}'_{12})$.
The technical details on how $p'(\mathcal {J})$ is computed from $p(\mathcal{J})$ will be given  shortly.  \\

According to the iterative construction of Eq. (\ref{1}), a new HEA is  then constructed by taking two realizations of the $2$-spin HEA. 
Each realization is given by drawing the coupling $\mathcal{J}'$ according to its probability distribution $p'(\mathcal{J}')$. 
Say that the first realization has spins $S'_1, S'_2$  and Hamiltonian ${H'}_{1}[S'_1, S'_{2}] = -   \mathcal{ J}'_{12}   S'_1S'_2$, while the second realization has spins $S'_3,  S'_{4}$  and Hamiltonian ${H'}_{1}[S'_3,S'_4]= -   \mathcal{ J}'_{34}   S'_3S'_4$. We make these two copies  interact and form a $2^2$-spin HEA with Hamiltonian
\be \label{comp2} 
{H'}_{1}[S'_1, S'_2]  
 + {H'}_{1}[S'_3,S'_4]
 -  \frac{C^2}{2} [ \mathcal{J}'_{13}S'_1S'_3 +  \mathcal{J}'_{14}S'_1S'_4 +  \mathcal{J}'_{23}S'_2S'_3 +  \mathcal{J}'_{24}S'_2 S'_4  ],
\ee
where $ \mathcal{J}'_{13}, \mathcal{J}'_{14}, \mathcal{J}'_{23},  \mathcal{J}'_{24}$ have been drawn independently from the distribution $p'(\mathcal{J'})$. 
Since each of the HEA models that we made interact represents a $2^2$-spin HEA, the model defined by Eq. (\ref{comp2}) represents a $2^3$-spin HEA. At the next step of the iteration, this HEA model is again  approximated by a $2$-spin HEA with coupling, say, $\mathcal{J}''_{12}$, and the probability distribution $p''(\mathcal{J}''_{12})$ of  $\mathcal{J}''_{12}$ is computed from $p'(\mathcal{J}')$. 
 Two realizations of such a $2$-spin HEA are then taken and  coupled again, to obtain a system representing a $2^ {4}$-spin HEA.   This step is repeated $k$-times, and a system representing  a $2^{2+k}$-spin HEA is obtained.\\

 Setting $p_0(\mathcal{J})\equiv p(\mathcal{J}), p_1(\mathcal{J}) \equiv p'(\mathcal{J}), p_2(\mathcal{J}) \equiv  p''(\mathcal{J} ), \cdots$, this procedure establishes a relation between $p_{k}(\mathcal{J})$ and $p_{k+1}(\mathcal{J})$. Since at each step $k$ of this procedure we double the system size, this physically represents the RG flow of the probability distribution of the  coupling $p_k(\mathcal{J})$ under reparametrization of the unit length $2^k \rightarrow 2^{k+1}$. \\

A $2^2$-spin HEA has been approximated by a  $2$-spin HEA as follows.  Consider a physical observable $O_{2}(\{\beta  \mathcal{J}_{ij} \}_{ij} )$ of the $2^ 2$-spin HEA, depending on the $6$  couplings $\{ \mathcal{J}_{ij} \}_{ij}$ and $\beta$ through the dimensionless quantities $\{ \beta \mathcal{J}_{ij} \}_{ij}$. Consider also an observable $O_{1}( \beta \mathcal{J}'_{12}  )$ of the $2$-spin HEA, depending on the  coupling $\mathcal{J}'_{12}$ and $\beta$ through the dimensionless quantity $\beta \mathcal{J}'_{12}$. 
We recall that in the real-space approach for DHM with $k_0=2$ we built up the observables (\ref{obs_ferr}) as products of the magnetizations inside the left and right-half of the $2^2$-spin DHM and of the $2$-spin DHM. In a sense  that choice was natural, because we know that the magnetization is the correct order parameter of DHM. When one tries to generalize that choice to the HEA, one has to face the fact that the order parameter for non-mean-field spin glasses is not known, and so the choice of the observables is more difficult and ambiguous. Inspired by the fact that the order parameter in the mean-field case is the overlap \cite{parisi1979infinite,parisi1980order,parisi1983order,MPV,NishimoriBook01}, here we will build up $O_2, O_1$ as thermal averages of products of spin overlaps. We want to stress that to a certain extent this choice relies on the underlying assumption that the overlap is still the good quantity describing the physics of the system, and should be thus be verified a posteriori. In particular, there is no guarantee that this choice is correct for values of $\sigma$ lying in the non-mean-field region $2/3<\sigma<1$. A detailed analysis of the predictions of this real-space approach for the critical exponents in this region will be exposed in the following. \\

   To build up $O_2$ and $O_1$, consider two \textit{real} replicas $\vec{S}^1, \vec{S}^2$ of the spins of the $2^2$-spin model, and  two real replicas $\vec{S'}^1, \vec{S'}^2$ of the spins of the $2$-spin model. The normalized overlap between $\vec{S}^1$ and $\vec{S}^2$ on the left and on the right leaf of the $2^2$-spin HEA  are
   \bea \label{155}\no 
Q_L &=&    \frac { \frac{S^ 1_1 S^ 2_1 +S^ 1_2 S^ 2_2}{2} }{\sqrt{\mathbb{E}_{\vec{S}^1,\vec{S}^2}  \left[ \left( \frac{S^ 1_1 S^2_1 +S^1_2 S^2_2}{2} \right) ^2  \right] }},\\
Q_R &= &    \frac { \frac{S^ 1_3 S^ 2_3 +S^ 1_4 S^ 2_4}{2} }{\sqrt{\mathbb{E}_{\vec{S}^1,\vec{S}^2}  \left[ \left( \frac{S^ 1_3 S^2_3 +S^1_4 S^2_4}{2} \right) ^2  \right] }}
   \eea
respectively,  where $\mathbb{E}_{\vec{S}}$ denotes the thermal average at fixed disorder $\{ \mathcal{J} \}_{ij}$ and temperature, performed  with the Hamiltonian (\ref{156}).   The normalized overlap between $\vec{S'}^1$ and $\vec{S'}^2$ on the left and on the right leaf of the $2$-spin HEA are
   \bea \label{157}\no 
Q'_L & = &S'^ 1_1 S'^ 2_1,\\
  Q'_R & = &S'^ 1_2 S'^ 2_2
   \eea
  respectively. 
   Following Kadanoff's decimation rule, we want to  map the $2^ 2$-spin HEA into the $2$-spin HEA  by imposing that the spins $S_1, S_2$ correspond to the spin $S'_1$, and that the spins $S_3, S_4$ correspond to the spin $S'_2$. This mapping results in a correspondence between the overlaps in Eq. (\ref{155}) and those in Eq. (\ref{157}), which can be qualitatively written as 
\be \label{kadanoff2}
Q_L = Q'_L, \, Q_R = Q'_R. 
\ee
Making the following choice for the observables   
    \bea \label{obs_hea}
    O_{2}(\{ \beta \mathcal{J}_{ij} \} ) & \equiv & \mathbb{E}_{\vec{S}^1, \vec{S}^ 2} \left[ Q_L Q_R\right]  ,  \\ \no
    O_{1}(\beta \mathcal{J}'_{12}) & \equiv & \mathbb{E}_{\vec{S}'^1, \vec{S}'^ 2} \left[  Q'_L Q'_R \right], 
\eea
Kadanoff's decimation rule encoded in Eq. (\ref{kadanoff2}) can be practically implemented by enforcing the constraint   
\be \label{158}
 O_{2}(\{ \beta \mathcal{J}_{ij} \} ) = O_{1}(\beta \mathcal{J}'_{12}),
\ee
where $\mathbb{E}_{\vec{S'}}$ denotes the thermal average  at fixed disorder $\mathcal{J}' _{12}$ and temperature, performed  with the Hamiltonian (\ref{300}).
For any realization of the couplings $\{ \mathcal{J}_{ij} \}_{ij}$, Eq. (\ref{158}) determines $\mathcal{J}'_{12}$ as a function of  $\{ \mathcal{J}_{ij} \}_{ij}$ in such a way that the $2$-spin HEA yields the best-possible approximation of the $2^2$-spin HEA. 
The mapping (\ref{158}) results into a mapping between the probability distribution $p(\mathcal{J})$ of each of the couplings $\{ \mathcal{J}_{ij} \}_{ij}$ and $p'(\mathcal{J}'_{12})$. 
Indeed, Eq. (\ref{158}) can be easily rewritten as 
\[
 O_{2}(\{ \beta \mathcal{J}_{ij} \} ) = \tanh^2(\beta \mathcal{J}'_{12}). 
\]
and thus the mapping between $p(\mathcal{J})$ and $p'(\mathcal{J}')$ is
\bea \label{159}  \no
p'(\mathcal{J}') &= &\int \Bigg[ \prod_{i<j}p(\mathcal{J}_{ij})  d \mathcal{J}_{ij} \Bigg] \frac{1}{2}\Bigg[ \delta \left(  \mathcal{J}' - \frac{1}{\beta} \operatorname{arctanh}\left(\sqrt{O_{2}(\{ \beta  \mathcal{J}_{ij} \})} \right)    \right) + \\
&& + \delta \left(  \mathcal{J}' + \frac{1}{\beta} \operatorname{arctanh} \left(\sqrt{O_{2}(\{ \beta \mathcal{J}_{ij} \}} )\right)    \right) \Bigg].
\eea
Eq. (\ref{159}) is the RG flow relating the distribution  of the disorder $p(\mathcal{J})$ at  length scale $2^k$ with the distribution  of the disorder $p'(\mathcal{J}')$ at length scale $2^{k+1}$, and is the generalization of Eq. (\ref{rg_equation_ferro}) holding for DHM.
We recall that the RG equation (\ref{rg_equation_ferro}) for DHM yields a flow for a number $J$, while  Eq. (\ref{159}) yields a flow for a function $p(\mathcal{J})$, which can be seen  as a set  $\{ p (\mathcal{J}) \}_{\mathcal{J}}$  of an infinite number of degrees of freedom, each of which is the value of $p(\mathcal{J})$ at a point $\mathcal{J}$. 
Accordingly, the solution of Eq. (\ref{159}) is slightly more complicated than that of Eq. (\ref{rg_equation_ferro}), and it has been worked out with two independent techniques. The first one  transforms Eq. (\ref{159}) into a recursion equation relating the moments of $p(\mathcal J)$ to the moments of $p'(\mathcal{J}')$, which is built up perturbatively by means of a high-temperature expansion, and is presented in Section \ref{pert_hea}. The second one is purely numerical, and solves Eq. (\ref{159}) by means of the population dynamics algorithm, as illustrated in Section \ref{pop_hea}. In the following two Sections we thus illustrate these solution techniques for Eq. (\ref{159}), and analyze the resulting fixed-points structure and show how the critical exponents can be calculated in the $k_0=2$-approximation.
 Then, we illustrate how the very same techniques can be implemented in better approximations of the real-space  approach, i. e. $k_0>2$, and analyze the predictions of the real-space approach for the critical exponents as a function of $k_0$.

\subsubsection{Solution of the real-space RG equations with the high-temperature expansion}\label{pert_hea}

Since it is not easy to handle analytically the continuous set of degrees of freedom $\{ p (\mathcal{J}) \}_{\mathcal{J}}$ in Eq. (\ref{159}), it is better to transform the latter into an equation for the moments of $p (\mathcal{J}) , p' (\mathcal{J}')$. Since there is $\pm \mathcal J$-symmetry, $p (\mathcal{J}) , p' (\mathcal{J}')$ are even functions of $\mathcal{J}, \mathcal{J}'$ respectively. Hence, setting
\beas
m_{a}& \equiv & \int d \mathcal{J} p(\mathcal{J}) \mathcal{J}^a, \\
m'_{a}& \equiv & \int d \mathcal{J}' p'(\mathcal{J}') (\mathcal{J}')^a, 
\eeas
one has $m_{2a+1}=m'_{2a+1}=0$, and Eq. (\ref{159}) can be transformed into an equation relating $\{ m_{2a} \}_{a}$ to $\{ m'_{2a} \}_{a}$. Indeed, let us
call the $6$ couplings $\mathcal{J}_{12}, \mathcal{J}_{13}, \ldots, \mathcal{J}_{34}$ \hspace{2mm} $\mathcal{J}_1, \cdots, \mathcal{J}_6$ respectively, and integrate both sides of Eq. (\ref{159}) with respect to $\mathcal{J}'$
 \bea \label{160}\no 
 m'_{2a}  &=& \int \left[ \prod_{\alpha=1}^ {6} d\mathcal{J}_\alpha p(\mathcal{J}_\alpha)  \right] \frac{1}{\beta^{2a}} \left[ \operatorname{arctanh}{^{2a}} \left( \sqrt{O_2( \{ \beta  \mathcal{J}_\gamma \} _\gamma )}\right)  \right] \\ 
 & \equiv & F_{2a}[\{ m_ {2 b} \}_{b}].
 \eea
The function $F_{2a}$  depends in a complicated way on the even moments $\{ m_ {2 b} \}_{b}$, and we have not been able to compute it explicitly. Still, this dependence can be systematically worked out by expanding in powers of $\beta$ the  square brackets in the right-hand side of Eq. (\ref{160}). 
If we truncate the expansion at a given order $\beta^{2 m}$, the right-hand side of Eq. (\ref{160}) becomes a linear combination of $\{m_{2b} \}_{b=1, \ldots, m}$ which can be computed explicitly. Hence, if we take Eq. (\ref{160}) for $a=1, \ldots, m$, we obtain a set of equations relating $\{ m'_{2a} \}_{a=1,\ldots,m}$ to  $\{m_{2b} \}_{b=1, \ldots, m}$. This set of equations is nothing but the flow $p(\mathcal{J}) \rightarrow p'(\mathcal{J}')$ represented with the discrete set of degrees of freedom $\{ m_{2a} \}_{a=1,\ldots,m} \rightarrow \{m'_{2b} \}_{b=1, \ldots, m}$. \medskip\\

In Appendix \ref{real_high_temp} we show explicitly these equations for $m=2$,  Eq. (\ref{161}), and discuss their solution.  In particular, we show that 
if we consider only the leading terms in the $\beta$-expansion, the RG equations relating $\{ m'_{2a} \}_{a=1,\ldots,m}$ to  $\{m_{2b} \}_{b=1, \ldots, m}$ reproduce the  condition $\sigma>1/2$ that has been previously derived in Eq. (\ref{sigma_hea}). On the contrary, they do not reproduce the condition $\sigma<1$. This  is presumably due to the fact that the present approach implements the lowest-order approximation $k_0=2$ of the real-space method.  Indeed, in Section \ref{k_0_larger_2} we will show that the numerical implementation of the real-space method for $k_0>2$ yields a better description of the region where $\sigma$ significantly deviates from $1/2$, and the condition $\sigma <1$ should be recovered for $k_0$ large enough.  \\

It is easy to see that Eq.  (\ref{161}) has an attractive high-temperature fixed point $m_{2a} = 0 \, \forall a$, and an attractive low-temperature fixed point  $m_{2a} =  \infty \, \forall a$. These fixed points are separated by a repulsive critical fixed point $\{ m_{2a}^\ast \}_{a=1, \ldots, m}$.
In order to investigate the latter, one can introduce a critical temperature $\beta_c^{RS}$ such that  Eq. (\ref{161}) converges to   $\{ m^{\ast}_{2a} \}_{a=1, \ldots, m}$ for $\beta =\beta_c^{RS}$. This fixed point is determined by means of an expansion in powers of $\sigma-1/2$, physically representing the distance from the purely mean-field regime $\sigma=1/2$ of the model. 
The critical exponent $\nu$ defined by Eq. (\ref{xi_hea})
is expressed in terms of the largest eigenvalue $\Lambda_{RS}$ of the matrix linearizing  Eq. (\ref{161}) in the neighborhood of $\{ m^{\ast}_{2a} \}_{a=1, \ldots, m}$ through the relation
\[
\nu = \frac{\log 2}{\log \Lambda_{RS}},
\]
and is given by Eq. (\ref{166}). \\

This computation can be performed to higher orders. If the expansion is done up to  $O(\beta^{2m})$,  $\Lambda_{RS}$ can be computed to order $(\sigma-1/2)^{m-1}$. The calculation has been done for $m=5$ by means of a symbolic manipulation program \cite{wolfram1996mathematica}, and the result is 
\bea  \label{lambda_rs_high_temp} \no
\Lambda_{RS} &=& 1+ 2 \log 2 \,  (\sigma-1/2)  -\frac{219  (\log 2)^2}{20}  (\sigma-1/2) ^2+ \\ \no 
&&-\frac{113453 (\log 2)^3}{1200}  (\sigma-1/2) ^3+\frac{56579203 (\log 2)^4}{403200} (\sigma-1/2) ^4 +\\
&& + O((\sigma-1/2)^5). 
\eea

Even though  only the first four terms of the expansion are available, Eq. (\ref{lambda_rs_high_temp}) yields an accurate estimate of $\Lambda_{RS}$ in a relatively wide range of values of $\sigma$. This is shown in Table \ref{tab2}, where the values of  $\Lambda_{RS}^{(i)}$ obtained by truncating the expansion (\ref{lambda_rs_high_temp}) to order $(\sigma-1/2)^{i}$ are listed for different values of  $0.54 \leq \sigma  \leq 0.62$ and $i$. Since  $\Lambda_{RS}^{(i)}$ increases by less than $1 \%$ when increasing $i$ from $3$ to $4$, in this region we can extract the exact value of $\Lambda_{RS}$ with good accuracy.\medskip\\ 
\begin{table}
\caption{$\Lambda_{RS}^{(i)}$ as a function of $i$ for different values of $0.54 \leq \sigma\leq 0.62$. The relative change in $\Lambda_{RS}^{(i)}$ obtained as one increases the order $i$ from $3$ to $4$ is less than $1 \%$, and yields an estimate of the error on the critical exponent $\nu$.  }
\label{tab2}
\centering
\begin{tabular}{lcccc}
\toprule
$\sigma$ & $\Lambda_{RS}^{(1)}$ & $\Lambda_{RS}^{(2)}$ & $\Lambda_{RS}^{(3)}$ & $\Lambda_{RS}^{(4)}$ \\
\midrule
0.54 & 1.05545 & 1.04703 & 1.04502 & 1.0451 \\
0.58 & 1.1109& 1.07723& 1.06111& 1.06244\\
0.62 & 1.16636& 1.0906& 1.03619& 1.04291 \\ 
\bottomrule
\end{tabular}
\end{table}

It turns out that the high-temperature expansion cannot be implemented for $k_0>2$, because the symbolic manipulations become too difficult.
Still, the values of the critical exponent $\nu$ computed with the high-temperature expansion for $k_0=2$ will serve as an important test of a different, purely numerical implementation of the RG equations (\ref{159}) for $k_0=2$, that will be illustrated in the following Section.
Once the agreement between the numerical and analytical approach will be established for $k_0=2$, the numerical method will be easily implemented for $k_0>2$, yielding an estimate of the exact value of the exponents obtained as $k_0$ is increased.

\subsubsection{Solution of the real-space RG equations with the population-dynamics method}\label{pop_hea}

The RG equations (\ref{159}) are nonlinear integral equations, and it is difficult to solve them analytically and determine $p'(\mathcal{J}')$ as a functional of $p(\mathcal{J})$. Accordingly,  one can  use some numerical methods. Here we describe a stochastic approach known as population dynamics, yielding an extremely simple and powerful  solution  of Eq. (\ref{159}). Historically, population dynamics appeared first in the theory of localization of electrons in disordered systems \cite{chacra1973self}, and was later  developed for spin glasses \cite{mezard2001bethe} and constraint-satisfaction problems \cite{0}. \\

In population dynamics one represents the function $p(\mathcal{J})$ as a population of $P$ numbers $\{ \mathcal{J}_i \}_{i=1, \ldots, P}$, where each $\mathcal{J}_i$ has been drawn with probability  $p(\mathcal{J}_i)$. Accordingly if $P$ is large enough, once we know  $p(\mathcal{J})$ we can compute the population $\{ \mathcal{J}_i \}_{i}$, while once we know the population $\{ \mathcal{J}_i \}_{i}$ we can compute $p(\mathcal{J})$ from the relation 
\be \label{binning}
p(\mathcal{J}) dJ \sim \frac{1}{P} \sum_{i=1}^P \mathbb{I}[\mathcal{J}_i \in (J, J+dJ)],
\ee
where the function $ \mathbb{I}[\mathcal{J}_i \in (J, J+dJ)]$ equals one if $\mathcal{J}_i \in (J, J+dJ)$ and zero otherwise, and $dJ$ is a suitably chosen small binning interval. Thus, there is  a one-to-one correspondence between $p(\mathcal{J})$ and $\{ \mathcal{J}_i \}_{i=1, \ldots, P}$, and these are  different representations of the same object
\be \label{pop_p}
p(\mathcal{J}) \leftrightarrow \{ \mathcal{J}_i \}_{i=1, \ldots, P}. 
\ee

In practice, once one knows the population $\{ \mathcal{J}_i \}_{i}$ the corresponding $p(\mathcal{J})$ is computed by setting 
\be \label{j_max}
\mathcal{J}_{\text{MAX}} \equiv \max_i (\lvert \mathcal{J}_i \rvert),
\ee
 choosing $dJ = 2 \mathcal{J}_{\text{MAX}}/B$, and using Eq. (\ref{binning}), where $B$ is a suitably chosen large integer number.  \\

The mapping $p(\mathcal{J}) \rightarrow p'(\mathcal{J}')$ given by Eq. (\ref{159}) yields a mapping between $ \{ \mathcal{J}_i \}_{i=1, \ldots, P}$ and the population $\{ \mathcal{J}'_i \}_{i}$ representing $p'(\mathcal{J}')$
\be \label{pop_pp}
p'(\mathcal{J}') \leftrightarrow \{ \mathcal{J}'_i \}_{i=1, \ldots, P}. 
\ee\\

Indeed, once $ \{ \mathcal{J}_i \}_{i}$ is known, it is easy to show that one can compute $ \{ \mathcal{J}'_i \}_{i}$ by means of the pseudocode illustrated in Routine \ref{pop_dyn_routine}. 

\begin{algorithm}

\begin{description}  
\item [{for}] $i=1,\ldots,P$
\begin{description}
\item [{for}] $\alpha=1,\ldots, 6$
\begin{description}
\item [{draw}] uniformly a random number $j$ in $\{1,\ldots, P\}$. 
\item [{set}] $\mathcal{J}^{temp}_\alpha = \mathcal{J}_r$. 
\end{description}
\item [{end}]~
\item [{draw}] uniformly a random sign $s= \pm 1$. 
\item [{set}] $\mathcal{J}'_i = s \frac{1}{\beta} \operatorname{arctanh}\left(\sqrt{O_{2}(\{ \beta  \mathcal{J}^{temp}_{\alpha} \}_ \alpha )} \right) $.
\end{description}
\item [{end}]~
\item [{return}] $\{ \mathcal{J}'_i \}_i$.\medskip{}
\end{description}
\caption{Population-dynamics routine}\label{pop_dyn_routine}
\end{algorithm}

The reader should notice that the population dynamics Routine is extremely simple and versatile to implement, and that it requires no evaluation of the integrals in the right-hand side of Eq. (\ref{159}). Once  $\{ \mathcal{J}'_i \}_i$ is known, this routine is iterated to compute $\{ \mathcal{J}''_i \}_i$ from $\{ \mathcal{J}'_i \}_i$, and so on. In particular, by iterating $k$ times Routine \ref{pop_dyn_routine} one can compute the population $\{ \mathcal{J}_{k\, i} \}_i$ representing the probability distribution $p_k(\mathcal{J}_k)$. Accordingly, the algorithm is named population dynamics after the fact that  $k$ is analogous to the dynamical-evolution time of the population $\{ \mathcal{J}_{k\, i} \}_i$. \\

The structure of the fixed points of Eq. (\ref{159}) can be now investigated numerically. Indeed, by iterating the population-dynamics Routine at fixed   $\beta$ and by computing $p_k(\mathcal{J})$ as a function of $k$, it is easy to show that there is a finite value of $\beta = \beta_c^{RS}$ such that for $\beta < \beta_c^{RS}$ $p_k(\mathcal{J})$ converges to  $\delta(\mathcal{J})$ as $k$ is increased, while  for $\beta > \beta_c^{RS}$ $p_k(\mathcal{J})$ broadens, i. e. its variance is an increasing function of $k$. 
The reader should observe that this flow to weak and strong coupling for  $p(\mathcal{ J})$ in the high and in the low-temperature phase respectively is the analog of the flow to weak and strong coupling for the number $J$ in the real-space approach for DHM, depicted in Fig. (\ref{fig9}). 
The physical interpretation of these two temperature regimes is that for $\beta < \beta_c^{RS}$ $p_k(\mathcal{J})$ flows to an attractive high-temperature fixed point with $\mathcal{J}=0$ where spins are decorrelated, while for $\beta > \beta_c^{RS}$ it flows to an attractive low-temperature fixed point with $\mathcal{J}=\infty$ where spins are strongly correlated. 
This fact implies that as the temperature is lowered below $T_c$ a phase transition occurs, and this transition yields a collective and strongly interacting behavior of spins in the low-temperature phase. Even though this result has been derived in the  $k_0=2$-approximation, the implementations of the real-space method for $k_0>2$ will confirm this picture.
The existence of a finite-temperature phase transition for a diluted version of the HEA model has been established heretofore with MC simulations by means of finite-size scaling techniques \cite{franz2009overlap}. Since the critical properties of such a diluted version of the HEA should  be the same \cite{franz2009overlap}  as those of the HEA defined here, the real-space approach confirms the picture on the criticality of the system given by MC simulations. \\

An important feature of the population-dynamics approach is that for $\sigma<1/2$ the thermodynamic limit is ill-defined, which has been discussed in Part \ref{hea}, Eq. (\ref{sigma_hea}). Indeed, the numerics show that  for  $\sigma\rightarrow 1/2$ $\beta_c^{RS} \rightarrow 0$, in such a way that the system is always in the low-temperature phase, i. e.  the variance of $p_k(\mathcal{J})$  is an increasing function of $k$, and the thermodynamic limit $k \rightarrow \infty$ is ill-defined. On the other hand, according to  Eq. (\ref{sigma_hea}) one should have that $\beta_c^{RS} \rightarrow \infty$ as $\sigma \rightarrow 1$, because for $\sigma>1$ no finite-temperature phase transition occurs. Unfortunately, this condition  is not reproduced by the real-space approach. As discussed in Section \ref{pert_hea}, this is presumably due to the fact that $k_0$ is small, i. e. that Eq. (\ref{159}) implements only the lowest-order approximation of the real-space method ($k_0=2$). This hypothesis is supported by the fact that the estimate of the critical exponents that we will give in what follows  significantly improve as $k_0$ is increased in the region where $\sigma$  differs significantly from $1/2$, while they hardly change in the region $\sigma \approx 1/2$, implying that the closer $\sigma $ to $1$, the larger the values of $k_0$  needed.   Accordingly, for $\sigma \rightarrow 1$ a significantly better description would be obtained if larger values of $k_0$ were accessible, and the $\sigma<1$-limit would be recovered.\\

The numerical implementation  of Routine \ref{pop_dyn_routine}   also shows that there is a repulsive critical fixed point, that we will call $p_\ast(\mathcal J)$, which is reached by iterating  Routine \ref{pop_dyn_routine} with $\beta = \beta_c^{RS}$
\bea \label{p_ast}  \no
p_\ast(\mathcal{J}) &= &\int \Bigg[ \prod_{\alpha=1}^6 p_\ast(\mathcal{J}_{\alpha})  d \mathcal{J}_{\alpha} \Bigg] \frac{1}{2}\Bigg[ \delta \left(  \mathcal{J} - \frac{1}{\beta_c^{RS}} \operatorname{arctanh}\left(\sqrt{O_{2}(\{ \beta_c^{RS}  \mathcal{J}_{\alpha} \}_\alpha)} \right)    \right) + \\
&& + \delta \left(  \mathcal{J} + \frac{1}{\beta_c^{RS}} \operatorname{arctanh} \left(\sqrt{O_{2}(\{ \beta_c^{RS} \mathcal{J}_{\alpha} \}_\alpha} )\right)    \right) \Bigg].
\eea

This critical fixed point is analogous to the critical value of the coupling $K_c$ in  the real-space method for DHM, depicted in Fig. (\ref{fig9}). 
In  the numerical implementation both $\beta_c^{RS}$ and $p_\ast(\mathcal J)$ are computed by iterating Routine \ref{pop_dyn_routine} and by dynamically adjusting $\beta$  at each step to its critical value, which is approximately determined as the value of $\beta$ such that $\int d \mathcal{J'} p'(\mathcal {J}') (\mathcal{J}')^2 = \int d \mathcal{J} p(\mathcal {J}) \mathcal{J}^2$. In order to do so, one starts with two values of the temperature $T_{\text{min}}, T_{\text{MAX}}$ such that  
\be \label{bracket}
T_{\text{min}} < T_c^{RS}<  T_{\text{MAX}},
\ee
and then iterates the bisection Routine \ref{bisection_routine}.

\begin{algorithm}
\begin{description}
\item[{for}]  $k=1, \ldots, k_{\text{MAX}}$
\begin{description}
\item[{set}]  $T=(T_{\text{min}}+ T_{\text{MAX}})/2$.
\item[{set}]  $\varsigma^2 =  \frac{1}{P} \sum_{i=1}^P \mathcal{J}_i^2$.
\item[{compute}] $\{ \mathcal{J}'_{ i} \}_i$ from Routine \ref{pop_dyn_routine} and set $ \mathcal{J}_{ i} = \mathcal{J}'_{ i} \, \forall i=1, \ldots, P$. 
\item[{set}]  ${\varsigma'}^2 =  \frac{1}{P} \sum_{i=1}^P {\mathcal{J}'}_i^2$.
\item[{if}] ${\varsigma'}^2 > \varsigma^2$
\textbf{set} $T_{\text{min}} = T_{\text{min}} + x (T_{\text{MAX}}-T_{\text{min}})$. 
\item[{else}] ~
\textbf{set} $T_{\text{MAX}} = T_{\text{MAX}} - x  (T_{\text{MAX}}-T_{\text{min}})$. 
\end{description}
\item[{end}]
\item[{return}] $T_c^{RS}=T$ and $\{ \mathcal{J}_i \}_i \leftrightarrow p_\ast(\mathcal{J})$
\end{description}
\caption{Bisection routine}
\label{bisection_routine}
\end{algorithm}

Each iteration $k$ of Routine \ref{bisection_routine} is one step of the RG transformation  performed at temperature $T=(T_{\text{min}}+ T_{\text{MAX}})/2$, and if the second moment of $p'(\mathcal{J}')$ is smaller  than that of  $p(\mathcal{J})$,  $T$ is in the high-temperature phase, and the interval $[T_{\text{min}}, T_{\text{MAX}}]$ is reduced by  $0<x<1$ by lowering  the upper limit  $T_{\text{MAX}}$, and vice versa if $T$ is in the low-temperature phase. By iterating this procedure $k_{\text{MAX}} \gg 1$ times, the Routine returns an estimate of the critical temperature $T_c^{RS}=T$ and of the critical fixed point $\{ \mathcal{J}_i \}_i \leftrightarrow p_\ast(\mathcal{J})$. \\
Since population dynamics is a stochastic approach, this bisection Routine is not deterministic, and might give slightly different results when one runs it several times if the population size $P$ is not large enough. In particular, such a stochastic character can introduce some instabilities when approaching the repulsive fixed point, and these might let the bisection Routine flow towards the low or high-temperature fixed point, far away from $p_\ast(\mathcal{J})$.  Accordingly, the parameter $x$ has been chosen by hand in order to minimize this instability, and the numerical implementation of the bisection Routine has shown that $x \sim 0.1$ yields a good estimate of the critical fixed point.  

\begin{figure}
\begin{centering}
\includegraphics[width=10cm]{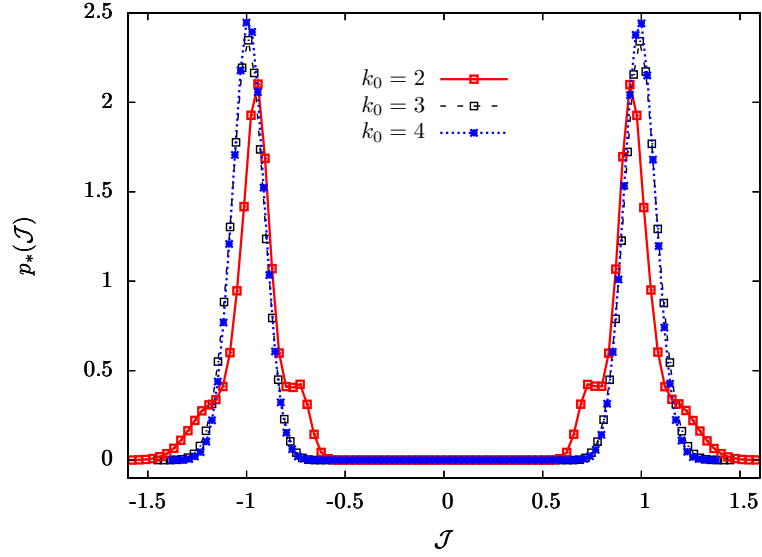}
\caption{Fixed point $p_\ast(\mathcal{J})$ as a function of $\mathcal{J}$ for $\sigma=0.5146$ and $k_0=2,3,4$, obtained by iterating Routine \ref{bisection_routine} starting with $p_{0}(\mathcal{J})= 1/\sqrt{2 \pi} \textrm{e}^{-\mathcal{J}^2/2}$.
For $k_0=2$, $k_\text{MAX}=50, P=2 \times 10^  6, x=0.1$ and  $B=96$. 
For $k_0=3$, $k_{\text{MAX}}=50, P= 10^  6, x=0.1$ and  $B=96$.
For $k_0=4$, $k_{\text{MAX}}=20, P=2\times 10^4, x=0.1$ and $B=96$. 
$p_{\ast}(\mathcal{J})$ has  compact support and a  convergent behavior as $k_0$ is increased. This convergence indicates that $k_0$ is large enough in such a way that the real-space approach is asymptotically  exact for this value of $\sigma$.  }
\label{fig13}
\end{centering}
\end{figure}

The critical fixed point $p_\ast(\mathcal{J})$ obtained with this bisection method is depicted in Fig. \ref{fig13} for a given $\sigma$-value,
where  we also represent  $p_\ast(\mathcal{J})$ obtained with the $k_0=3,4$-approximations  that will be described in what followins.\\

Once the  fixed point $p_\ast(\mathcal{J})$ has been computed numerically, the critical exponents can be determined by linearizing the transformation (\ref{159}) in the neighborhood of $p_\ast(\mathcal J)$.  Being the RG equations (\ref{159}) a flow for a continuous set of degrees of freedom $\{ p(\mathcal{J}) \}_{\mathcal{J}}$, the matrix linearizing the RG transformation in the neighborhood of $p_\ast(\mathcal{J})$ has continuous indices $\mathcal{J}, \mathcal{J}'$, and is defined as 
\bea \label{matrix_2}\no 
\mathscr{M}_{\mathcal{J}, \mathcal{J}'} ^{RS} &\equiv&   \left.  \frac{\delta p'(\mathcal{J})} {\delta p(\mathcal{J}')}  \right| _{p = p_\ast,\, \beta = \beta_c^{RS}}\\ \no 
& = & \sum_{\gamma=1}^{6} 
 \int \left[ \prod_{\lambda \neq \gamma =1}^ {6} p_\ast (\mathcal{J}_\lambda)  d\mathcal{J}_\lambda \right]  \times \\ \no 
&&\times \frac{1}{2}\Bigg[ \delta \left(  \mathcal{J} - \frac{1}{\beta_c^{RS}} \operatorname{arctanh}\left(\sqrt{O_{2} ( \{\beta_c^{RS} \mathcal{J}_\lambda\}_{\lambda}^{ \gamma, \mathcal{J}'  }  )} \right)    \right) + \\  
&& + \delta \left(  \mathcal{J} + \frac{1}{\beta_c^{RS}} \operatorname{arctanh}\left(\sqrt{O_{2}(  \{\beta_c^{RS} \mathcal{J}_\lambda\}_{\lambda}^{ \gamma, \mathcal{J} ' }  )} \right)    \right) \Bigg],
\eea
where $  \{\beta_c^{RS} \mathcal{J}_\lambda\}_{\lambda}^{ \gamma, \mathcal{J} ' } \equiv \{ \beta_c^{RS} \mathcal{J}_1, \ldots, \beta_c^{RS} \mathcal{J}_{\gamma-1}, \beta_c^{RS} \mathcal{J}', \beta_c^{RS} \mathcal{J}_{\gamma+1}, \ldots, \beta_c^{RS} \mathcal{J}_6 \}$ and in the second line of Eq. (\ref{matrix_2}) Eq. (\ref{159}) has been used. \\

\begin{centering}
  \begin{figure}[htb] 
\centering
\includegraphics[width=14cm]{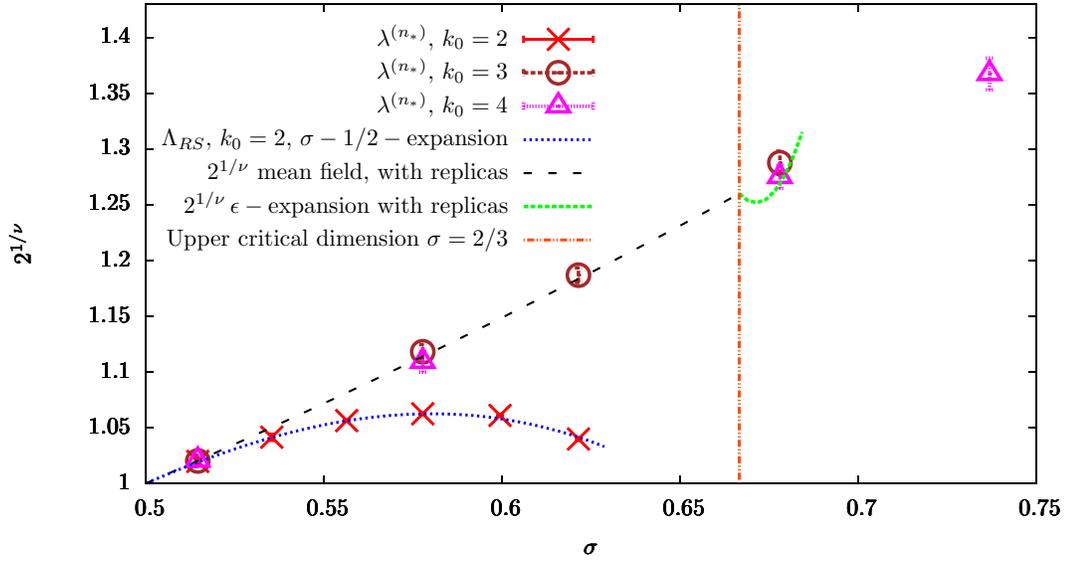}
\caption{
$2^{1/\nu}$ as a function of $\sigma$. The red points represent $\lambda^{(n_\ast)}$  computed with the population-dynamics implementation of the real-space RG equations with $k_0=2$ and $48 \leq B \leq 96, k_{\text{MAX}}=50, 10^6 \leq P \leq 2 \times 10^7, x=0.1$ and initial distribution $p_0(\mathcal{J}) = 1/\sqrt{2 \pi} \text{e}^{-\mathcal{J}^2/2}$. 
The brown points represent $\lambda^{(n_\ast)}$  computed with the population-dynamics implementation of the real-space RG equations with $k_0=3$ and $48 \leq B \leq 100, k_{\text{MAX}}=50, 4 \times 10^4 \leq P \leq  10^6, x=0.1$ and initial distribution $p_0(\mathcal{J}) = 1/\sqrt{2 \pi} \text{e}^{-\mathcal{J}^2/2}$. 
The violet points represent $\lambda^{(n_\ast)}$  computed with the population-dynamics implementation of the real-space RG equations with $k_0=4$ and $48 \leq B \leq 96, k_{\text{MAX}}=20,  5 \times 10^3 \leq P \leq   2 \times  10^4, x=0.1$ and initial distribution $p_0(\mathcal{J}) = 1/\sqrt{2 \pi} \text{e}^{-\mathcal{J}^2/2}$. 
For any fixed $k_0$ one cannot compute $\nu$ for too large  $\sigma$, because the bisection routine used to determine the critical fixed point is unstable for large $\sigma$. 
The blue dashed curve represents $\Lambda_{RS}$ computed with the high-temperature expansion of the real-space RG equations to fourth order in  $\sigma-1/2$, see Eq. (\ref{lambda_rs_high_temp}). The black  dashed curve and the green dashed curve represent $2^{1/\nu}$ obtained with the replica approach presented in Chapter \ref{replica_approach}: the black dashed curve represents the mean-field value of $2^{1/\nu}$ for $\sigma \leq 2/3$ given by the first line in Eq. (\ref{nu}), while the green dashed curve represents the two-loop result (\ref{nu_2})  for $\sigma>2/3$.  The orange dashed curve represents the upper critical dimension  $\sigma=2/3$  resulting from the replica approach and discussed in Chapter \ref{replica_approach}. 
}
\label{fig15}
\end{figure}
  \end{centering}

In the rigorous treatment \cite{collet1977epsilon} of the $\epsilon_F$-expansion for DHM, the linearization of the  transformation $\mathfrak{p}_k(m) \rightarrow \mathfrak{p}_{k+1}(m)$ in the neighborhood of $\mathfrak{p}_\ast(m)$ is formulated in terms of a linear functional acting on a suitably defined space of functions  $\phi(m)$, and the critical exponents are extracted from the eigenvalues  of this functional. These eigenvalues are determined by the theory of linear functionals, in particular by the theory of Hermite polynomials.  Similarly, here the matrix $\mathscr{M}^{RS}$ defines a linear functional $\mathcal{L}$ acting on a suitably defined space of functions $\phi(\mathcal{J})$, and yielding a function $\mathcal{L}[\phi](\mathcal{J})$ 
\[
\mathcal{L}[\phi](\mathcal{J}) \equiv \int d \mathcal{J}' \mathscr{M}^{RS}_{\mathcal{J}, \mathcal{J}'} \phi(\mathcal{J}'). 
\]
Unfortunately, the complicated form (\ref{matrix_2}) of $\mathscr{M}^{RS}$ did not allow for an analytic treatment, and in particular the  spectrum of $\mathscr{M}^{RS}$, and so the critical exponents, could not be determined in terms of the spectrum of well-known linear functionals. Hence, a completely numerical analysis of the spectrum of $\mathscr{M}^{RS}$ has been done. 
This analysis is illustrated in  Appendix \ref{app_discretization}. As a result, the critical exponent $\nu$ defined by Eq. (\ref{xi_hea}) is determined from the spectrum of $\mathscr{M}^{RS}$. 
In Fig. \ref{fig15} we depict $\lambda^{(n_\ast)}$ obtained with this  $k_0=2$-approximation,  
$\lambda^{(n_\ast)}$ obtained   with the $k_0=3,4$-approximations that will be discussed in what follows,
 and $\Lambda_{RS}$ obtained with the high-temperature expansion as a function of $\sigma$.  We also depict the prediction for $2^{1/\nu}$ of the replica approach discussed in Chapter \ref{replica_approach}, in both the mean-field region $\sigma\leq 2/3$ and the non-mean-field region $\sigma>2/3$.  Fig. \ref{fig15} shows that the prediction $\Lambda_{RS}$ for $2^{1/\nu}$  obtained with the high-temperature expansion  and the prediction $\lambda^{(n_\ast)}$ for $2^{1/\nu}$  obtained with  population dynamics are in excellent agreement,  confirming the validity of both methods. Even though, the agreement between $\Lambda_{RS}, \lambda^{(n_\ast)}$ and $2^{1/\nu}$ obtained with the replica approach is good only if $\sigma$ is sufficiently close to $1/2$. As we will see shortly, this discrepancy progressively disappears when implementing approximations with $k_0>2$. Since these have been developed along the lines of the $k_0=2$-approximation, in the following Section we will sketch  only  the main steps of the derivation of the real-space RG equations for $k_0>2$, and of the resulting computation of the critical exponents. 

\subsection{Improved approximations of the real-space method}\label{k_0_larger_2}

In the real-space approach developed in Section \ref{k_0_equal_2}   a $2^{k_0}$-spin HEA   is approximated by a $2^{k_0-1}$-spin HEA, with $k_0=2$. This approximation can be implemented for larger $k_0$ and, as discussed above, it becomes asymptotically exact for $k_0 \rightarrow \infty$.  In order to generalize the real-space approach to $k_0>2$, let us consider a $2^{k_0}$-spin HEA with spins $S_1, \ldots, S_{2^{k_0}}$ and Hamiltonian $H_{k_0}[\vec S]$, where $H_{k_0}$ is obtained by iterating the recursion equations (\ref{1}), (\ref{redefinition}). Let us set $C^2/2 J_{ij}  \equiv \mathcal{J}_{ij}$, where  $J_{ij}$ are the couplings defined in Eq. (\ref{1}), (\ref{redefinition}).

Let us then consider a $2^{k_0-1}$-spin HEA with spins $S'_1, \ldots, S'_{2^{k_0}-1}$ and Hamiltonian $H_{k_0-1}[\vec{S}']$, where $H_{k_0-1}$ is obtained by iterating the recursion equations (\ref{1}), (\ref{redefinition}). Let us set $C^2/2 J'_{ij}  \equiv \mathcal{J}'_{ij}$, where $J'_{ij}$ are the couplings defined in Eq. (\ref{1}), (\ref{redefinition}).
Let us also call the $M\equiv 2^{k_0}(2^{k_0}-1)/2$ couplings $\mathcal{J}_{12}, \mathcal{J}_{13}, \ldots, \mathcal{J}_{2^{k_0}-1 \; 2^{k_0}}$ \hspace{2mm} $\mathcal{J}_1, \cdots, \mathcal{J}_{M}$ respectively, and let us call the $M'\equiv 2^{k_0-1}(2^{k_0-1}-1)/2$ couplings $\mathcal{J}'_{12}, \mathcal{J}'_{13}, \ldots, \mathcal{J}'_{2^{k_0-1}-1 \; 2^{k_0-1}}$ \hspace{2mm} $\mathcal{J}'_1, \cdots, \mathcal{J}'_{M'}$ respectively. \\

\begin{centering}
  \begin{figure} 
\centering
\includegraphics[width=12cm]{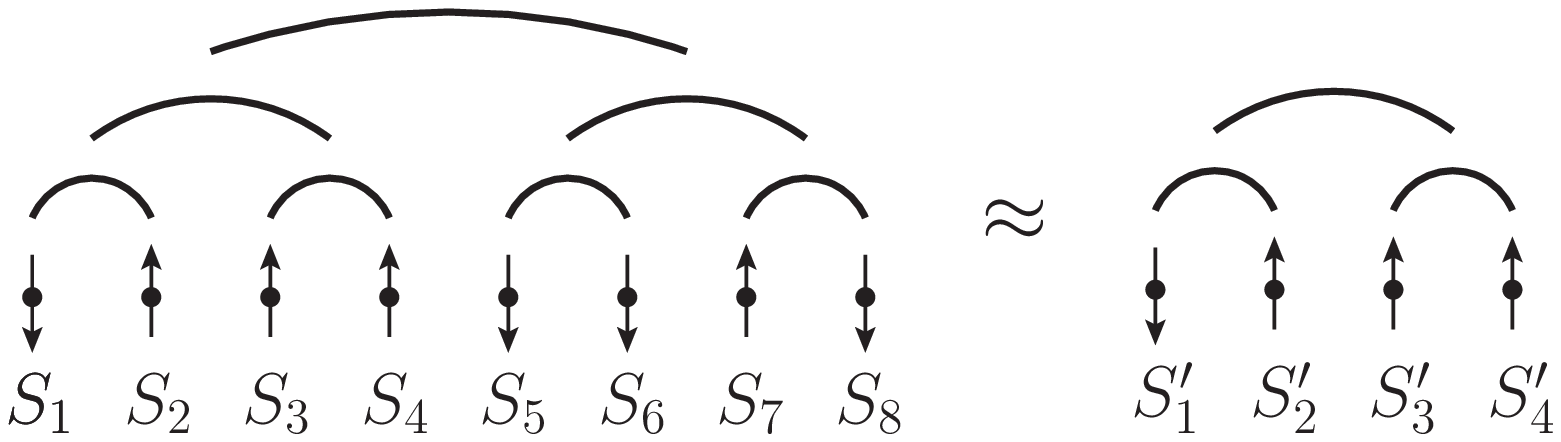}
\caption{Approximation of the real-space approach for $k_0=3$. A $2^3$-spin HEA with spins $S_1,\cdots, S_8$ and Hamiltonian $H_{8}[\vec S]$ is approximated by a $2^2$-spin HEA with spins $S'_1,\ldots, S'_4$ and Hamiltonian $H_2[\vec {S}']$. 
}
\label{fig17}
\end{figure}
  \end{centering}

According to the analysis of Section \ref{k_0_equal_2}, for each sample of the couplings $\{ \mathcal{J}_{\alpha} \}_{\alpha}$ we choose $\{ \mathcal{J}'_{\alpha}\}_\alpha$ as a function of $\{ \mathcal{J}_{\alpha} \}_{\alpha}$  in such a way that the $2^{k_0-1}$-spin HEA yields the best-possible approximation of the $2^{k_0}$-spin HEA. 
This procedure is qualitatively depicted in Fig. \ref{fig17} for $k_0=3$.
 \\

Since in this case the number of couplings of the $2^{k_0-1}$-spin HEA is $M'$, we need a set of $M'$ equations to determine the optimal couplings $\{ \mathcal{J}' _\alpha \}_\alpha$. 
This set of equations ensuring that the $2^{k_0-1}$-spin HEA approximates the $2^{k_0}$-spin HEA can be obtained  by   considering $M'$ physical observables $\{ O_{k_0}^{\alpha}(\{\beta  \mathcal{J}_{\gamma} \}_{\gamma} ) \} _{\alpha=1, \ldots, M'}$ of the $2^ {k_0}$-spin HEA, depending on the $M$  couplings $\{ \mathcal{J}_{\alpha} \}_{\alpha}$ and  on $\beta$ through the dimensionless quantities $\{ \beta \mathcal{J}_{\alpha} \}_{\alpha }$. Consider also $M'$ observables $\{ O_{k_0-1}^\alpha (\{  \beta \mathcal{J}'_{\gamma} \}_\gamma ) \}_{\alpha=1, \ldots, M'}$ of the $2^{k_0-1}$-spin HEA, depending on the  couplings $ \{ \mathcal{J}'_{\alpha} \}_\alpha$ and  on $\beta$ through the dimensionless quantity $\{  \beta \mathcal{J}'_{\alpha} \}_\alpha$. \\

   In order to build up $O_{k_0}^\alpha$ and $O_{k_0-1}^\alpha$, consider two real replicas $\vec{S}^1, \vec{S}^2$ of the spins of the $2^{k_0}$-spin HEA, and  two real replicas $\vec{S'}^1, \vec{S'}^2$ of the spins of the $2^{k_0-1}$-spin HEA. Now consider the $2^{k_0-1}$ pairs of contiguous spins $(S_1,S_2),(S_3,S_4),\ldots, (S_{2^{k_0}-1},S_{2^{k_0}})$  of the $2^{k_0}$-spin HEA, and the 
normalized overlaps between $\vec{S}^1$ and $\vec{S}^2$ on each of the leaves defined by these pairs 
   \bea \no 
Q_\alpha & = &    \frac { \frac{S^ 1_{2 \alpha -1} S^ 2_{2 \alpha -1} +S^ 1_{2 \alpha } S^ 2_{2 \alpha }}{2} }{\sqrt{\mathbb{E}_{\vec{S}^1,\vec{S}^2}  \left[ \left( \frac{S^ 1_{2 \alpha -1} S^ 2_{2 \alpha -1} +S^ 1_{2 \alpha } S^ 2_{2 \alpha }}{2} \right) ^2  \right] }},
\eea
where $\alpha = 1, \ldots, 2^{k_0-1}$, and $\mathbb{E}_{\vec{S}^1,\vec{S}^2}$ denotes the expectation value with respect to $\vec{S}^1, \vec{S}^2$ with Boltzmann weight $\textrm{e}^{-\beta H_{k_0}}$. 
Now consider the normalized overlaps between $\vec{S}'^1$ and $\vec{S}'^2$ on each of the  spins of the $2^{k_0-1}$-spin HEA
 \bea \no 
Q'_\alpha & = &S'^ 1_\alpha S'^ 2_\alpha, 
  \eea
 where $\alpha=1, \ldots, 2^{k_0-1}$. 

   Kadanoff's decimation rule is implemented by imposing the correspondence between $Q_\alpha$ and $Q'_\alpha$ $\forall \alpha=1, \ldots, 2^{k_0-1}$. This can be done by choosing the following observables 
\bea \no
O_{k_0}^{1}(\{\beta  \mathcal{J}_{\gamma} \}_{\gamma} ) &=& \mathbb{E}_{\vec{S}^1, \vec{S}^2}[Q_1 Q_2], \\ \no 
O_{k_0}^{2}(\{\beta  \mathcal{J}_{\gamma} \}_{\gamma} ) &=& \mathbb{E}_{\vec{S}^1, \vec{S}^2}[Q_1 Q_3], \\ \no 
&\ldots& \\ 
O_{k_0}^{M'}(\{\beta  \mathcal{J}_{\gamma} \}_{\gamma} ) &=& \mathbb{E}_{\vec{S}^1, \vec{S}^2}[Q_{2^{k_0-1}-1} Q_{2^{k_0-1}}],
\eea
\bea \no
O_{k_0-1}^{1}(\{\beta  \mathcal{J}'_{\gamma} \}_{\gamma} ) &=& \mathbb{E}_{\vec{S}'^1, \vec{S}'^2}[Q'_1 Q'_2], \\ \no 
O_{k_0-1}^{2}(\{\beta  \mathcal{J}'_{\gamma} \}_{\gamma} ) &=& \mathbb{E}_{\vec{S}'^1, \vec{S}'^2}[Q'_1 Q'_3], \\ \no 
&\ldots& \\ 
O_{k_0-1}^{M'}(\{\beta  \mathcal{J}'_{\gamma} \}_{\gamma} ) &=& \mathbb{E}_{\vec{S}'^1, \vec{S}'^2}[Q'_{2^{k_0-1}-1} Q'_{2^{k_0-1}}],
\eea
where $\mathbb{E}_{\vec{S}'^1, \vec{S}'^2}$ denotes the expectation value with respect to $\vec{S}'^1, \vec{S}'^2$ with Boltzmann weight $\textrm{e}^{-\beta H_{k_0-1}}$.
The $2^{k_0}$-spin HEA can now be approximated by the $2^{k_0-1}$-spin HEA by enforcing the constraints 
\be \label{170}
O_{k_0}^{\alpha}(\{\beta  \mathcal{J}_{\gamma} \}_{\gamma} )  = O_{k_0-1}^{\alpha}(\{\beta  \mathcal{J}'_{\gamma} \}_{\gamma} ) \; \forall \alpha=1, \ldots, M'.
\ee
In what follows we assume that Eq. (\ref{170}) can be written as 
\be \label{172}
\mathcal{J}'_\alpha = 1/\beta f_\alpha(\{\beta  \mathcal{J}_{\gamma} \}_{\gamma}),  
\ee
where in Eq. (\ref{172}) we have used the fact that the right-hand side of Eq. (\ref{170}) depends on $\{ \mathcal{J}'_\alpha\}_\alpha$ through the dimensionless products $\{ \beta \mathcal{J}'_\alpha\}_\alpha$. 
\\

Now suppose that the couplings  $\{ \mathcal{J}_\alpha\}$ are independent and that each of them is distributed according to a given $p(\mathcal{J})$.
 According to Eq. (\ref{172}), the couplings $\{ \mathcal{J}'_\alpha\}_\alpha$ are not independent, because the entangled structure of Eq. (\ref{172}) introduces some correlation between them. Accordingly, the joint probability distribution of $\{ \mathcal{J}'_\alpha\}_\alpha$ is 
\be \label{173}
p'_{C}(\{ \mathcal{J}'_\alpha\}_\alpha) = \int \left[ \prod_{\alpha=1}^{M'} p(\mathcal{J}_\alpha) d \mathcal{J}_\alpha  \right] \prod_{\alpha=1}^{M'} \delta \left( \mathcal{J}'_\alpha - \frac{1}{\beta} f_\alpha(\{\beta  \mathcal{J}_{\gamma} \}_{\gamma}) \right) ,
\ee
where the label $C$ stands for correlated.
 Hence, starting with a set of uncorrelated couplings $\{ \mathcal{J}_\alpha \}_\alpha$, after one step of the RG transformation we generate some correlation between the couplings. 
This situation occurs in many cases where one performs an effective reduction of the degrees of freedom of a system under reparametrization of the length scale. Indeed, it is a quite general fact that if one starts with a set of degrees of freedom at a given length scale $L$, additional degrees of freedom are generated at the length scale $2 L$. A typical example is the RG flow for the Ising model in dimensions $d\approx4$, where in the effective $\phi^4$-theory one generates additional $\phi^6, \phi^ 8, \cdots$-terms  after one RG step (see \cite{wilson1971renormalization1,wilson1971renormalization2,wilson1974renormalization,wilson1983renormalization,wilson1974critical} for  a general discussion in the approach \`a la Wilson and \cite{zinnjustin} for  a discussion in the field-theory approach for the Ising model). Notwithstanding this, it turns out that in the field-theory RG approach \cite{zinnjustin} these terms are finite, i. e. non singular in $1/(4-d)$,  so they do not need to be absorbed into the renormalization constants, and do not contribute to the critical exponents. This fact is intrinsically related to the perturbative renormalizability  of the $\phi^ 4$-theory \cite{zinnjustin}, and implies that the RG equations can be written in  closed form.
\\

It is not easy to tell if the above correlation can be consistently neglected in this real-space approach, and the answer to this question could  be intrinsically related to the renormalizability of the theory.  In our approach we will not address this delicate point, and we will simply get rid of the above correlation between the $ \mathcal{J}'_\alpha $s by assuming that they are independent, and that each of them is distributed according to a distribution $p'(\mathcal{J}')$
 given by the average of $M'$ marginalized distributions, each of which is obtained by integrating $p'_{C}(\{ \mathcal{J}'_\alpha\}_\alpha)$ over $M'- 1 $ couplings
\be \label{174}
p'(\mathcal{J}')   =  \frac{1}{M'} \sum_{\alpha=1}^ {M'} \int \left[ \prod_{\gamma=1,\gamma\neq \alpha}^ {M'} d\mathcal{J}'_\gamma \right]  p'_{C}( \mathcal{J}'_1, \ldots, \mathcal{J}'_{\alpha-1}, \mathcal{J}', \mathcal{J}'_{\alpha+1}, \ldots, \mathcal{J}'_{M'} ). 
\ee

By plugging Eq. (\ref{173}) into Eq. (\ref{174}) we obtain the RG equation relating $p(\mathcal{J})$ to $p(\mathcal{J}')$
\bea \label{175}
p'(\mathcal{J}')  & = & \frac{1}{M'} \sum_{\alpha=1}^ {M'}  \int \left[ \prod_{\gamma=1}^{M'}p(\mathcal{J}_\gamma)  d \mathcal{J}_\gamma \right]  \delta  \left(\mathcal{J}' - \frac{1}{\beta	} f_\alpha(\{\beta  \mathcal{J}_{\gamma} \}_\gamma ) \right).
\eea

A procedure along the lines of that presented in Section \ref{k_0_equal_2} can be applied to Eq. (\ref{175}): once a $2^{k_0}$-spin HEA has been approximated by a $2^{k_0-1}$-spin HEA, one takes two realizations of the latter and couples them to obtain a system representing a $2^{k_0+1}$-spin HEA, and iterates this procedure. In this way one obtains a sequence $p(\mathcal{J}) \equiv p_0(\mathcal{J}), p'(\mathcal{J}') \equiv p_1(\mathcal{J}'), p''(\mathcal{J}'') \equiv p_2(\mathcal{J}'') , \ldots$ physically representing the RG flow of the disorder distribution.  \bigskip\\

In the numerical implementation $p(\mathcal{J})$ and $p'(\mathcal{J}')$ are represented by two populations, and the RG equations (\ref{175}) are implemented in population dynamics by generalizing Routine \ref{pop_dyn_routine}. 
The numerical implementation of Eq. (\ref{175}) shows that the main qualitative features of the $k_0=2$-case stay the same here. In particular, there is a finite value of $\beta$, that we will call $\beta_c^{RS}$, such that for $\beta< \beta_c^{RS}$ $p_k(\mathcal{J})$ converges to   the high-temperature fixed point $\delta(\mathcal{J})$ as $k$ is increased, while   for $\beta> \beta_c^{RS}$ the width of $p_k(\mathcal{J})$ is an increasing function of $k$, and $p_k(\mathcal{J})$ converges to the low-temperature fixed point. If one iterates Eq. (\ref{175}) with $\beta = \beta_c^{RS}$, the iteration converges to a finite critical-fixed-point $p_\ast(\mathcal{J})$. \\

The critical fixed point $p_\ast(\mathcal J)$ is obtained by implementing Routine \ref{bisection_routine}, and is depicted in Fig. \ref{fig13} as a function of $\mathcal{J}$ for a given $\sigma$-value and $k_0=3,4$.

The matrix $\mathscr{M}^{RS}_{\mathcal{J},\mathcal{J}'}$
 linearizing the RG transformation in the neighborhood of $p_\ast(\mathcal{J})$  is defined in the same way as in Eq. (\ref{matrix_2}), and its eigenvalue $\lambda^{(n_\ast)}$  defined by Eq. (\ref{lambda_n_ast}) 
yields the critical exponent $\nu$ defined by Eq. (\ref{xi_hea})
according to  Eq. (\ref{nu_lambda_ast}). 
\\

$\lambda^{(n_\ast)}$  for $k_0=3,4$ is depicted in Fig \ref{fig15} as a function of $\sigma$. 
\mnote{The real-space RG method makes precise predictions for  the critical exponents of the Hierarchical Edwards-Anderson model, which are in agreement with those of the replica method in the classical region. In the non-classical region these predictions cannot be compared to those of the replica method. }
 Fig. \ref{fig15} shows that even though for $k_0=2$ $\lambda^{(n_\ast)}$  is significantly different from the mean-field value obtained with the replica approach, as $k_0$ is increased both $\lambda^{(n_\ast)}$ for $k_0=3$ and $\lambda^{(n_\ast)}$ for $k_0=4$ agree very well with the mean-field value obtained with the replica approach in the whole mean-field region $1/2<\sigma \leq 2/3$.  This is an important a posteriori test of the whole real-space RG framework presented here. The situation  subtler in the non-mean-field region $\sigma > 2/3$. In this region the $\epsilon$-expansion based on the replica approach is non-predictive, because the first few orders (\ref{nu_2}) of the series have a nonconvergent behavior. Still, the data of the real-space approach can be compared to that of MC simulations \cite{franz2009overlap} performed on a diluted version of the HEA,  where $\nu$ is an increasing function of $\sigma$ in the non-mean-field region $\sigma > 2/3$, which disagrees with the results of the real-space approach, Fig. \ref{fig15}. \\

There might be several reasons for this disagreement \cite{castellana2011real}. A first issue might be the smallness of $k_0$ in the real-space approach: it is plausible that for larger  $k_0$ the derivative of $\lambda^{(n_\ast)}$ at $\sigma ={ 2/3}_+$ turns out to be negative, in agreement with the MC results. Another issue might be that the exponent $\nu$ is not universal: the exponent $\nu$ of the HEA model studied here  might be different from that of the diluted version of the HEA studied in MC simulations.  Indeed, universality in non-mean-field spin glasses has never been established rigorously \cite{huse2011private}. On the one hand, universality violation  in finite-dimensional systems \cite{bernardi1995violation,bernardi1996ordering} resulted from numerical studies done heretofore, even though more recent analyses based on MC simulations \cite{katzgraber2006universality}  and high-temperature expansion \cite{daboul2004test} suggest that universality holds. A third important issue might be that the couplings  correlation has been neglected in the real-space approach for $k_0>2$, see Eq. (\ref{174}). Indeed,  this couplings  correlation might play a vital role in the non-mean-field region $2/3 <\sigma < 1$, and it might yield a radically different critical behavior of the system if one took it into account. \vspace{2cm}\\

These issues could be investigated in some future directions of this real-space method.
For instance, it would be interesting to find a  way to handle couplings correlation in the real-space approach. One could then investigate the relevance of this correlation in both the mean and the non-mean-field region,  and compare the resulting values of the critical exponents to those obtained with MC simulations.
Another interesting future direction would be to implement this approach in the presence of an external magnetic field. Indeed, by analyzing the existence of a critical fixed point one could establish whether there is a phase transition in the non-mean-field region, which has been a hotly debated issue in the last years \cite{katzgraber2005probing,leuzzi2009ising}, and could give some insight into the correct picture describing the low-temperature phase of the system \cite{fisher1986ordered}.  Finally, it would be interesting to implement the present approach for more realistic spin-glass systems, like the three-dimensional EA model. Indeed, a simple analysis shows that this method can be easily generalized to models with short-range interactions built  on a finite-dimensional hypercube.

\part{Conclusions}\label{conc}

This thesis has investigated the implementation of renormalization-group (RG) techniques in finite-dimensional glassy systems, in order to shed light on the critical behavior of spin and structural glasses beyond mean field. In finite dimensions  the existence of a phase transition in structural glasses is not well-established, and the  structure of the low-temperature phase  for both spin and structural glasses is fundamentally unknown and controversial. Since Wilson's RG equations emerge in a natural and simple way in ferromagnetic spin models with a hierarchical interaction structure, in this work we considered two finite-dimensional models of spin and structural glasses built on Dyson's hierarchical lattice \cite{dyson1969existence}. \vspace{1cm}\\

After giving a brief introduction on spin and structural glasses in Part \ref{introduction}, in Part \ref{hrem} we focused on a structural-glass model built on a hierarchical lattice, the Hierarchical Random Energy Model (HREM) \cite{castellana2010hierarchical,castellana3}. In this study, we showed the first evidence of a non-mean-field model of a supercooled liquid undergoing a Kauzmann phase transition.
On the one hand, the  features of the phase transition are different from the mean-field case \cite{derrida1980random}. The free energy is found to be nonanalytical at the critical point, and our study of the correlation length of the system in the critical region shows that the data   is consistent with the existence of a diverging correlation length. 
On the other hand, by  investigating the properties of the low-temperature phase we  showed that the free energy has a one-step replica-symmetry-breaking (RSB) saddle point in the low-temperature phase, describing a fragmentation of the free-energy landscape into disconnected components.\\

A first future direction of this work would be to generalize it to more realistic models, like $1$-RSB  models with $p$-spin  interactions and $p \geq 3$, which have an entropy-crisis transition in the mean-field case: it would be interesting to build up a non-mean-field version of these models on a hierarchical lattice, and to implement a suitable generalization of the RG techniques used for the HREM to study their critical behavior. 
A second future direction would be to study the dynamics of the HREM and of these hierarchical $1$-RSB models. In particular, $1$-RSB models have a dynamical phase transition in the mean-field case mimicking the dynamical arrest in glass-formers at the glass-transition temperature predicted by the Mode Coupling Theory. If one could check whether this transition persists or is smeared out in such  hierarchical counterparts of  mean-field $1$-RSB models, one could test directly some of the building blocks of the Random First Order Transition Theory. \vspace{1cm}\\

In Part \ref{hea} we presented the study of a spin-glass model built on a hierarchical lattice, the Hierarchical Edwards-Anderson Model (HEA). Differently from the HREM, in the HEA the spins are not mere labels for the energy variables, but they are physical degrees of freedom. This fact allowed for an implementation of a RG decimation protocol, which has been implemented with two different approaches. The first approach is based on the replica method \cite{castellana2010renormalization,castellana2011renormalization}, while the second one does not rely on the replica formalism, but on a real-space picture \cite{castellana2011real}. \\

In the replica RG approach the infrared (IR) limit of the theory has been taken with two different methods, and both of them yield the same two-loop prediction  for the $\epsilon$-expansion of the critical exponents. This shows that the IR limit of the theory is well defined, and suggests the existence of a diverging correlation length in the system. Unfortunately, this approach makes predictions for the critical exponents only in the classical region, i. e. in the parameter region where the mean-field approximation is exact. In the non-classical region the two-loop $\epsilon$-expansion has a  nonconvergent behavior, in such a way that no conclusion can be drawn on the actual values of the critical exponents. 

In the real-space RG approach a generalization of Kadanoff's block-spin decimation is implemented in spin glasses, and the resulting RG equations are worked out by means of a series of approximation steps. These equations have been solved by means of the high-temperature expansion and of the population-dynamics method, yielding consistent results.    Similarly to the replica RG method, the real-space approach shows that a phase transition occurs in the HEA and that the correlation length diverges at the critical point. At variance with the replica RG method, this method makes precise predictions for the critical exponents  in both the classical region, where the critical exponents are in excellent agreement with those of the replica method, and in the non-classical region. The real-space predictions in this region are in disagreement with Monte Carlo (MC) simulations done for a diluted version of the HEA. There might be several possible reasons for this disagreement: the discrepancy  should disappear if better approximation steps were considered in the real-space approach, or  the critical exponents of the HEA model defined here might be different from those of the diluted HEA, i. e. universality might be violated. \\

There are several future directions for the replica and for the real-space approach. As far as the replica approach is concerned, the two-loop calculation that we have done here is  a base of departure for an automated computation of the $\epsilon$-expansion for the critical exponents. Despite its highly technical nature, the  calculation of  high orders of the $\epsilon$-series  would clarify whether the $\epsilon$-expansion can be resummed and made convergent, and so whether the non-mean-field physics of the system can be considered as a small perturbation of the mean-field one.  As far as the real-space approach is concerned, it would be interesting to improve the approximation scheme  to check whether the disagreement with MC simulations disappears. If the disagreement does not disappear, one could test  the universality of the exponents by  checking directly whether the critical indices  stay the same when changing the details of the quenched-disorder probability distribution. Moreover, in order to shed light on the structure of the low-temperature phase, it would be interesting to implement the real-space method in the presence of an external magnetic field and to verify the existence of a phase transition in the non-classical region, by searching for a critical fixed point of the RG equations. Indeed, the existence of a phase transition in a field is one of the fundamental elements discriminating between the RSB and the droplet picture for finite-dimensional spin glasses, and it would shed light on the features of the low-temperature phase of these systems.


\appendix

\chapter{Properties of Dyson's Hierarchical Model}\label{app_field_dhm}

\section{Derivation of Eq. (\ref{rec_dhm})} \label{app_field_dhm_1}

Let us derive Eq. (\ref{rec_dhm}) first. We start from the definition (\ref{def_pcont_dhm}), omit  any $m$-independent multiplicative constant to simplify the notation, and we have
\bea \label{138} \no 
p_{k+1}(m) & = & \sum_{\vec{S}} \textrm{e}^{-\beta(H_{k}^F[\vec{S}_1] +  H_{k-1}^F[\vec{S}_2]) + \beta J C_F^{k+1} \left( \frac{1}{2^{k+1}} \sum_{i=1}^{2^k} S_i \right)^2} \times \\ \no 
&& \times \delta\left( \frac{1}{2^{k+1}} \sum_{i=1}^{2^{k+1}} S_i - m \right) 
\int dm_1 dm_2 \delta\left( \frac{1}{2^{k}} \sum_{i=1}^{2^{k}} S_i - m_1 \right) \times \\ \no 
&& \times \delta\left( \frac{1}{2^{k}} \sum_{i=2^k+1}^{2^{k+1}} S_i - m_2 \right) \\ \no 
& =&   \textrm{e}^{\beta J C_F^{k+1} m ^2} \int dm_1 dm_2  \delta\left( \frac{m_1+m_2}{2} - m \right)  \times \\ \no 
&& \times \left[ \sum_{\vec{S}_1}  \textrm{e}^{-\beta H_{k}^F[\vec{S}_1] }   \delta\left( \frac{1}{2^{k}} \sum_{i=1}^{2^{k}} S_i - m_1 \right) \right] \times \\ \no 
&& \times \left[    \sum_{\vec{S}_2}  \textrm{e}^{-\beta H_{k}^F[\vec{S}_2] } \delta\left( \frac{1}{2^{k}} \sum_{i=2^k+1}^{2^{k+1}} S_i - m_2 \right) \right] \\ 
& = &   \textrm{e}^{\beta J C_F^{k+1} m ^2} \int dm_1 dm_2  \delta\left( \frac{m_1+m_2}{2} - m \right)   p_{k}(m_1) p_k(m_2),
\eea 
where we set $\vec{S}_1 \equiv \{ S_1, \ldots, 2_{2^k} \}, \vec{S}_2 \equiv \{ S_{2^k+1}, \ldots, 2_{2^{k+1}} \}$,  in the first line we used Eq. (\ref{20}) and multiplied by  a factor equal to one, and in the third line we used the fact  that the quantities in square brackets in the second line are equal to $p_{k}(m_1), p_{k}(m_2)$ because of the definition (\ref{def_pcont_dhm}). By changing the  variables of integration, Eq. (\ref{138}) leads to Eq. (\ref{rec_dhm}).

\section{Structure of the fixed points of Eq. (\ref{rec2_dhm})} \label{app_field_dhm_2}

Taking $H_0^F[S]=0$, the Hamiltonian $H_k^F[\vec S ]$ has $\pm \vec S$ symmetry, and thus according to Eqs. (\ref{def_pcont_dhm}), (\ref{redef_dhm}), $\mathfrak{p}_k(m)$ is an even function of $m$. Hence, the simplest approximation is to assume that it is a Gaussian 
\be \label{p_dhm_gauss}
\mathfrak{p}_k(m) = \textrm{e}^{-r_k^F m^2}, 
\ee
by neglecting higher powers of $m$ in the exponential. Non-Gaussian terms will be later added in the argument of the exponential of Eq. (\ref{p_dhm_gauss}). 
For the integral of $\mathfrak{p}_{k}(m)$ with respect to $m$ to be finite, we have $r_k^F>0$. 

By plugging Eq. (\ref{p_dhm_gauss}) into Eq. (\ref{rec2_dhm}), we obtain a recursion equation for $r_k^F$
\be \label{rec_r}
r^F_{k+1} = \frac{2 r^F_{k}}{C_F} - \beta J.
\ee
Eq. (\ref{rec_r}) can be solved explicitly, and yields
\be \label{140}
r_k^F = \left( \frac{2}{C_F} \right)^k \left( r_0^F  - \frac{\beta J }{\frac{2}{C_F}-1} \right) +  \frac{\beta J }{\frac{2}{C_F}-1}. 
\ee
It follows that if $\beta = \beta_{c\, F}$, with
\be \label{141}
 \beta_{c \, F} \equiv r_0^F/J(2/C_F-1),
\ee 
$r_k^F$ has a finite limit $r_\ast = r_0$ for $k \rightarrow \infty$. 
This fixed point  will be called the critical fixed point. The critical fixed point is unstable, because any small deviation of $\beta$ from $\beta_{c\, F}$ would let $r_k^F$ flow away from $r_{\ast}^F$ towards another fixed point, as one can see from Eq. (\ref{140}). If $\beta < \beta_{c \, F}$ or $\beta > \beta_{c \, F}$, this attractive fixed point is called the high or low-temperature fixed point respectively. According to Eq. (\ref{141}), the critical temperature depends on the value $r^F_0$ at the initial step of the iteration, i. e. at microscopic length scales $2^k \sim 1$. This fact is in agreement with the very general picture occurring in ferromagnetic systems  like the Ising model \cite{wilson1974renormalization}, where the critical temperature is not universal, because it depends on the microscopic properties of the lattice, like the nearest-neighbor couplings between spins. \\

Another important feature of the solution of Eq. (\ref{rec_r}) is the following. Take $\beta < \beta_{c\, F}$ and $k \gg 1$. According to Eqs. (\ref{redef_dhm}), (\ref{p_dhm_gauss}) and (\ref{140}), one has
\be  \label{153}
p_k(m) = \exp \left[ - 2^k \left( r_0^F  - \frac{\beta J }{\frac{2}{C_F}-1} \right) m^2\right]. 
\ee
Eq. (\ref{153}) shows that for large $k$ the variable $m = 1/2^k \sum_{i=1}^{2^k}S_i$ is distributed according to a Gaussian distribution with variance proportional to $1/2^k$. This is the result that one could have guessed with the central limit theorem, supposing the spins $S_i$ to be independent. Now take $\beta = \beta_{c\, F}$.  According to Eqs. (\ref{redef_dhm}), (\ref{p_dhm_gauss}) and to the fact that $r_k = r_\ast  \, \forall k$, for large $k$
one has
\be  \label{154}
p_k(m) =\textrm{e}  ^{  - r_\ast C_F^ k m^2 }. 
\ee
 Eq. (\ref{154}) shows that the distribution of $m$ is still Gaussian, but its  variance is proportional to $1/C_F^{k}$, while supposing the spins to be independent by means of the central limit theorem, one would predict the variance to be proportional to $1/2^{k}$. The physical interpretation of these facts resulting from Eqs. (\ref{153}), (\ref{154}) is the following. For $\beta<\beta_{c\,F}$  spins can be  considered as  independent, and so $m$ has the $k$-dependence predicted by the central limit theorem. On the contrary, at the critical point strong correlations are developed, resulting in a collective behavior of spins which cannot be considered as independent anymore, and yielding  a magnetization $m$ with a $k$-dependence different from that predicted with the independence  hypothesis. \bigskip \\

Let us now seek for a more accurate approximation of $\mathfrak{p}_k$, by adding a quartic term in the exponential in the right-hand side of Eq. (\ref{p_dhm_gauss})
\be \label{142}
\mathfrak{p}_k(m) = \textrm{e}^{-(r_k^F m^2 + w_k^F m^4)}, 
\ee
where for the integral of $\mathfrak{p}_{k}(m)$ with respect to $m$ to be finite, one has $w_k^F>0$.

In the following we will suppose that the non-Gaussian term $w_k^F$ is  small for every $k$, i. e. that an expansion in powers of $w_k^F$ is meaningful. Practically speaking, this is equivalent to supposing that the above  qualitative picture resulting from the Gaussian ansatz (\ref{p_dhm_gauss}) is  slightly modified from the introduction of non-Gaussian terms in Eq. (\ref{142}).
The correctness of this assumption will be tested a posteriori by checking if the resulting perturbative series for physical quantities is convergent, or if it can be made convergent with some suitable resummation technique \cite{zinnjustin}. Plugging Eq. (\ref{142}) into Eq. (\ref{rec2_dhm}) and developing in powers of $w_k^F$, one has
\bea \label{143}\no 
\mathfrak{p}_{k+1}(m)  &=&  \textrm{e}^{-\frac{2 r^F_{k}}{C_F} m^2 + \beta J m ^2} \Bigg\{ 1 - \frac{1}{8 (r_k^F)^2}\left[ 3 + \frac{8 m^2 r_k ^F (3 C_F+2 m^2 r_k^F)}{C_F^2}   \right]w_k^F + \\ \no 
&& + \frac{1}{128 C_F^4 (r_k^F)^4}  [ 105 C_F^4 + 720 C_F^3 r_k^F m^2 + 1824 C_F^2 (r_k^F)^2 m^4 + \\
&& +  768 C_F (r_k^F)^3 m^6 +  256 (r_k^F) ^4 m^8 ]( w_k^F)^2  + O((w_k^F)^3)   \Bigg\}. 
\eea

The right-hand side of Eq. (\ref{143}) does not have the same form as the ansatz (\ref{142}). Notwithstanding this,  one can rewrite the term in braces in the right-hand side of Eq. (\ref{143}) as an exponential up to order $(w_k^F)^2$, and obtain  
\bea \label{144}\no 
\mathfrak{p}_{k+1}(m)  &=&  \exp \Bigg\{ - \left[ \frac{2 r^F_{k}}{C_F} - \beta J  + \frac{3 }{C_F r_k^F} w_k ^F - \frac{9 }{2 C_F (r_k^F)^3} (w_k^F)^2 \right]  m^2 + \\ 
&&- \left[ \frac{2}{C_F^2} w_k^F - \frac{9}{C_F^2(r_k^F)^2}(w_k^F)^2 \right] m^4+  O((w_k^F)^3) 
  \Bigg\}. 
\eea

Comparing Eq. (\ref{144}) to Eq. (\ref{142}), one has
 \be \label{145}
\left\{
\begin{array}{lcl}
r_{k+1}^F  &= &  \frac{2 r^F_{k}}{C_F} - \beta J  + \frac{3 }{C_F r_k^F} w_k ^F - \frac{9 }{2 C_F (r_k^F)^3} (w_k^F)^2 + O((w_{k}^F)^3),\\ 
w_{k+1}^F &= & \frac{2}{C_F^2} w_k^F - \frac{9}{C_F^2(r_k^F)^2}(w_k^F)^2   +O((w_{k}^F)^ 3).
\end{array}
\right . 
\ee

\begin{centering}
  \begin{figure}[htb] 
\centering
\includegraphics[width=10cm]{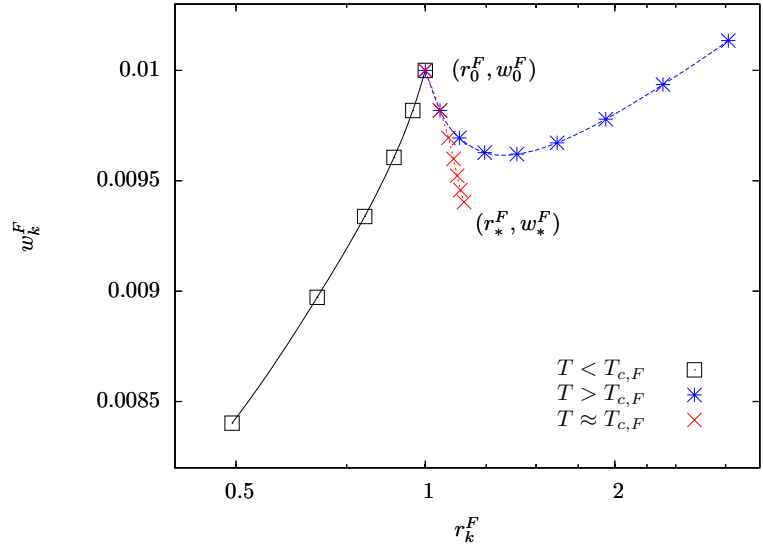}
\caption{Parametric plot of  $(r_k^F, w_k^F)$ as a function of $k$ with $\epsilon_F=0.01, J=1$. The black points represent $(r_k^F, w_k^F)$ for increasing  $k$  from top to bottom, with $T<T_{c\, F}$.  The blue points represent $(r_k^F, w_k^F)$ for increasing  $k$  from left to right, with $T>T_{c\, F}$.  The red points represent $(r_k^F, w_k^F)$ for increasing $k$ from top to bottom: here  $T$  has been dynamically adjusted as follows  at each step $k$. At the step $k=0$ of the iteration we consider two initial values of the temperature $T_{m\, F},T_{M\, F}$ such that $T_{m\, F}<T_{c\, F}<T_{M\, F}$. Then, we iterate the following procedure. We choose $T=(T_{m\, F}+T_{M\, F})/2$ and compute $r_{k+1}^F, w_{k+1}^F$ as a function of $r_{k}^F, w_{k}^F$ with Eqs. (\ref{145}). If $r_{k+1}^F>r_k^F$, we are in the high-temperature phase, and thus we lower the upper bound on $T$ by setting $T_{M \, F}\rightarrow (T_{m\, F}+T_{M\, F})/2$, otherwise we are in the low-temperature phase, and we raise the lower bound  by setting $T_{m \, F}\rightarrow (T_{m\, F}+T_{M\, F})/2$. Then we set $k\rightarrow k+1$ and repeat. By iterating this procedure many times, we obtain an estimate of the critical temperature $T_{c\, F}$ by successive bisections of the interval $[T_{m\, F}, T_{M\,F}]$, and we also obtain an estimate of the critical fixed point $r^F_\ast, w^F_\ast$, which is depicted in the Figure. 
This bisection procedure is analog to the bisection Routine \ref{bisection_routine}, illustrated in Section \ref{pop_hea} for the RG approach in real space. 
}
\label{fig11}
\end{figure}
  \end{centering}

The second line of Eq. (\ref{145}) can be rewritten as 
\[
w_{k+1}^F - w_k^F=  \left(\frac{2}{C_F^2} -1\right)w_k^F - \frac{9}{C_F^2(r_k^F)^2}(w_k^F)^2   +O((w_{k}^F)^ 3). 
\]
If follows that  $\sigma_F \lessgtr 3/4$ implies $2/C_F^2 \lessgtr 1$, and so $w_{k+1}^F \lessgtr w_k^F$. Hence, for $\sigma \leq 3/4$ $w_k^F \rightarrow  0$ for $k \rightarrow \infty$, and any fixed point is Gaussian,  while for $\sigma_F>3/4$ a non-Gaussian fixed point arises for $k \rightarrow \infty$.
Historically, the analysis of  Gaussian fixed points of Eq. (\ref{redef_dhm}) was first done in \cite{bleher1973investigation,bleher1973investigation3,bleher1973investigation2}, while  non-Gaussian fixed points have been studied  first  in \cite{bleher1975critical2}, and their analysis was later developed in \cite{collet1977numerical, collet1977epsilon,collet1978renormalization,jona2001renormalization}. 
 According to the above assumption that an expansion in powers of $w_k^F$ is meaningful,  the structure of the fixed points of Eq. (\ref{rec2_dhm}) discussed  in the Gaussian case must still hold. 
Let us set 
\be \label{epsilon_f}
\epsilon_F \equiv \sigma_F - \frac{3}{4}
\ee
and show this explicitly in Fig. \ref{fig11}, where a parametric plot of  $(r_k^F, w_k^F)$ as a function of $k$ is depicted for three different  temperatures $T<T_{c\, F}, T>T_{c\, F}$ and $T\approx T_{c\, F}$ for a given $\epsilon_F$ and $J$-value, with $\epsilon_F>0$. The Figure shows the existence of two attractive fixed points, the high-temperature fixed point and the low-temperature fixed point, and of an unstable critical fixed point separating them, as illustrated in the Gaussian case. Given an initial condition, the parameters $r_k^F, w_k^F$ converge to the critical fixed point only if $\beta$ is equal to its critical value  $\beta_{c\, F}$.
\\

Let us now focus on  the critical fixed point. This is obtained by setting $\beta = \beta_{c \, F}$ and by requiring  that $r_{k+1}^F=r_k^F= r^F_\ast,\,  w_{k+1}^F=w_k^F= w^F_\ast$ in Eq. (\ref{145})
 \be \label{146}
\left\{
\begin{array}{lcl}
r_{\ast}^F  &= &  \frac{2 r^F_{\ast}}{C_F} - \beta_{c \, F} J  + \frac{3 }{C_F r_\ast^F} w_\ast ^F - \frac{9 }{2 C_F (r_\ast^F)^3} (w_\ast^F)^2 + O((w_{\ast}^F)^3),\\ 
w_{\ast}^F &= & \frac{2}{C_F^2} w_\ast^F - \frac{9}{C_F^2(r_\ast^F)^2}(w_\ast^F)^2   +O((w_{\ast}^F)^ 3).
\end{array}
\right . 
\ee
If $w_\ast^F$ is nonzero, one can divide the second equality of Eq. (\ref{146}) by $w_\ast^F$, and get 
\be \label{147}
1 -  \frac{2}{C_F^2} =  - \frac{9}{C_F^2(r_\ast^F)^2}w_\ast^F   +O((w_{\ast}^F)^ 2). 
\ee
Since $w_\ast^F$ must be positive, we have $1 -  2/C_F^2<0$, i. e. $\epsilon_F>0$. As anticipated above, a non-Gaussian critical fixed point $w_\ast^F\neq 0$ exists only if $\epsilon_F>0$, and according to Eq. (\ref{147}) it is proportional to $\epsilon_F$. Accordingly, if $\epsilon_F \leq 0$ one has $w_\ast^F=0$.  

\section{Calculation of $\nu_F$} \label{app_field_dhm_3}

The critical exponent $\nu_F$ can be calculated \cite{wilson1974renormalization} by linearizing Eq. (\ref{146}) in the neighborhood of the critical fixed point $r_\ast^F, w_\ast^F$. Let us  introduce the  $2 \times 2$ matrix 
\be \label{149} 
\mathscr{M}^F_{ij} \equiv \left. \frac{\partial  (r^F_{k+1}, w^F_{k+1}) }   {\partial (r^F_{k}, w^F_{k})  } \right|_{ r_k^F  = r_\ast^F, w_k^F = w_\ast^F}. 
\ee
One can show \cite{wilson1974renormalization}, that the critical exponent $\nu_F$ is related to the largest eigenvalue $\Lambda_F$ of $\mathscr{M}^F$ 
\be \label{nu_f}
\nu_F = \frac{ \log 2} {\log \Lambda_F}. 
\ee \\ 

If $\epsilon_F\leq 0$, one has $r_\ast ^F = \beta_{c \, F} J/(2/C_F-1), w_\ast ^F=0$. $\mathscr{M}^F$ can be directly computed from Eq. (\ref{145}), and one has
\be \label{150}
\Lambda_F = 2^{2 \sigma_F-1}. 
\ee 
Even though Eq. (\ref{150}) has been derived by using  the approximate ansatz (\ref{142}), one can show that any ansatz including $m^6, m^8, \cdots$-terms in $\mathfrak{p}_k(m)$ would lead to Eq. (\ref{150}). Hence, Eq. (\ref{150}) is  exact.
\\ 

If $\epsilon_F>  0$, the critical fixed point $r_\ast^F, w_\ast^ F$ can be computed perturbatively as a power series in $\epsilon_F$ by observing from Eq. (\ref{147}) that $w_\ast^F = O(\epsilon_F)$, and by expanding the left and the right-hand side of Eq. (\ref{146}) in powers of $\epsilon_F$. The result is
\bea \label{148}
\no 
r_\ast^F & = & (1+\sqrt{2}) \beta_{c\, F} J - \frac{10}{3}(4+3 \sqrt{2}) \beta_{c \, F} J \log 2 \cdot \epsilon_F + O(\epsilon_F^2), \\ 
w_\ast^F & = & \frac{8}{9}(3+2 \sqrt{2})(\beta_{c \, F}J)^2 \log 2 \cdot  \epsilon_F + O(\epsilon_F^2). 
\eea
Using  Eqs. (\ref{149}), (\ref{145}), (\ref{148}), one has
\be \label{151}
\Lambda_F = \sqrt 2 \left( 1 + \frac{2}{3} \log 2 \cdot \epsilon_F + O(\epsilon_F^2) \right) . 
\ee 
Differently from Eq. (\ref{150}), Eq. (\ref{151}) is not exact, because it is the first order of a series in $\epsilon_F$. 

\chapter{Calculation of $\phi_{0}$}\label{app0}

In order to calculate $\phi_0$, let us consider  Eq. (\ref{29}) for $C=0$.  One has

\bea \label{41}
\mathbb{E}_\epsilon[Z[T,\{ \epsilon \}]^ n] & = & \sum_{ \{ \vec{S}_a \}_{a=1,\ldots, n}  } \exp  \left(  \frac{\beta^2}{4}  \sum_{i=1}^ {2^ {k}} \sum_{a,b=1}^ {n} \delta_{S_{a, i} , S_{b, i}}\right)  \\ \no
 & = &   \sum_{ \{ \vec{S}_a \}_{a=1,\ldots, n}  } \exp  \left(  \frac{\beta^2}{4}  \sum_{i=1}^ {2^ {k}} \sum_{a,b=1}^ {n} \frac{1+S_{a, i}  S_{b, i}}{2} \right) \\ \no 
  & = &  \sum_{ \{ \vec{S}_a \}_{a=1,\ldots, n}  } \exp  \left(  \frac{\beta^2}{8}  \sum_{i=1}^ {2^ {k}} \sum_{a,b=1}^ {n} S_{a, i}  S_{b,i } \right) [1+O(n^2)]\\ \no
  & = &  \sum_{ \{ \vec{S}_a \}_{a=1,\ldots, n}  } \exp  \left[  \frac{\beta^2}{8} \sum_{i=1}^ {2^k} \left(  \sum_{a=1}^ {n} S_{a, i} \right)^2   \right] \\ \no
& = &  \sum_{ \{ \vec{S}_a \}_{a=1,\ldots, n}  }    \prod_{i=1}^ {2^ k} \sqrt{\frac{2}{\pi}}  \int_{-\infty}^{\infty} dx_i \exp  \left( -2 x_i^2 + \beta   x_i \sum_{a=1}^ {n} S_{a, i}    \right) \\ \no
   & = &  \prod_{i=1}^ {2^ k} \sqrt{\frac{2}{\pi}}  \int_{-\infty}^{\infty} dx_i e^{ -2 x_i^2    } \left[ 2 \cosh( \beta x_i) \right] ^n   \\ \no
      & = &  \left\{   \sqrt{\frac{2}{\pi}}  \int_{-\infty}^{\infty} dx e^{ -2 x^2    } \left[ 1 + n \log  \left[ 2 \cosh( \beta x) \right]  + O(n^2) \right]   \right\}^ {2^ k} \\ \no
            & = &  1 + n 2^ k   \sqrt{\frac{2}{\pi}}  \int_{-\infty}^{\infty} dx e^{ -2 x^2    }  \log  \left[ 2 \cosh( \beta x) \right]  + O(n^2). 
\eea

In the second line of  Eq. (\ref{41}) we write explicitly  $\delta_{S_{a,i},S_{b,i}}$ in terms of the spins. In the third line we observe that the first addend in the exponential is  $O(n^2)$, and so we don't have to calculate it  because, according to Eq. (\ref{30}), in order to compute $f$ we need only  the terms in $\mathbb{E}_\epsilon[Z[T,\{ \epsilon \}]^ n]$ that are linear in $n$.  In the fifth line we write the exponential in terms of a Gaussian integral, according to the Hubbard-Stratonovich transformation \cite{zinnjustin}. In the sixth line we sum over the spins. The expression obtained in the sixth line is an explicit function of $n$, in such a way that  in the seventh and eighth line we develop such an expression in powers of $n$ up to  linear terms.\\

Plugging Eq. (\ref{41}) in Eq. (\ref{30}) and using Eq. (\ref{31}), one has
\[
\phi_0(T) = -   \frac{1}{\beta }\sqrt{\frac{2}{\pi}}  \int_{-\infty}^{\infty} dx e^{ -2 x^2    }  \log  \left[ 2 \cosh( \beta x) \right]. 
\]

 This  calculation has been automated with a symbolic manipulation program \cite{wolfram1996mathematica}, and the coefficients $\phi_{i}(T)$ have been computed  for $0\leq i \leq10$.  

\chapter{Calculation of $\Upsilon_{m,0}$}\label{app1}

From the last line of Eq. (\ref{104}) with $C=0$, one has
\bea \label{112}
\Upsilon_{m,0}(T) & = & \lim_{n\rightarrow 0} \sum_{ \{ \vec{S}_a \}_{a=1,\ldots, n}  } \exp  \left(  \frac{\beta^2}{4}  \sum_{i=1}^ {2^ {k}} \sum_{a,b=1}^ {n} \delta_{S_{a, i} , S_{b, i}}\right) \prod_{i=1}^ {2^m}  \delta_{S_{1,i},S_{2,i}}\\ \no
 & = & \lim_{n\rightarrow 0} \sum_{ \{ \vec{S}_a \}_{a=1,\ldots, n}  } \exp  \left(  \frac{\beta^2}{4}  \sum_{i=1}^ {2^ {k}} \sum_{a,b=1}^ {n} \frac{1+S_{a, i}  S_{b, i}}{2} \right) \prod_{i=1}^ {2^m}  \delta_{S_{1,i},S_{2,i}}\\ \no 
  & = & \lim_{n\rightarrow 0} \sum_{ \{ \vec{S}_a \}_{a=1,\ldots, n}  } \exp  \left(  \frac{\beta^2}{8}  \sum_{i=1}^ {2^ {k}} \sum_{a,b=1}^ {n} S_{a, i}  S_{b,i } \right) \prod_{i=1}^ {2^m}  \delta_{S_{1,i},S_{2,i}}[1+O(n^2)]\\ \no
  & = & \lim_{n\rightarrow 0} \sum_{ \{ \vec{S}_a \}_{a=1,\ldots, n}  } \exp  \left[  \frac{\beta^2}{8} \sum_{i=1}^ {2^k} \left(  \sum_{a=1}^ {n} S_{a, i} \right)^2   \right] \prod_{i=1}^ {2^m}  \delta_{S_{1,i},S_{2,i}}\\ \no
       & = & \lim_{n\rightarrow 0} \sum_{ \{ \vec{S}_a \}_{a=1,\ldots, n}  }  \left[ \prod_{i=1}^ {2^ k} \sqrt{\frac{2}{\pi}}  \int_{-\infty}^{\infty} dx_i \exp  \left( -2 x_i^2 + \beta   x_i \sum_{a=1}^ {n} S_{a, i}    \right) \right]\times \\ \no 
       && \times \prod_{i=1}^ {2^m}  \delta_{S_{1,i},S_{2,i}} \\ \no
            & = & \lim_{n\rightarrow 0} \Bigg\{ \sum_{\{ \vec{S}_{I}^ a \}_{a=1}^ {n}}  \left[ \prod_{i=1}^ {2^ m} \sqrt{\frac{2}{\pi}}  \int_{-\infty}^{\infty} dx_i \exp  \left( -2 x_i^2 + \beta   x_i \sum_{a=1}^ {n} S_{a, i}    \right) \right]  \times \\ \no 
         &&   \prod_{i=1}^ {2^m} \delta_{S_{1,i},S_{2,i}}\Bigg\}   \times \\ \no 
         &&\times \Bigg\{ \sum_{\{ \vec{S}_{O}^ a \}_{a=1}^ {n}}  \left[ \prod_{i=2^ m+1}^ {2^ k} \sqrt{\frac{2}{\pi}}  \int_{-\infty}^{\infty} dx_i \exp  \left( -2 x_i^2 + \beta   x_i \sum_{a=1}^ {n} S_{a, i}    \right) \right] \Bigg\}    \\ \no
             \eea
   \beas
              & = & \lim_{n\rightarrow 0} \Bigg\{ \sum_{\{ \vec{S}_{I}^ a \}_{a=1}^ {n}}  \left[ \prod_{i=1}^ {2^ m} \sqrt{\frac{2}{\pi}}  \int_{-\infty}^{\infty} dx_i \exp  \left( -2 x_i^2 + \beta   x_i \sum_{a=1}^ {n} S_{a, i}    \right) \right] \times \\ \no 
              && \times \prod_{i=1}^ {2^m}  \delta_{S_{1,i},S_{2,i}}\Bigg\} \times \\ \no
            && \times  \Bigg\{ \prod_{i=2^ m+1}^ {2^ k} \sqrt{\frac{2}{\pi}}  \int_{-\infty}^{\infty} dx_i e^{ -2 x_i^2}  [2 \cosh(\beta x_i)]^ n \Bigg\}  \\ \no
   & = & \lim_{n\rightarrow 0} \sum_{\{ \vec{S}_{I}^ a \}_{a=1}^ {n}}  \left[ \prod_{i=1}^ {2^ m} \sqrt{\frac{2}{\pi}}  \int_{-\infty}^{\infty} dx_i \exp  \left( -2 x_i^2 + \beta   x_i \sum_{a=1}^ {n} S_{a, i}    \right) \right] \times \\  \no 
   && \times \prod_{i=1}^ {2^m}  \delta_{S_{1,i},S_{2,i}}  [1+O(n)]\\ \no
   & = & \lim_{n\rightarrow 0} \sum_{\{ \vec{S}_{I}^ a \}_{a=2}^ {n}} \prod_{i=1}^ {2^ m} \sqrt{\frac{2}{\pi}}  \int_{-\infty}^{\infty} dx_i \exp  \left[ -2 x_i^2 + \beta   x_i (2 S_{2,i}+ S_{3,i} +  \cdots + S_{n,i})    \right]\\ \no
   & = & \lim_{n\rightarrow 0}   \Bigg\{ \prod_{i=1}^ {2^ m} \sqrt{\frac{2}{\pi}}  \int_{-\infty}^{\infty} dx_i e^{ -2 x_i^2    } \frac{2 \cosh(2 \beta x_i)}{[2 \cosh(\beta x_i)]^ {2-n}}\Bigg\}  \\ \no
   & = & \left[    \sqrt{\frac{2}{\pi}}  \int_{-\infty}^{\infty} dx e^{ -2 x^2    } \frac{2 \cosh(2 \beta x)}{[2 \cosh(\beta x)]^ {2}}  \right] ^ {2^ m}
\eeas
 In the second line of Eq. (\ref{112}) we write explicitly  $\delta_{S_{a,i},S_{b,i}}$ in terms of the spins. In the third line we observe that the first addend in the exponential is $O(n^2)$, and so we don't have to compute it in the limit $n\rightarrow 0$.  In the fifth line we write the exponential in terms of a Gaussian integral, according to the Hubbard-Stratonovich transformation \cite{zinnjustin}. In the sixth line we split the sum over the spins into a sum involving  spins $ \vec{S}_I^ a \equiv S_{a,1},\cdots,S_{a, 2^ m}$, and a sum involving spins  $\vec{S}_O^ a \equiv S_{a,2^ m+1},\cdots,S_{a, 2^ k}$. In the seventh line we explicitly calculate the latter sum, and in the eighth line we observe that this is given by $1+O(n)$. In the ninth line we drop the $O(n)$ terms, and sum over $\vec{S}_{I}^ 1$, while in the tenth line we sum  over the remaining spins. The expression obtained in the tenth line is an explicit function of $n$, in such a way that  in the last line the limit $n\rightarrow 0$ can be taken.\\

 This calculation has been automated with a symbolic manipulation program \cite{wolfram1996mathematica}, and the coefficients $\Upsilon_{m,i}(T)$ have been computed for $0 \leq i \leq 9$.  
 
 \chapter{\label{app_rec}Derivation of the recurrence equations (\ref{56})}

  Plugging Eq. (\ref{55}) into  Eq.  (\ref{51}), one finds
 
\bea\label{app_rec100}
\mathcal{\mathscr{P}}_k[Q]&=&
\exp\left\{ -\left[ {\left(  \frac{2 r_{k-1}}{C^4}  -\frac{\beta^2}{4}  \right)\text{Tr}[ Q^2 ]} + \frac{2 w_{k-1}}{3 C^{6}}  \text{Tr}[Q^3] \right] \right\} \times \\ \no 
&& \times \int \left[ d P \right] \exp\left[ -S_{k-1}[P,Q]\right],\\ \no 
S_{k-1}[P,Q] & \equiv & \frac{2 r_{k-1}}{C^4}  \text{Tr} [ P^2 ] + \frac{2 w_{k-1}}{ C^{6}}  \text{Tr}[Q P^ 2]. 
\eea

The integral  in Eq. (\ref{app_rec100}) is Gaussian, and thus it can be calculated exactly by using standard formulas \cite{zinnjustin}. Indeed, defining $\forall a>b$ the index $A \equiv (a,b) $,  the  $n(n-1)/2$ integration variables $\{ P_{ab} \}_{a<b=1, \ldots, n}$ can be labeled with the  index $A$: $\{ P_{ab} \}_{a<b=1, \ldots, n} \rightarrow \{ P_A \}_{A=1,\ldots, n(n-1)/2}$. In order to compute the Gaussian integral in Eq.  (\ref{app_rec100}), we observe that 
\[
\frac{\partial ^ 2 S_{k-1}[P,Q]  }{\partial P_A \partial P_B}  =    \frac{8 r_{k-1}}{C^4}\delta_{AB} + \frac{ 4 w_{k-1}}{C^ {6}}M_{AB}[Q],
  \]
  where
  \bea  
M_{ab, cd}[Q] & \equiv & N_{ab, cd}[Q]  + N_{ab, dc}[Q] , \label{app_rec210}\\
N_{ab, cd}[Q] & \equiv & \delta_{bc} Q_{da} + \delta_{ac} Q_{db} . \label{app_rec211}
\eea
 Calculating the Gaussian integral, Eq. (\ref{app_rec100}) becomes 
\bea \label{app_reca} \no 
\mathcal{\mathscr{P}}_k[Q] &= & \exp\left\{ -\left[ {\left(  \frac{2 r_{k-1}}{C^4}  -\frac{\beta^2}{4}  \right)\text{Tr} [ Q^2  ]} + \frac{2 w_{k-1}}{3 C^{6}}  \text{Tr}[Q^3] \right] \right\}\times \\ 
&& \times  \left[ \det  \left(      \frac{8 r_{k-1}}{C^4}\delta_{AB} + \frac{ 4 w_{k-1}}{C^ {6}}M_{AB}[Q]  \right) \right]^ {-\frac{1}{2}} , 
\eea
 where in Eq. (\ref{app_reca}) and in the following, $Q$-independent multiplicative constants are  omitted.

 Supposing  that $w_k$ is small for every $k$, the determinant in the right-hand side of Eq. (\ref{app_reca}) can now be expanded in powers of $w_{k-1}$. Calling   $\textbf{Tr}$ the trace over $A$-type indices, we use the  relation $\log \det = \textbf{Tr} \log$ for the matrix in round brackets in Eq. (\ref{app_reca})
\bea \label{53}
&& \left[ \det  \left(      \frac{8 r_{k-1}}{C^4}\delta_{AB} + \frac{ 4 w_{k-1}}{C^ {6}}M_{AB}[Q]  \right) \right]^ {-\frac{1}{2}} = \\ \no 
&& \exp \Bigg\{ - \frac{1}{2}\Bigg[  
\frac{  w_{k-1}}{2C^ {2} r_{k-1}} \textbf{Tr}[M[Q]]  - \frac{1}{2} \left( \frac{w_{k-1}}{2C^ {2} r_{k-1}} \right)^2 \textbf{Tr}[ M[Q]^2 ] + \\ \no 
&& + \frac{1}{3} \left( \frac{w_{k-1}}{2C^ {2} r_{k-1}} \right)^3 \textbf{Tr}[ M[Q]^3 ] +O(w_{k-1}^ 4)
\Bigg] \Bigg\}. 
\eea

By using the definitions (\ref{app_rec210}) and (\ref{app_rec211}), one has $\textbf{Tr}[M[Q]] =0$. Then, by using Eqs. (\ref{app_rec210}), (\ref{app_rec211}), one has 
 
\bea \label{app_rec212} \no
\textbf{Tr}[M[Q]^2]&=& \sum_{AB} M[Q]_{AB} M[Q]_{BA}   \\ \no 
&=&\sum_{a>b, c>d} \left( N_{ab, cd}[Q]  + N_{ab, dc}[Q]\right)  \left( N_{cd, ab}[Q]  + N_{cd, ba}[Q]\right)\\  \no 
&=&\sum_{a\neq b, c\neq d}  N_{ab, cd}[Q]   N_{cd, ab}[Q] \\  \no 
&=&\sum_{a\neq b, c\neq d}  (\delta_{bc} Q_{da} + \delta_{ac} Q_{db} ) ( \delta_{da} Q_{bc} + \delta_{ca} Q_{bd} ) \\  \no 
&=&\sum_{a\neq b, c\neq d}   \delta_{ca} Q_{bd}^2  \\  \no 
&=&\sum_{a bcd} (1- \delta_{ab})(1-\delta_{cd})   \delta_{ca} Q_{bd}^2  \\ \no 
&=&(n-2) \sum_{ab} Q^ 2_{ab}\\  
&=&(n-2) \text{Tr}[Q^2].   
\eea
 
 In the second line of Eq. (\ref{app_rec212}) we write the sum over the indices $A,B,\ldots $ in terms of a sum over the replica indices $a,b,\ldots$. In the third line we use the symmetry of $N_{ab,cd}[Q]$ with respect to $a \leftrightarrow b$ and rewrite the sum over $a>b,\, c>d$ in terms of a sum with $a\neq b,\, c\neq d$. In the fifth line we find out that only one of the terms stemming from  the product $(\delta_{bc} Q_{da} + \delta_{ac} Q_{db} ) ( \delta_{da} Q_{bc} + \delta_{ca} Q_{bd} ) $  does not vanish, because of the constraints $a\neq b,\, c\neq d, \, Q_{aa}=0$ (see Eq. (\ref{overlap})), and because of the Kronecker $\delta$s in the sum. Once we are left with the nonvanishing term, in the sixth line we write explicitly the sum over $a\neq b,\, c\neq d$ in terms of an unconstrained sum over $a,b,c,d$ by adding the constraints $(1-\delta_{ab})(1-\delta_{ cd})$. In the seventh line we calculate explicitly the sum over the replica indices, and write everything in terms of the replica invariant $I^{(2)}_1[Q] \equiv \text{Tr}[Q^2]$ (see Table \ref{tab1}).\medskip\\

By following the steps shown Eq. (\ref{app_rec212}), all the other tensorial operations can be done. In particular, one finds  
\bea \label{app_rec213}
 \textbf{Tr}[M[Q]^3]&=& (n-2) \text{Tr}[Q^3].
\eea\smallskip\\

By plugging Eqs. (\ref{app_rec212}), (\ref{app_rec213}) into Eq. (\ref{53}), and then substituting Eq. (\ref{53}) into the recursion relation (\ref{app_reca}), one finds 
\bea\label{app_rec4}
\mathscr{P}_k[Q]&=& \exp \Bigg\{ -\Bigg[ \left( \frac{2 r_{k-1}}{C^4} - \frac{\beta^2}{4}-\frac{n-2}{4}\left( \frac{w_{k-1}}{2 C^{2} r_{k-1} }\right)^2   \right) \text{Tr}[Q^2] + \\ \nonumber
&&+\frac{1}{3} \left( \frac{2 w_{k-1}}{C^ {6}} +\frac{n-2}{2}\left( \frac{w_{k-1}}{2 C^ {2} r_{k-1} }\right)^3   \right) \text{Tr}[Q^3] + O(w_{k-1}^4)    \Bigg] \Bigg\}.
\eea
 
Comparing Eq. (\ref{app_rec4}) to the ansatz  (\ref{55}) for $\mathscr{P}_k$, one finds the recursion relations (\ref{56}) for the coefficients $r_k, w_k$. 
 
\chapter{Results of the two-loop RG calculation  \`a la Wilson} \label{app_p}

Here we sketch the results of the perturbative calculation to the order $w_k^5$ mentioned in Section \ref{a_la_wilson}.  The invariants $I^{(j)}_l[Q]$ yielding $\mathscr{P}_k[Q]$ as a fifth-degree polynomial in $Q$ are given  in Table \ref{tab1}. 

\begin{centering}
\begin{table}
\caption{\label{tab1}  Invariants generated to the order $p=5$. In each line of the table we show  the invariants $I^ {(j)}_1[Q],\ldots,I^ {(j)}_{n_j}[Q]$ from left to right.}
\centering
\begin{tabular}{ccccc}
\toprule
$j$ & \multicolumn{4}{c}{$I_{l}^{(j)}[Q]$}\\
\midrule 
$2$ & \multicolumn{4}{c}{$\text{Tr}[Q^{2}]$ } \tabularnewline
\midrule
$3$ & \multicolumn{4}{c}{$\text{Tr}[Q^{3}]$} \tabularnewline
\midrule
$4$ & $\text{Tr}[Q^{4}]$ & $\text{Tr}[Q^{2}]^{2}$ & $\sum_{a\neq  c}Q_{ab}^{2}Q_{bc}^{2}$ & $\sum_{a b}Q_{ab}^{4}$\tabularnewline
\midrule
$5$ & $\text{Tr}[Q^{5}]$ & $\text{Tr}[Q^{2}]\text{Tr}[Q^{3}]$ & $\sum_{a b c d}Q_{ab}^{2}Q_{bc}Q_{bd}Q_{cd}$ & $\sum_{a b c}Q_{ab}^{3}Q_{ac}Q_{bc}$\tabularnewline
\bottomrule
\end{tabular}
\end{table}
\end{centering}

The recurrence RG equation (\ref{51}) relating $\mathscr{P}_{k-1}[Q]$ to $\mathscr{P}_k[Q]$ yields a set of equations relating the coefficients  $\{ c^{(j)}_{l,\, k-1}\}_{j,l}$   to $\{ c^{(j)}_{l,\, k}\}_{j,l}$. After a quite involved calculation, one finds that these are 
\bea \no
c^{(2)}_{1, \, k}&=& \frac{2 c^{(2)}_{1, \, k-1}}{C^4}-\frac{\beta^ 2}{4}-\frac{n-2}{4}\left(\frac{c^{(3)}_{1, \, k-1} }{2 C^ {2} c^{(2)}_{1, \, k-1}} \right)^2+ (2n-1)  \frac{c^{(4)}_{1, \, k-1}}{8 C^4 c^{(2)}_{1, \, k-1}}+\label{220}\\  \no 
&&+  \frac{c^{(4)}_{2, \, k-1}}{2 C^4 c^{(2)}_{1, \, k-1}}\left[ 1 + \frac{n(n-1)}{4} \right]+(n-2) \frac{c^{(4)}_{3 ,\, k-1}}{8 C^4 c^{(2)}_{1, \, k-1}} + 
 \frac{3 c^{(4)}_{4, \, k-1}}{8 C^4 c^{(2)}_{1, \, k-1}}+ \\ \label{80}
&& + O\left((c^{(3)}_{1, \, k-1} )^ 6\right), 
\eea
\bea \no 
c^{(3)}_{1, \, k} & = & \frac{2 c^{(3)}_{1, \, k-1}}{C^ {6}} + \frac{n-2}{2}\left(\frac{c^{(3)}_{1, \, k-1}}{2 C^ {2} c^{(2)}_{1, \, k-1}} \right)^3 + \frac{3 n c^{(5)}_{1, \, k-1}}{4 C^{6} c^{(2)}_{1, \, k-1}}+\\ \no 
&& + (n+3)\frac{3 c^{(5)}_{2, \, k-1}}{20 C^ {6} c^{(2)}_{1, \, k-1}} + \frac{9 c^{(5)}_{3, \, k-1}}{20 C^ {6} c^{(2)}_{1, \, k-1}}+ \\    \no 
& & + \frac{3 c^{(5)}_{4, \, k-1} }{20 C^ {6} c^{(2)}_{1 ,\, k-1}}[12 + n(n-1)]+ \\ \no 
&& - \frac{3 c^{(3)}_{1, \, k-1}}{4 C^ {6} c^{(2)}_{1, \, k-1}} \left[  \frac{(n-1) c^{(4)}_{1, \, k-1}}{2 C^4 c^{(2)}_{1 ,\, k-1}}  +  \frac{2 c^{(4)}_{2, \, k-1}}{ C^4 c^{(2)}_{1, \, k-1}} +  \frac{c^{(4)}_{3, \, k-1}}{2 C^4 c^{(2)}_{1, \, k-1}} \right]+\\ \label{81}
&&+ O\left((c^{(3)}_{1, \, k-1} )^ 7\right),\\
 \label{82}
c^{(4)}_{1 ,\, k} & = & \frac{2 c^{(4)}_{1, \, k-1}}{C^8}- \frac{n}{2}\left(\frac{ c^{(3)}_{1, \, k-1} }{2 C^ {2} c^{(2)}_{1 ,\, k-1}} \right)^4+ O\left((c^{(3)}_{1, \, k-1} )^ 6\right),  \\
 \label{83}
c^{(4)}_{2, \, k} & = & \frac{2 c^{(4)}_{2, \, k-1}}{C^8}- \frac{3}{2}\left(\frac{ c^{(3)}_{1 ,\, k-1} }{2 C^ {2} c^{(2)}_{1 ,\, k-1}} \right)^4+ O\left((c^{(3)}_{1, \, k-1} )^ 6\right),\\
 \label{84}
c^{(4)}_{3, \, k} & = & \frac{2 c^{(4)}_{3, \, k-1}}{C^8}+ 8\left(\frac{ c^{(3)}_{1, \, k-1} }{2 C^ {2} c^{(2)}_{1, \, k-1}} \right)^4+ O\left((c^{(3)}_{1 ,\, k-1} )^ 6\right), \\
 \label{85}
c^{(4)}_{4, \, k} & = & \frac{2 c^{(4)}_{4, \, k-1}}{C^8}+ 4\left(\frac{ c^{(3)}_{1, \, k-1} }{2 C^ {2} c^{(2)}_{1, \, k-1}} \right)^4+ O\left((c^{(3)}_{1, \, k-1} )^ 6\right), \\
\label{86}
c^{(5)}_{1 ,\, k} & = & \frac{2 c^{(5)}_{1 ,\, k-1}}{C^{10}}+ \frac{n+6}{2}\left(\frac{ c^{(3)}_{1 ,\, k-1} }{2 C^ {2} c^{(2)}_{1 ,\, k-1}} \right)^5+ O\left((c^{(3)}_{1 ,\, k-1} )^ 7\right), \\
\label{87}
c^{(5)}_{2 ,\, k} & = & \frac{2 c^{(5)}_{2, \, k-1}}{C^{10}}- 40\left(\frac{ c^{(3)}_{1 ,\, k-1} }{2 C^ {2} c^{(2)}_{1 ,\, k-1}} \right)^5+ O\left((c^{(3)}_{1, \, k-1} )^ 7\right), \\
  \label{88}
c^{(5)}_{3, \, k} & = & \frac{2 c^{(5)}_{3 ,\, k-1}}{C^{10}}+ 30\left(\frac{ c^{(3)}_{1 ,\, k-1} }{2 C^ {2} c^{(2)}_{1, \, k-1}} \right)^5+ O\left((c^{(3)}_{1 ,\, k-1} )^ 7\right), \\
 \label{89}
c^{(5)}_{4, \, k} & = & \frac{2 c^{(5)}_{4, \, k-1}}{C^{10}}+5\left(\frac{ c^{(3)}_{1, \, k-1} }{2 C^ {2} c^{(2)}_{1 ,\, k-1}} \right)^5+ O\left((c^{(3)}_{1 ,\, k-1} )^ 7\right). 
\eea

\chapter{One-loop RG calculation in the field-theory approach}\label{app_field_theor}

In this Appendix we present the computation of the RG functions $Z_g, Z_{Q^2}$ to order  $g_r^2$. \\

In the bare theory, $1$PI correlation functions are defined by the action (\ref{93}), and they can be obtained as the derivative of the bare $1$PI generating functional $\Gamma[\mathcal{Q}]$  with respect to $\mathcal{Q}_{i,\, ab}$. Similarly, the renormalized $1$PI correlation functions are the derivatives with respect to $\mathcal{Q}$ of the generating functional $\Gamma_r[\mathcal{Q}]$ of the renormalized theory, which depends on the renormalized parameters $m_r, g_r$. Accordingly, $\Gamma_r[\mathcal{Q}]$ can be expanded  in powers of $g_r$ by means of the loop expansion
 \bea  \label{106}
\Gamma_r[\mathcal{Q}] & = & \frac{1}{2} \sum_{i ,j=0}^{2^ k-1} \Delta_{ij} \text{Tr}[\mathcal{Q}_i \mathcal{Q}_j ] + 
\frac{m_r^ {3 \epsilon} g_r}{3!} \sum_i \text{Tr}[\mathcal{Q}_i^3] \Bigg( Z_g +\\ \no
&& + \frac{n-2}{8} m_r^ {\frac{6\epsilon}{2 \sigma-1}} \mathscr{I}_7 g_r^2\Bigg)   +  O(g_r^ 5).
\eea
 
The Feynman diagram $\mathscr{I}_7$ is depicted in Fig. \ref{fig8}, and is equal to
\bea \label{99}
\mathscr{I}_7 &= &  \frac{1}{2^ k} \sum_{p=0}^ {2^ k-1} \frac{1}{\left(m_r +\delta m+ |p|_2^ {2 \sigma-1}\right)^ 3} . 
\eea

It is easy to show that $\mathscr{I}_7$ has a finite limit for $k\rightarrow \infty$. Indeed, the propagator (\ref{prop}) in the sum in the right-hand  side of Eq. (\ref{99}) depends on $p$ through its dyadic norm. Hence,  the sum over $p$ in the right-hand side of Eq. (\ref{99}) can be easily transformed into a sum over all the possible values of $|p|_2$. In order to do so,  we recall  \cite{parisi2000p} that the number of integers $0 \leq p \leq 2^ k -1 $  which satisfy $|p|_2=2^ {-j}$ is given by  $ 2^ {-j+k-1} $. This number is the volume of a shell in a space of integer numbers $p$, where the distance between two integers $p_1,p_2$ is given by the dyadic norm $|p_1-p_2|_2$. Hence, Eq. (\ref{112}) becomes 
\bea \label{100}
\mathscr{I}_7 & = & \sum_{j=0}^ {k-1} 2^ {-j-1}  \frac{1}{\left[m_r   +\delta m+ 2^ {-j(2 \sigma-1)}\right]^ 3}\\ \no 
& \overset{k \rightarrow \infty}{=} & \sum_{j=0}^ {\infty} 2^ {-j-1}  \frac{1}{\left[m_r  +\delta m+ 2^ {-j(2 \sigma-1)}\right]^ 3},
\eea
where in the second line of Eq. (\ref{100}) the $k \rightarrow \infty$ limit has been taken, because the sum in the first line is convergent in this limit. 
One can also show that  
$
\delta m = O(g_r^2),
$
 and thus rewrite (\ref{100}) as 
\be \label{101}
\mathscr{I}_7  =  \sum_{j=0}^ {\infty} 2^ {-j-1}  \frac{1}{\left[m_r   +   2^ {-j(2 \sigma-1)}\right]^ 3}+ O(g_r^2).
\ee
Looking at Eq. (\ref{101}), we observe  that  $\mathscr{I}_7$ is divergent for $m_r \rightarrow 0$. In particular, the smaller $m_r$, the larger the values of $j$ dominating the sum. 
It follows that in the IR limit $m_r \rightarrow 0$ the sum in the right-hand side of Eq. (\ref{101}) can be approximated by an integral, because in the region $j\gg 1$ dominating the sum the integrand function is almost constant in the interval $[ j,j+1]$. Setting $q\equiv 2^ {-j}$,  for $m_r \rightarrow 0$ we have $- q   \log 2 \, dj = dq$, and
\bea \label{103}\no
\mathscr{I}_7 & = & \frac{1}{2 \log 2}  \int_0 ^ 1  \frac{dq}{\left[m_r  + q^ {2 \sigma-1}\right]^ 3}  + O(g_r^2)\\ \no 
& = & \frac{m_r^ {-\frac{6 \epsilon}{2\sigma -1}}}{2 \log 2}  \int_0 ^ {m_r^{-\frac{1}{2 \sigma -1}}}  \frac{dx}{\left(1+ x^ {2 \sigma-1}\right)^ 3} + O(g_r^2)\\
& = & \frac{m_r^ {-\frac{6 \epsilon}{2\sigma -1}}}{2 \log 2}  \int_0 ^ {\infty}  \frac{dx}{\left(1+ x^ {2 \sigma-1}\right)^ 3} + O(g_r^2),
\eea
where in the last line of Eq. (\ref{103}) the $m_r \rightarrow 0$-limit has been taken. 
By considering the asymptotic behavior for $x \rightarrow \infty$ of the integrand function in the last line of Eq. (\ref{103}), one finds that its integral  is convergent for $\epsilon>0$ and divergent for $\epsilon < 0$, in such a way that it  has a singularity for $\epsilon \rightarrow 0^+$. Its $\epsilon$-divergent part can be easily evaluated
\bea \label{105} \no 
\mathscr{I}_7 & = &  \frac{m_r^ {-\frac{6 \epsilon}{2\sigma -1}}}{4 \log 2}  \Gamma\left( 3 + \frac{1}{1-2 \sigma}\right)   \Gamma\left(  1 + \frac{1}{1-2 \sigma} \right) + O(g_r^2)\\ 
& = & m_r^ {-\frac{6 \epsilon}{2\sigma -1}} \left[\frac{1}{12\epsilon \log 2  } + O_{\epsilon}(1) \right]+ O(g_r^2),
\eea
where $\Gamma$ is the Euler's Gamma function,  and in the second line of Eq. (\ref{105}) we developed the right-hand side in the first line in powers of $\epsilon$  around $\epsilon=0$, where $O_{\epsilon}(1)$  denotes terms which stay finite as $\epsilon \rightarrow 0$. \\
We now plug Eq. (\ref{105}) into Eq. (\ref{106}), and require that  $\Gamma_r[\mathcal{Q}]$, the generating functional of the renormalized theory, is finite, i. e. that it has no terms singular in $\epsilon$. Accordingly, we require  that the $\epsilon$-singular part of $\mathscr{I}_7$ is canceled by $Z_g$: this is the minimal subtraction scheme.
  Taking $n=0$, this subtraction implies that 
\be \label{109}
Z_g = 1 + \frac{1}{48   \epsilon \log 2}g_r^2 + O(g_r^ 4).
\ee \\\medskip

A very similar  calculation can be done by considering the generating  functional  $\Gamma[\mathcal{Q},K] $, whose derivatives with respect to $\mathcal{Q}_{i,\,ab}$ and $K_j$ yield $1$PI correlation functions with $Q_{i, \, ab}$ and $\textrm{Tr}[Q_j^2]$-insertions, and by introducing the corresponding functional of the renormalized theory $\Gamma_r[\mathcal{Q},K]$. By  requiring that $\Gamma_r[\mathcal{Q},K]$ is finite, we obtain
\be \label{108}
Z_{Q^2} = 1 + \frac{1}{24   \epsilon \log 2}g_r^2 + O(g_r^ 4).
\ee

Eqs.  (\ref{109}), (\ref{108}) are the one-loop renormalization constants $ Z_g, Z_{Q^2}$. 

\chapter{Computation of the observables \ref{obs_ferr} in Dyson's Hierarchical\\ Model}\label{app_dhm}

In order to compute the observables (\ref{obs_ferr}), it is convenient to introduce for any $k$ a discrete magnetization variable $\mu$ taking  $2^k+1$ possible values $\{ -1, -1 + 2/2^k, \cdots, 0, \cdots, 1- 2/2^k, 1\}$, and its probability distribution
\be \label{def_pi}
\pi_k ( \mu) \equiv \frac{\sum_{\vec S} \textrm{e}^{-\beta H^F_k[\vec S]} \delta\left( \frac{1}{2^k} \sum_{i=1}^{2^k} S_i = \mu\right) }{Z_k},
\ee 
where
\[ 
Z_k \equiv \sum_{\vec S} \textrm{e}^{-\beta H^F_k[\vec S]},
\]
and  $\delta$ denotes the Kronecker delta. Eq. (\ref{20}) implies \cite{dyson1969existence} that $\pi_k(\mu)$ satisfies a recursion equation analogous to Eq. (\ref{rec_dhm})
\be  \label{rec_discrete}
\pi_{k+1} (\mu) = \textrm{e}^{J \beta C_F^{k+1} \mu^2} \sum_{\mu_1, \mu_2} \pi_k(\mu_1) \pi_k (\mu_2) \delta\left( \frac{\mu_1 + \mu_2}{2} = \mu \right),
\ee 
where a $\mu$-independent multiplicative constant has been omitted in the right-hand side of Eq. (\ref{rec_discrete}). \\

Given $\beta J$, the recursion equation (\ref{rec_discrete}) can be iterated numerically $k_0-1$ times in $2^{k_0}$ operations. Once $\pi_{k_0}(\mu)$ is known, 
the observable $ O^ F_{k_0}(\beta J )$ can be easily computed. Indeed, according to Eqs. (\ref{obs_ferr}), (\ref{m}), we have
\be \label{130}
  O^ F_{k_0}(\beta J )  =  \frac{ \mathbb{E}_{\vec{S}} \left[  \left(    \frac{1}{2^ {k_0-1}} \sum_{i=1}^ {2^ {k_0-1}} S_i \right) \left( \frac{1}{2^ {k_0-1}} \sum_{i=2^{k_0-1}+1}^ {2^ {k_0}} S_i 
\right) \right] } {\mathbb{E}_{\vec{S}} \left[  \left(    \frac{1}{2^ {k_0-1}} \sum_{i=1}^ {2^ {k_0-1}} S_i \right)^2 \right] }. 
\ee

The numerator of Eq. (\ref{130}) is 
\bea \no  \label{131}
&& \mathbb{E}_{\vec{S}} \left[  \left(    \frac{1}{2^ {k_0-1}} \sum_{i=1}^ {2^ {k_0-1}} S_i \right) \left( \frac{1}{2^ {k_0-1}} \sum_{i=2^{k_0-1}+1}^ {2^ {k_0}} S_i 
\right) \right]  =\\ \no 
&=& \frac{1}{Z_{k_0}} \sum_{\vec S} \exp\left[- \beta \left( H_{k_0-1}^F[\vec S_1] + H_{k_0-1}^F[\vec S_2] \right) + \beta J C_F^{k_0} \left( \frac{1}{2^{k_0}} \sum_{i=1}^{2^{k_0}} S_i \right)^2 \right] \times \\ \no 
&& \times  \left(    \frac{1}{2^ {k_0-1}} \sum_{i=1}^ {2^ {k_0-1}} S_i \right) \left( \frac{1}{2^ {k_0-1}} \sum_{i=2^{k_0-1}+1}^ {2^ {k_0}} S_i \right) \times \\ \no
&& \times \sum_{\mu_1, \mu_2}  \delta\left( \frac{1}{2^{k_0-1}} \sum_{i=1}^{2^{k_0-1}} S_i = \mu_1\right)   \delta\left( \frac{1}{2^{k_0-1}} \sum_{i=2^{k_0-1}+1}^{2^{k_0}} S_i = \mu_2\right) \\ 
& = & \frac{Z_{k_0-1}^2}{Z_{k_0}} \sum_{\mu_1, \mu_2} \textrm{e}^{ \beta J C_F^{k_0} \left( \frac{\mu_1 + \mu_2}{2} \right)^2}  \mu_1 \mu_2 \pi_{k_0-1}(\mu_1) \pi_{k_0-1}(\mu_2), 
\eea
where $S_1 \equiv \{ S_1, \cdots , S_{2^{k_0-1}} \}, S_2 \equiv \{ S_{2^{k_0-1}+1}, \cdots , S_{2^{k_0}} \}$,  and in the first line of Eq. (\ref{131}) we used the recurrence relation (\ref{20}), and we multiplied by a factor equal to one, while in the second line we used the definition (\ref{def_pi}). By following the same steps as in Eq. (\ref{131}), the denominator in Eq. (\ref{130}) is 
\be \label{132}
\mathbb{E}_{\vec{S}} \left[  \left(    \frac{1}{2^ {k_0-1}} \sum_{i=1}^ {2^ {k_0-1}} S_i \right)^2 \right] =    \frac{Z_{k_0-1}^2}{Z_{k_0}} \sum_{\mu_1, \mu_2} \textrm{e}^{ \beta J C_F^{k_0} \left( \frac{\mu_1 + \mu_2}{2} \right)^2}  \mu_1^2 \pi_{k_0-1}(\mu_1) \pi_{k_0-1}(\mu_2). 
\ee
By dividing Eq. (\ref{131}) by Eq. (\ref{132}), the multiplicative constants cancel out, and we are left with 
\be \label{133}
  O^ F_{k_0}(\beta J ) = \frac{ \sum_{\mu_1, \mu_2} \textrm{e}^{ \beta J C_F^{k_0} \left( \frac{\mu_1 + \mu_2}{2} \right)^2}  \mu_1 \mu_2 \pi_{k_0-1}(\mu_1) \pi_{k_0-1}(\mu_2) } {\sum_{\mu_1, \mu_2} \textrm{e}^{ \beta J C_F^{k_0} \left( \frac{\mu_1 + \mu_2}{2} \right)^2}  \mu_1^2 \pi_{k_0-1}(\mu_1) \pi_{k_0-1}(\mu_2) }. 
\ee
 The right-hand side of Eq. (\ref{133}) can be computed in $2^{k_0}$ operations.

\chapter{Solution of the real-space RG equations with the \\ high-temperature expansion}\label{real_high_temp}

In this Appendix we show how the RG equations (\ref{160}) can be solved with a systematic expansion in powers of $\beta$, by illustrating an explicit example where this expansion is performed up to  order $\beta^{4}$. \\

Expanding in powers of $\beta$ the $\operatorname{arctanh}$ term in the right-hand side of Eq. (\ref{160}) we have

\be  \label{161}
\left\{ 
 \begin{array}{lcl}
m'_{2} & = &   \frac{C^4}{2} m_2 + \frac{C^4}{2}  \beta^2 (m_2)^2 + \frac{C^8}{8}  \beta^2( m_2)^2 -  \frac{C^8}{24}  \beta^2 m_4,\\
m'_4  & = &  \frac{3 C^8}{16} (m_2)^2 + \frac{C^8}{16} m_4 .
\end{array} 
\right. 
\ee\\

An important feature of Eq. (\ref{161}) is that it reproduces the fact that for $\sigma<1/2$ the thermodynamic limit is ill-defined, as we discussed in Part \ref{hea}, Eq. (\ref{sigma_hea}). In order to see this, let us look at the first line of Eq. (\ref{161}). The first addend in the right-hand side is the  $O(\beta^2)$-term resulting from the $\beta$-expansion of the term in square brackets  in the right-hand side of Eq. (\ref{160}), while the other addends are  $O(\beta^4)$-terms. Keeping only the first term and using Eq. (\ref{def_c}), we have $m'_{2}  =    2^{1-2 \sigma} m_2$ . Accordingly, if $\sigma < 1/2$ we have $m_{2}' > m_2$, i. e. the variance of the coupling $\mathcal{J}$  increases at each RG step, in such a way  that  in the thermodynamic limit $k \rightarrow \infty$ the interaction energy diverges, and the model is ill-defined.  
On the other hand, Eq. (\ref{161}) does not reproduce the condition $\sigma < 1$.  This fact will emerge also in the numerical solution of the RG equations (\ref{159}), and it will be discussed in Section \ref{pert_hea}.\\

\begin{centering}
  \begin{figure}[htb] 
\centering
\includegraphics[width=10cm]{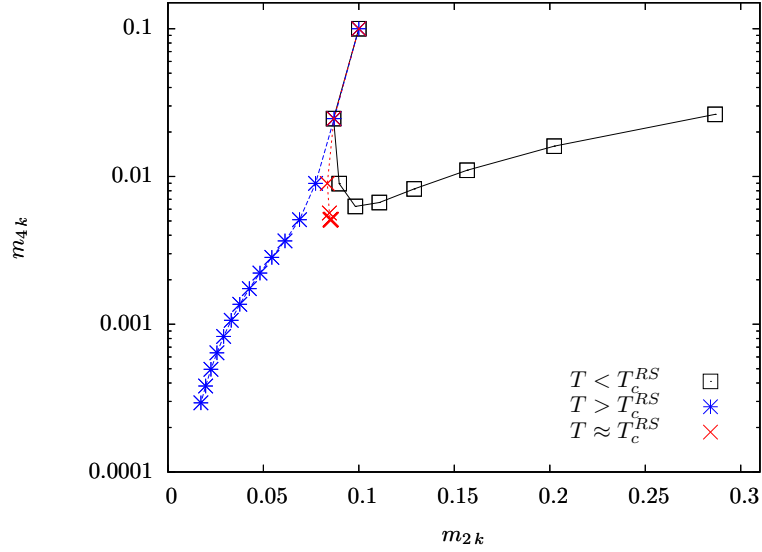}
\caption{Parametric plot of  $(m_{2\, k}, m_{4\, k})$ as a function of $k$ with $\sigma=0.6$. The black points represent $(m_{2\, k}, m_{4\, k})$ for increasing  $k$  from left  to right, with $T<T_c^{RS}$.  The blue points represent $(m_{2\, k}, m_{4\, k})$ for increasing  $k$  from top to bottom, with $T>T_c^{RS}$.  The red points represent $(m_{2\, k}, m_{4\, k})$ for increasing $k$ from top to bottom: here  $T$  has been dynamically adjusted  to $T_c^{RS}$ at each step $k$ with the same procedure as that described in the Caption of Fig \ref{fig11} for DHM.
}
\label{fig12}
\end{figure}
  \end{centering}

Eq. (\ref{161}) is formally analogous to Eq. (\ref{145}) for DHM, and to Eq. (\ref{56}), (\ref{80}), (\ref{81}) for the HEA. Accordingly, it is easy to see that Eq.  (\ref{161}) has a stable  high-temperature fixed point $m_2 = m_4 = 0$, and a stable low-temperature fixed point  $m_2= m_4 = \infty$. These fixed points are separated by an unstable critical fixed point $m_2 = m_2^\ast,m_4 = m_4^\ast$.
By iterating $k$ times Eq. (\ref{161}) we generate the sequence $m_{2\, k}, m_{4\, k}$, and we depict  the flow $ \{ m_{2\,k}, m_{4\, k} \}_k$ in Fig. \ref{fig12} for different values of the temperature.  Fig. \ref{fig12} shows that there is a value of the temperature $T_{c }^ {RS}$ such that for $T = T_c^{RS}$ $ \{ m_{2\,k}, m_{4\, k} \}_k$ converges to $m^\ast_2,m^\ast_4$, while for $T \gtrless T_c^{RS}$ $ \{ m_{2\,k}, m_{4\, k} \}_k$ converges to the high or low-temperature fixed point respectively. 

We now use the same procedure as that illustrated in Section \ref{app_field_dhm_2} to calculate  $m^\ast_2,m^\ast_4$, by taking  $\beta = \beta_{c}^{RS}$ in such a way that Eq. (\ref{161}) has a nontrivial fixed point $m_a = m'_a = m^\ast_a$
\be  \label{162}
\left\{ 
 \begin{array}{lcl}
m^\ast _{2} & = &   \frac{C^4}{2} m^{\ast}_2 + \frac{C^4}{2}  (\beta_c^{RS})^2 (m^{\ast}_2)^2 + \frac{C^8}{8}  (\beta_c^{RS})^2( m^{\ast}_2)^2 -  \frac{C^8}{24} ( \beta_c^{RS})^2 m^{\ast}_4,\\
m^{\ast}_4  & = &  \frac{3 C^8}{16} (m^{\ast}_2)^2 + \frac{C^8}{16} m^{\ast}_4 .
\end{array} 
\right. 
\ee

The second line in Eq. (\ref{162}) yields 
\be \label{164} 
m^{\ast}_4 = 3 C^8  (m^{\ast}_2)^2/[16(1-C^8/16)],
\ee
and by plugging Eq. (\ref{164}) into the first line of Eq. (\ref{162}) we obtain
\be \label{163}
m_2^\ast = \frac{C^4}{2} m_2^\ast \left\{ 1+\left[1+\frac{C^4}{4}+\frac{C^{12}}{2^6(-1+C^8/16)}\right] (\beta_c^{RS}) ^2 m^\ast_2\right\}. 
\ee
Eq. (\ref{163}) has a solution $m_2^\ast=0$ which is ruled out, and a nonzero solution $m_2^\ast \propto 1-C^4/2$. 
In the following we will compute this solution with an expansion in the neighborhood of $\sigma=1/2$. 
According to Eq. (\ref{def_c}) one has $ 1-C^4/2 = O(\sigma-1/2)$, thus we have $m_2^\ast = O(\sigma-1/2)$, and $m_4^\ast = O((\sigma-1/2)^2)$. More precisely, from Eqs. (\ref{163}), (\ref{164}) we have
\bea \label{167} \no 
m_2^\ast & = &\frac{3 \log 2 } {2 (\beta_{c} ^{RS}) ^2} (\sigma -1/2)+O((\sigma -1/2)^2),\\
m_4^\ast &= & \frac{9 (\log 2)^2}{4 (\beta_c^{RS})^4}(\sigma-1/2)^2+O((\sigma -1/2)^3). 
\eea\\

Once the critical fixed point has been found, we can linearize the RG transformation (\ref{161}) in the neighborhood of $m^\ast _2, m^\ast_4$  to extract the critical exponents. To this end, we introduce the $2 \times 2$ matrix
\be \label{165}
\textrm{M}_{ij} \equiv
 \left.
  \pder{m'_{2i}}{m_{2j}}  
\right|_{\vec m = \vec{m}^\ast}
\ee
and its largest eigenvalue $\Lambda_{RS}$. From Eqs. (\ref{165}), (\ref{161}), (\ref{167}) one finds
\be \label{166}
\Lambda_{RS} = 1 + 2 \log 2 \, (\sigma-1/2) + O((\sigma-1/2)^2). 
\ee\medskip\\

This high-temperature expansion  can be implemented to higher orders in $\beta$. More precisely, it turns out that if the expansion of the term in square brackets in Eq. (\ref{160}) is done up to order $\beta^{2m}$, one obtains a set of $m$ RG equations analogous to Eq. (\ref{161}), relating $\{ m_{2a} \}_{a=1, \ldots, m}$ to $\{ m'_{2a} \}_{a=1, \ldots, m}$. The critical fixed point and the matrix $\textrm{M}$ linearizing the RG transformation in its neighborhood are then extracted. The largest eigenvalue $\Lambda_{RS}$ of  $\textrm{M}$ can be computed as a power series in $\sigma-1/2$ up to  order $(\sigma-1/2)^{m-1}$. This computation has been done for $m\leq 5$ by means of a symbolic manipulation program \cite{wolfram1996mathematica}, and the result is given in Eq. (\ref{lambda_rs_high_temp}).

\chapter{Numerical discretization of the matrix $\mathscr{M}^{RS}$ in the \\ $k_0=2$-approximation} \label{app_discretization}

In this Section we describe the numerical computation of the matrix $\mathscr{M}^{RS}$ and of its spectrum through a discretization of the continuous variable $\mathcal J$. \\

Suppose that by iterating Routine \ref{bisection_routine} we computed the critical fixed point $p_\ast(\mathcal{J}) \leftrightarrow \{ \mathcal{J}_i \}_i$. As shown in Fig. \ref{fig13}, $p_\ast(\mathcal{J})$ has a compact support $[-\mathcal{J}_{\text{MAX}}, \mathcal{J}_{\text{MAX}}]$, where $\mathcal{J}_{\text{MAX}}$ is defined by Eq. (\ref{j_max}). This feature of the critical fixed point suggests a rather natural way to compute $\mathscr{M}^{RS}_{\mathcal{J}, \mathcal{J}'}$, based on a discretization of the continuous variable 
$\mathcal{J}$ in  the compact  interval $[-\mathcal{J}_{\text{MAX}}, \mathcal{J}_{\text{MAX}}]$. Let us consider 
\be \label{j_discrete}
\mathcal{J}(i)\equiv \left[\frac{1}{2}+(i-1)\right] \frac{2 \mathcal{J}_{\text{MAX}}}{B} -  \mathcal{J}_{\text{MAX}}, \, \forall i=1, \ldots, B, 
\ee
and the $B \times B$ matrix 
\be \label{m_discrete}
\textfrak{M}_{ij} \equiv \mathscr{M}^{RS}_{\mathcal{J}(i), \mathcal{J}(j) }. 
\ee
The matrix $\textfrak{M}$, and  so $\mathscr{M}^{RS}$,  can be easily computed numerically with a population-dynamics routine which is quite similar to Routine \ref{pop_dyn_routine}.\\

\begin{figure}
\begin{centering}
\includegraphics[width=12cm]{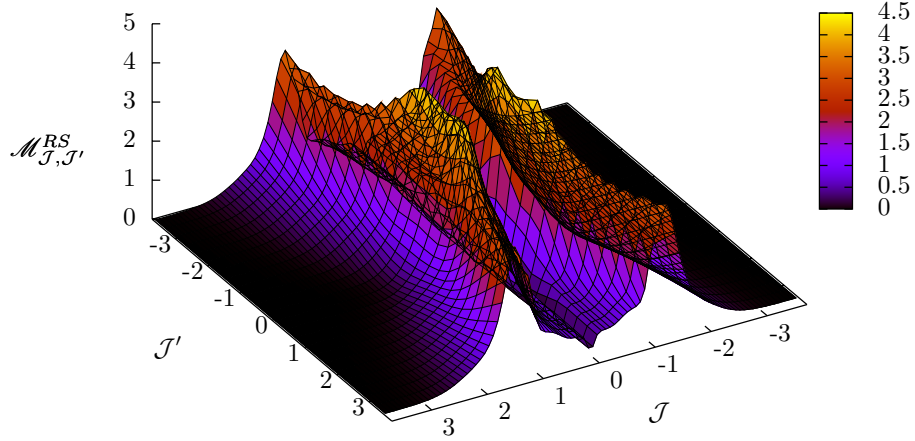}
\caption{Matrix $\mathscr{M}^{RS}_{\mathcal{J},\mathcal{J}'}$ as a function of $\mathcal{J},\mathcal{J}'$ in the $k_0=2$-approximation obtained with the discretization method (\ref{m_discrete}) with $\sigma = 0.6215$, $P=10^7, B=96, k_{\text{MAX}}=50,x=0.1$. 
The matrix is not symmetric, and thus some eigenvalues are complex. Notwithstanding this, the explicit numerical computation of the spectrum of $\mathscr{M}^{RS}$ shows that the eigenvalue yielding the critical exponent $\nu$ is real.}
\label{fig14}
\end{centering}
\end{figure}

In Fig. \ref{fig14} we depict $\mathscr{M}^{RS}_{\mathcal{J}, \mathcal{J}'}$ as a function of $\mathcal{J}, \mathcal{J}'$,  computed by means of Eq. (\ref{m_discrete})  for a given $\sigma$-value, and we show that by taking $B$ large enough the discretization method reconstructs a smooth function of $\mathcal{J}, \mathcal{J}'$. \\

 The eigenvalues of $\textfrak{M}$ can be  easily extracted numerically, and the  eigenvalues of $\mathscr{M}^{RS}$,  which are related to the critical exponents, can be easily obtained from those of $\textfrak{M}$ as follows. The $n$-th eigenvalue $\lambda^{(n)}$ and the left and right eigenfunctions $\phi^{L}_{n}(\mathcal{J}), \phi^{R}_{n}(\mathcal{J})$ of  $\mathscr{M}^{RS}$ are defined by
\bea \label{cont_right}
\int d \mathcal{J}' \mathscr{M}^{RS}_{\mathcal{J},\mathcal{J}'}  \phi^{R}_{n}(\mathcal{J}') &=& \lambda^{(n)}  \phi^{R}_{n}(\mathcal{J}),\\\label{cont_left}
\int d \mathcal{J} \phi^{L}_{n}(\mathcal{J}) \mathscr{M}^{RS}_{\mathcal{J},\mathcal{J}'}  &=& \lambda^{(n)}  \phi^{L}_{n}(\mathcal{J}'). 
\eea
 The $n$-th eigenvalue $\lambda^{(n)}_{D}$ and the left and right eigenvectors $\phi^{L}_{D\, n}(i), \phi^{R}_{D \, n}(i)$ of  $\textfrak{M}$ are defined by
\bea \label{discrete_right}
\sum_{j=1}^B \textfrak{M}_{ij} \phi^{R}_{D\, n}(j) &=& \lambda^{(n)}_{D}  \phi^{R}_{D\, n}(i),\\ \label{discrete_left}
\sum_{i=1}^B \phi^{R}_{L\, n}(i)  \textfrak{M}_{ij}  &=& \lambda^{(n)}_D  \phi^{L}_{D\, n}(j) . 
\eea
If we  multiply Eqs. (\ref{discrete_right}), (\ref{discrete_left}) by  $dJ \equiv  2 \mathcal{J}_{\text{MAX}}/B$, take the large-$B$ limit,  transform the sums in Eqs. (\ref{discrete_right}), (\ref{discrete_left}) into integrals and use the definition (\ref{m_discrete}),  by comparing   Eqs. (\ref{discrete_right}), (\ref{discrete_left}) to  Eqs. (\ref{cont_right}), (\ref{cont_left}) we obtain the following identifications holding in the $B \rightarrow \infty$-limit
\bea \label{lambda_lambda}
  \lambda^{(n)}_{D} \times dJ &=&  \lambda^{(n)},\\
 \phi^{R}_{D\, n}(i) & = &   \phi^{R}_{n}(\mathcal{J}(i)),\\
 \phi^{L}_{D\, n}(i) & = &   \phi^{L}_{n}(\mathcal{J}(i)).
\eea
 In particular, from  Eq. (\ref{lambda_lambda}) we can extract the eigenvalues of  $\mathscr{M}^{RS}$ from those of $\textfrak{M}$. \\

In order to extract the critical exponents, one should observe that   there is an eigenvalue, that we will call $\lambda^{(1)}$, which can be calculated analytically and which does not contribute to $\nu$ even though it is part of the spectrum of $\mathscr{M}^{RS}$. Indeed, by multiplying Eq.  (\ref{matrix_2}) by $p_\ast(\mathcal{J'})$, integrating with respect to $\mathcal{J}'$ and using Eq. (\ref{p_ast}),  we obtain 
\be \label{right_lambda_0}
\int d \mathcal{J}' \mathscr{M}^{RS}_{\mathcal{J}, \mathcal{J}'} p_\ast(\mathcal{J}') = 6 p_\ast(\mathcal{J}), 
\ee
while by multiplying Eq.  (\ref{matrix_2}) by a constant $A$  and integrating with respect to $\mathcal{J}$ we have
\be \label{left_lambda_0}
\int d \mathcal{J} A \, \mathscr{M}^{RS}_{\mathcal{J}, \mathcal{J}'}  = 6 A.
\ee
Comparing Eqs. (\ref{right_lambda_0}), (\ref{left_lambda_0}) to Eqs. (\ref{cont_right}), (\ref{cont_left}) we have 
\bea \label{lambda_1} \no 
\lambda^{(1)}&=&6,\\  \no 
 \phi^{R}_{1}(\mathcal{J}) &=& p_\ast(\mathcal{J}),\\ 
 \phi^{L}_{1}(\mathcal{J}) &=& A.
\eea
 Following the very same procedure as Wilson's \cite{wilson1974renormalization}, if we iterate the RG equations  (\ref{159}) $k$ times for $T \approx T_c^{RS}$ and then iterate $l$ times, the difference between $p_{k+l}$ and $p_\ast$ is given by 
\bea \label{departure} \no 
p_{k+l}(\mathcal{J}) - p_\ast(\mathcal{J}) & = & \int d \mathcal{J}' [(\mathscr{M}^{RS}) ^l]_{\mathcal{J}, \mathcal{J}'} [p_k(\mathcal{J}') - p_\ast(\mathcal{J}')] \\
& = &  \int d \mathcal{J}' \sum_{n} (\lambda^{(n)})^l  \phi^{R}_{n}(\mathcal{J})  \phi^{L}_{n}(\mathcal{J}') [p_k(\mathcal{J}') - p_\ast(\mathcal{J}')],
\eea
where $[(\mathscr{M}^{RS}) ^l]_{\mathcal{J}, \mathcal{J}'}$ is the $\mathcal{J}, \mathcal{J}'$-th component of the matrix $(\mathscr{M}^{RS}) ^l$, and in the second line of Eq. (\ref{departure}) the spectral representation of $\mathscr{M}^{RS}$ has been used. According to the third line of  Eq. (\ref{lambda_1}), the integral in the second line of Eq. (\ref{departure}) vanishes for $n=1$ because of the normalization condition $\int d \mathcal{J}' p_k(\mathcal {J}') = \int d \mathcal{J} 'p_\ast (\mathcal{J}') = 1$. Hence, the eigenvalue $\lambda^{(1)}$ does not contribute either to the exponential divergence of $p_{k+l}$ from $p_\ast$ nor  to $\nu$ \cite{wilson1974renormalization}. This fact is rather natural, because the eigenvalue $\lambda^{(1)}=6$ is an artefact of the $k_0=2$-approximation, where a $2^2$-spin HEA model with exactly six couplings $\{ \mathcal{J}_\alpha\}_\alpha$ is reduced to a $2$-spin HEA. This fact will be elucidated  further in Section \ref{k_0_larger_2} when illustrating the $k_0>2$-approximations. Indeed also for $k_0>2$ there is an eigenvalue $\lambda^{(1)} = 2^{k_0}(2^{k_0}-1)/2$ which depends explicitly on the approximation degree $k_0$, but which does not contribute to $\nu$. \\

The eigenvalue of $\mathscr{M}^{RS}$ determining $\nu$ is easily found by defining $n_\ast$ in such a way that 
\be \label{lambda_n_ast}
\lvert \lambda^{(n_\ast)} \rvert = \max_n (\lvert \lambda^{(n)} \rvert) \text{ and } \int d \mathcal{J} \phi^{L}_{n_\ast}(\mathcal J ) [p_k(\mathcal{J}) - p_\ast(\mathcal J)] \neq 0,  
\ee
and by observing that \cite{wilson1974renormalization} the critical exponent $\nu$ defined by Eq. (\ref{xi_hea})
 is given by 
\be \label{nu_lambda_ast}
\nu = \frac{\log 2} {\log \lambda^{(n_\ast)}}. 
\ee

\backmatter


\cleardoublepage
\bibliographystyle{sapthesis} 
\bibliography{bibliography} 

\end{document}